\pdfoutput=1
\documentclass[PAPER, cernpreprint, atlasdraft=false, coverpage=false, texlive=2020, UKenglish,  texmf, orcidlogo]{atlasdoc}
 
\usepackage{atlaspackage}
\usepackage{atlasbiblatex}
 
\usepackage[process,jetetmiss,other]{atlasphysics}

\graphicspath{{logos/}{figures/}}
 
\usepackage{ANA-EGAM-2019-01-PAPER-defs}
\usepackage{grffile}

\AtlasTitle{Tools for estimating fake/non-prompt lepton backgrounds with the ATLAS detector at the LHC}
 
\AtlasVersion{2.5}
 
\AtlasAbstract{
Measurements and searches performed with the ATLAS detector at the CERN LHC often involve signatures with one or more prompt leptons. Such analyses
are subject to `fake/non-prompt' lepton backgrounds, where either a hadron or a lepton from a hadron decay or an electron from a photon conversion satisfies the prompt-lepton
selection criteria. These backgrounds often arise within a hadronic jet because of particle decays in the showering process, particle misidentification or particle
interactions with the detector material. As it is challenging to model these processes with high accuracy in simulation, their estimation typically uses data-driven methods.
Three methods for carrying out this estimation are described, along with their implementation in ATLAS and their performance.
}
 
\author{The ATLAS Collaboration}
 
\AtlasRefCode{EGAM-2019-01}
 
\PreprintIdNumber{CERN-EP-2022-214}
 
\AtlasJournalRef{JINST 18 (2023) T11004}
\AtlasDOI{10.1088/1748-0221/18/11/T11004}

\AtlasDate{\today}

\AtlasJournal{JINST}

\hypersetup{pdftitle={ATLAS document},pdfauthor={The ATLAS Collaboration}}
 
\begin{document}
 
\maketitle
\tableofcontents
\clearpage

\section{Introduction}

Many measurements and searches for new physics performed with the ATLAS detector~\cite{PERF-2007-01} at the Large Hadron Collider~(LHC) at CERN require the presence of one or more
leptons~(electrons or muons) to indicate that a high-energy electroweak process occurred in the collision. Lepton candidates are reconstructed from signals in the inner tracker,
calorimeters, and muon spectrometer~\cite{EGAM-2018-01,MUON-2018-03}. Identification criteria are then applied to suppress candidates originating from physical objects other than
leptons from the hard-scattering process of the event. These background candidates fall into two categories: $i$)~\enquote{non-prompt leptons} from the semileptonic decay of hadrons,
or from photon~(\gam) conversions in detector material, and $ii$)~\enquote{fake leptons} where the reconstructed object is not, in fact, due to a lepton. In contrast to the
aforementioned categories, \enquote{real leptons} are defined as electrons or muons produced either directly in the hard-scattering process
or directly in the decay of a short-lived non-hadronic resonance (such as a \Wboson/\Zboson~boson).
 
The rates at which fake or non-prompt leptons are selected are difficult to model accurately from simulation. They can depend strongly on details of the physics simulation, including
in non-perturbative regions where the simulation would not be expected to be reliable. They also depend on the modelling of the material composition and response of the detector.
In addition, fake leptons sometimes occur with low probability as a result of misidentifying a hadronic jet in multi-jet events. The computing resources required to simulate these
processes with a sufficient sample size would be prohibitive. Therefore, \enquote{data-driven} approaches are commonly used to estimate these backgrounds.
 
To simplify the adoption of such methods, and to ensure that they are applied uniformly, a set of standard tools and prescriptions has been developed for use in ATLAS
physics analyses that are subject to fake/non-prompt lepton backgrounds. The principles and performance of these tools are described in this paper. The motivation and mathematical
basis of the methods are explained in Section~\ref{sec:methods}; a description of the relevant features of the ATLAS detector is given in Section~\ref{sec:ATLAS}; the criteria used
to select leptons are given in Section~\ref{sec:lepton_sel}; the simulated signal and background processes, as well as the different processes that can lead to fake/non-prompt leptons,
are discussed in Sections~\ref{sec:samples} and~\ref{sec:fake_sources}; the procedures used to measure the efficiencies for real and fake/non-prompt leptons are described in
Section~\ref{sec:eff_meas}; the systematic uncertainties associated with the methods are described in Section~\ref{sec:systematics}; and Section~\ref{sec:applications} provides
examples of the application of the fake/non-prompt lepton estimation methods in two published ATLAS physics analyses, with details that are not included in the existing publications.

\section{Methods}
\label{sec:methods}

The fake/non-prompt lepton estimation methods considered in this paper depend on defining two tiers of lepton selection criteria, called the \enquote{baseline} and \enquote{tight} criteria.
The tight criteria are used to select the signal leptons in a physics analysis,
while the baseline criteria accept all of the tight lepton candidates as well as an additional set of candidates with a
higher rate of fake/non-prompt contributions. Candidates that satisfy the baseline criteria but not the tight criteria are called \enquote{loose} leptons. If the two sets of criteria are
chosen well, the fraction of real leptons in the baseline sample that satisfy the tight criteria will be substantially higher than the corresponding fraction for fake/non-prompt leptons. These
fractions are called the \enquote{real efficiency}~($\varepsilon_\mathrm{r}$) and \enquote{fake efficiency}~($\varepsilon_\mathrm{f}$), respectively.
 
The $\varepsilon_\mathrm{r}$ values are generally taken from Monte Carlo~(MC) simulated events that are corrected to account for differences between data and the simulation, while the $\varepsilon_\mathrm{f}$ values
are typically measured in a data sample
consisting of different events from the ones that are used for the data analysis, as detailed in Section~\ref{sec:eff_meas}.
 
\subsection{Matrix method}
With the efficiencies known, a simple counting of the numbers of lepton candidates that satisfy the tight and loose criteria provides an estimate for the number of fake/non-prompt
leptons. In the simplest case, where an analysis selects signal events containing exactly one tight
lepton candidate  and no loose lepton candidates, the relationship between the numbers of tight and loose leptons observed in data and the composition of the
sample in terms of real and fake/non-prompt leptons is:
 
\begin{equation}
\label{eq:mm}
\left({\begin{array}{*{20}{c}}{{N^\mathrm{t}}}\\
{{N^\mathrm{l}}}\\
\end{array}} \right) = \left({\begin{array}{*{20}{c}}
\varepsilon_\mathrm{r} & \varepsilon_\mathrm{f} \\
{1-\varepsilon_\mathrm{r}} &{1-\varepsilon_\mathrm{f}} \\
\end{array}} \right)\left({\begin{array}{*{20}{c}}{N_\mathrm{r}^\mathrm{b}}\\
{N_\mathrm{f}^\mathrm{b}}
\end{array}} \right),
\end{equation}
where $N^\mathrm{t}$ and $N^\mathrm{l}$ are the numbers of events with tight and loose lepton candidates, and $N_\mathrm{r}^\mathrm{b}$ and $N_\mathrm{f}^\mathrm{b}$ are the unknown numbers of real and
fake/non-prompt leptons in the baseline sample. In matrix notation, the relationship is given by:
\begin{equation}
\label{eq:mm_matrix_form}
{\mathbf N}_\mathrm{tl} = {\mathbf M}_{\varepsilon}{\mathbf N}_\mathrm{rf}^\mathrm{b}.
\end{equation}
 
The fact that the unknown values ($N_\mathrm{r}^\mathrm{b}$ and $N_\mathrm{f}^\mathrm{b}$) and the observed yields ($N^\mathrm{t}$ and $N^\mathrm{l}$)  are related via the
matrix ${\mathbf M}_{\varepsilon}$ gives rise to the name of this method: the \enquote{matrix method}\cite{Abbott:1999tt}. Inversion of the matrix allows $N_\mathrm{f}^\mathrm{b}$ to be determined:
 
\begin{equation}
\label{eq:nfakes_b}
N_\mathrm{f}^\mathrm{b} = \frac{1}{ \varepsilon_\mathrm{r}-\varepsilon_\mathrm{f}}\left[ (\varepsilon_\mathrm{r}-1)N^\mathrm{t} + \varepsilon_\mathrm{r}N^\mathrm{l} \right].
\end{equation}

In the typical use case, the quantity of interest is the number of events in the tight sample where the lepton is fake/non-prompt, $N_\mathrm{f}^\mathrm{t}$.  This is related to the number of such events in the baseline sample, $N_\mathrm{f}^\mathrm{b}$, by:
 
\begin{equation}
\label{eq:nfakes_t_1lep}
N_\mathrm{f}^\mathrm{t} = \varepsilon_\mathrm{f}N_\mathrm{f}^\mathrm{b}.
\end{equation}
Similarly, the number of real leptons in the tight sample is
\begin{equation}
\label{eq:nreal_t_1lep}
N_\mathrm{r}^\mathrm{t} = \varepsilon_\mathrm{r}N_\mathrm{r}^\mathrm{b},
\end{equation}
and these can be treated as elements of a column matrix ${\mathbf N}_\mathrm{rf}^\mathrm{t}$.
 
The fact that $N^\mathrm{t}$ appears in Eq.~(\ref{eq:mm}) means that information about the content of the analysis signal region is used in the estimate, and therefore an analysis is not completely blinded when using this approach.
 
Generally, the values of $\varepsilon_\mathrm{r}$ and $\varepsilon_\mathrm{f}$ depend on factors such as the lepton candidate's momentum or its proximity to other objects.  Details of how these variations are accounted for in the estimation are given below.

\subsubsection{Asymptotic matrix method}
 
In this method, events in the baseline sample are considered one at a time, and a \enquote{fake weight} $w_i$ is defined for each event, corresponding to the two terms in Eq.~(\ref{eq:nfakes_b}) via Eq.~(\ref{eq:nfakes_t_1lep}):
 
\begin{equation*}
w_i = \left\{{\begin{array}{*{20}{c}}{\frac{\varepsilon_{\mathrm{f}i}}{ \varepsilon_{\mathrm{r}i}-\varepsilon_{\mathrm{f}i}}(\varepsilon_{\mathrm{r}i} -1)} & {\hbox{if lepton candidate $i$ is tight (so $N^\mathrm{t} = 1$ and $N^\mathrm{l}=0$)}} \\[5pt]
{\frac{\varepsilon_{\mathrm{f}i}}{ \varepsilon_{\mathrm{r}i}-\varepsilon_{\mathrm{f}i}}}\varepsilon_{\mathrm{r}i} & { \hbox{if lepton candidate $i$ is loose (so $N^\mathrm{t} = 0$ and $N^\mathrm{l}=1$)} }
\end{array}} ,  \right.
\end{equation*}
where $\varepsilon_{\mathrm{r}i}$ and $\varepsilon_{\mathrm{f}i}$ are the values of $\varepsilon_\mathrm{r}$ and $\varepsilon_\mathrm{f}$ that are appropriate for lepton $i$. Since $\varepsilon_{\mathrm{r}i} $ is always less than one, the weight for an event with a tight lepton candidate is negative. By extension, the total fake/non-prompt lepton background in the tight sample is then estimated by
\begin{equation*}
N_\mathrm{f}^\mathrm{t} = \sum_{\mathrm{events}} w_{i} .
\end{equation*}
This approach is convenient since the $w_{i}$ need only be calculated once and then can be stored with the event, allowing the distribution of the fake-lepton yield to be
binned in any variable of interest, as well as a simple re-computation of $N_{\mathrm{f}}^{\mathrm{t}}$ if the event selection is modified. One drawback is that since the value of $w_i$ may be negative, there is no guarantee that $N_\mathrm{f}^\mathrm{t}$ will be positive.  The value of $N_\mathrm{f}^\mathrm{t}$ is also sensitive to fluctuations in the input $\varepsilon_{\mathrm{r}i}$ and $\varepsilon_{\mathrm{f}i}$ values.
 
The statistical uncertainty of
$N_{\mathrm{f}}^{\mathrm{t}}$ is given by:
\begin{equation}
\sigma_{N_\mathrm{f}^\mathrm{t}} = \sqrt{\sum_{\mathrm{events}} w^2_i} .
\label{eq:per_event_uncertainty}
\end{equation}
The method that makes use of the $w_{i}$ is known as the \enquote{asymptotic matrix method}, since Eq.~(\ref{eq:per_event_uncertainty}) is only valid in the asymptotic limit
with a large number of events.
 
\subsubsection{Poisson likelihood matrix method}
\label{subsubsec:lhMM}

In this method~\cite{Erdmann:2021pqi}\footnote{An earlier variant of the likelihood matrix method is described in Ref.~\cite{Gillam:2014xua}.}  the elements of ${\mathbf N}_\mathrm{rf}^\mathrm{t}$ are treated as free parameters, which are varied to maximise the likelihood of the observed ${\mathbf N}_\mathrm{tl}$ values. By doing so, the Poisson-distributed nature of the ${\mathbf N}_\mathrm{tl}$ values is taken into account, so there is no need to use the asymptotic approximation.  In the fit, the ${\mathbf N}^\mathrm{t}_\mathrm{rf}$ values are converted to ${\mathbf N}^\mathrm{b}_\mathrm{rf}$ using Eqs.~(\ref{eq:nfakes_t_1lep}) and~(\ref{eq:nreal_t_1lep}), where the entries in the matrix ${\mathbf M}_\varepsilon$ are calculated using the averages of the prompt and fake/non-prompt lepton efficiencies in the baseline sample:

\begin{equation}{\mathbf M}_\varepsilon = \left({\begin{array}{*{20}{c}}
\langle \varepsilon_\mathrm{r} \rangle  & \langle \varepsilon_\mathrm{f} \rangle  \\
{1-\langle \varepsilon_\mathrm{r} \rangle } &{1-\langle \varepsilon_\mathrm{f} \rangle } \\
\end{array}} \right).
\label{eq:avg_eff_matrix}
\end{equation}
 
The resulting ${\mathbf N}^\mathrm{b}_\mathrm{rf}$ values are used to obtain the expectation values for ${\mathbf N}_\mathrm{tl}$ using Eq.~(\ref{eq:mm_matrix_form}). These expectation values are denoted by ${\mathbf N}_{\mathrm{tl},\mathrm{exp}}$. The ${\mathbf N}_\mathrm{rf}^\mathrm{t}$ parameters are adjusted (subject to the constraint that they must be non-negative) to maximise the joint Poisson likelihood
 
\begin{equation}
\label{eq:lhood}
L\left({\mathbf N}^\mathrm{t}_\mathrm{rf}\right)  \equiv \prod_i P\left[N_{\mathrm{tl}_i} | N_{\mathrm{tl},\mathrm{exp}_i}\left({\mathbf N}^\mathrm{t}_\mathrm{rf}\right)\right],
\end{equation}
 
where the product is over the elements of  ${\mathbf N}_\mathrm{tl}$; $N_{\mathrm{tl}_i}$ and $N_{\mathrm{tl},\mathrm{exp}_i}$ are the $i$th elements of ${\mathbf N}_\mathrm{tl}$ and ${\mathbf N}_{\mathrm{tl},\mathrm{exp}}$, respectively; and $P\left[n | \mu \right]$ is the Poisson probability for observing $n$ events when $\mu$ are expected. The output of the fit consists of an estimate of the number of fake/non-prompt leptons and the uncertainty in this quantity, which is obtained by noting the value for which $-\ln L$ exceeds its minimum by \num{0.5}.
 
The primary advantages of the Poisson likelihood approach are that the result is constrained to be non-negative, and the uncertainty is a better approximation to the range that
gives 68\% coverage, particularly in samples with few events. In addition, in some scenarios it can provide smaller statistical uncertainties than the asymptotic matrix method or the fake-factor method (described in Section~\ref{sec:ff}). The main drawback is that the estimated
yield must be calculated for the sample as a whole rather than from a sum of individual event weights, which complicates the process of producing a distribution of the fake-lepton
yield, as required for any differential measurement (to do so, the likelihood must be applied in every bin of the distribution).

\subsection{Fake-factor method}
\label{sec:ff}
The fact that real-lepton kinematic distributions and efficiencies are generally modelled well in simulation, and that scale factors can be applied to account for any differences observed between the values in simulation and in data control samples, leads to an alternative method that uses simulation, rather than the data, to measure the real-lepton contribution to the loose lepton sample.
 
The number of fake-lepton events in the loose sample is
 
\begin{equation*}
\nLF = \nL - \nLR = (1-\varepsilon_\mathrm{f})\nFB
\end{equation*}
 
and the number of fake-lepton events in the tight sample is
 
\begin{equation*}
\nTF = \varepsilon_\mathrm{f} \nFB = F(\nL - \nLR),
\end{equation*}
 
where the \enquote{fake factor}~$F$ is defined as $F \equiv \varepsilon_\mathrm{f}/(1 - \varepsilon_\mathrm{f})$. Thus, in the fake-factor method, the number of tight fake/non-prompt
leptons for a given analysis can be computed using the fake factor $F$, the total number of loose lepton candidates $\nL$, and the number of real leptons in the loose lepton sample $\nLR$, where the latter quantity can be estimated using MC simulated samples, and its contribution subsequently subtracted from the quantity $F\nL$ observed in the data. In practice, the calculation is performed on an event-by-event basis to account for potential variations in $F$ due to properties of the lepton:
 
\begin{equation*}
\nTF = \sum_{\mathrm{data}, i=1}^{N^\mathrm{l}}F_i - \sum_{\mathrm{MC}, j=1}^{N^\mathrm{l}_{\mathrm{MC}} }w_{\mathrm{MC}j}F_j,
\end{equation*}
 
where $F_i$ is the fake factor appropriate for lepton $i$, all sources of prompt leptons are considered in the sum over MC simulated events, $N^\mathrm{l}_{\mathrm{MC}}$ is the number of MC events in the loose sample, and $w_{\mathrm{MC}j}$ is the weight assigned to simulated event $j$, based on the cross-section of the simulated process and any corrections to the selection efficiency that may be needed to reflect the performance on data events.
 
The main advantage of the fake-factor method is that this result does not depend on $N^\mathrm{t}$, i.e.\ the yield in the
analysis signal region. This means that unlike the matrix method, the fake-factor method can be applied while remaining \enquote{blind} to the contents of the signal region.  However, the method does have some of the same drawbacks as the asymptotic matrix method, namely the possibility of $\nTF$ being negative, and sensitivity to fluctuations in the $\varepsilon_\mathrm{f}$ values.

\subsection{Generalisation for multi-lepton final states}
 
The above methods can be generalised to cases where multiple baseline lepton candidates are considered in each event.  For the matrix methods, this is done by increasing the dimensionality of ${\mathbf M_{\varepsilon}}$, ${\mathbf N}_\mathrm{tl}$, and ${\mathbf N}_\mathrm{rf}^\mathrm{(t,b)}$ to $2^{n_\mathrm{b}}$, where $n_\mathrm{b}$ is the number of baseline lepton candidates in each event. The estimated fake/non-prompt yield depends on the requirements of a particular analysis in three ways:
first, from the requirement placed on the desired number $n_\mathrm{t}$ of tight lepton candidates per event;
second, whether or not events with additional loose lepton candidates are vetoed;
and third, on the minimum number of fake/non-prompt leptons $m_\mathrm{f}$ defining the background to be evaluated with one of the data-driven methods.\footnote{An example of a case where $m_\mathrm{f}$ is greater than one would be a dilepton analysis where there are backgrounds from both $W(\ra \ell\nu)+j$ events where the jet forms a fake/non-prompt lepton, and dijet events where both jets form fake/non-prompt leptons.
The analysers may choose to estimate the first contribution from MC simulation, and use data-driven methods for the second contribution, and therefore setting $m_\mathrm{f} = 2$ for the data-driven approach is required to avoid double-counting.}
The consideration of $m_\mathrm{f}$ is reflected in the transition from the number of fake/non-prompt lepton events in the baseline sample to the number in the tight sample,
in a generalisation of Eq.~(\ref{eq:nfakes_t_1lep}):
\begin{equation}
N_\mathrm{f}^\mathrm{t} = \sum_i g_i\left( \varepsilon_\mathrm{r},\varepsilon_\mathrm{f} \right) N_{\mathrm{rf}i}^\mathrm{b}.
\label{eq:nfakes_t_general}
\end{equation}
 
Here, the sum is over all combinations of real and fake/non-prompt leptons that include at least $m_\mathrm{f}$ fake/non-prompt leptons, $g_{i}$ is a function of the real and fake/non-prompt efficiencies that will result in the required number of tight
lepton candidates for a given set of real and fake/non-prompt leptons, and $ N_{\mathrm{rf}i}^\mathrm{b} $ is the $i$th element of  $\mathbf{N_\mathrm{rf}^\mathrm{b}}$
 
To address the requirements on $n_\mathrm{t}$ and the possible presence of additional loose lepton candidates, the analysis must consider all events in the baseline sample with lepton multiplicities up to the sum of the allowed numbers
of tight and loose lepton candidates in the signal region.
 
As an example, consider the case where $n_\mathrm{b} = 2$. If an analysis selects signal events containing exactly two tight lepton candidates and no loose lepton
candidates, and the background with $m_\mathrm{f} \ge 1$ is being evaluated, then
\begin{equation*}
N_\mathrm{f}^\mathrm{t} = \varepsilon_{\mathrm{r}1}\varepsilon_{\mathrm{f}2}N_\mathrm{r_1f_2}^\mathrm{b} + \varepsilon_{\mathrm{f}1}\varepsilon_{\mathrm{r}2}N_\mathrm{f_1r_2}^\mathrm{b} + \varepsilon_{\mathrm{f}1}\varepsilon_{\mathrm{f}2}N_\mathrm{f_1f_2}^\mathrm{b},
\end{equation*}
where $N_\mathrm{r_1f_2}^\mathrm{b}$ is the number of events in the baseline sample where the first lepton candidate is real and the second is fake/non-prompt, and $N_\mathrm{f_1r_2}^\mathrm{b}$ and $N_\mathrm{f_1f_2}^\mathrm{b}$
are defined correspondingly.  The ordering of the lepton candidates is typically according to \pt, but the method does not depend on the ordering used.
 
Another example may be an analysis accepting dilepton events ($n_\mathrm{b}=2$)
with one tight lepton and another baseline (not necessarily tight) lepton.
If, again, the background being evaluated consists of events with at least one fake lepton
($m_\mathrm{f} \ge 1$),
the coefficients $g_i$ in Eq.~(\ref{eq:nfakes_t_general})
change such that the expression becomes:
\begin{equation*}
N_\mathrm{f}^\mathrm{t} =(\varepsilon_{\mathrm{r}1} + \varepsilon_{\mathrm{f}2} - \varepsilon_{\mathrm{r}1}\varepsilon_{\mathrm{f}2})N_\mathrm{r_1f_2}^\mathrm{b} + (\varepsilon_{\mathrm{f}1} + \varepsilon_{\mathrm{r}2} - \varepsilon_{\mathrm{f}1}\varepsilon_{\mathrm{r}2})N_\mathrm{f_1r_2}^\mathrm{b} + (\varepsilon_{\mathrm{f}1} + \varepsilon_{\mathrm{f}2} - \varepsilon_{\mathrm{f}1}\varepsilon_{\mathrm{f}2})N_\mathrm{f_1f_2}^\mathrm{b},
\end{equation*}
where the additional terms are needed to account for the additional ways that fake/non-prompt leptons might satisfy the signal selection (and the terms involving products of the efficiencies are subtracted to avoid double-counting).  For an analysis that imposes the same lepton candidate requirements  but only the background with $m_\mathrm{f} = 2$ is being evaluated, the expression is:
\begin{equation*}
N_\mathrm{f}^\mathrm{t} = (\varepsilon_{\mathrm{f}1} + \varepsilon_{\mathrm{f}2} - \varepsilon_{\mathrm{f}1}\varepsilon_{\mathrm{f}2})N_\mathrm{f_1f_2}^\mathrm{b}.
\end{equation*}

The methods can also be extended to cases where there are more than two levels of lepton selection criteria,
or distinct categories of fake/non-prompt leptons, as in Ref.~\cite{ATLAS-CONF-2010-073}.
 
As with the matrix method, the fake-factor method can be generalised to higher lepton candidate multiplicities. In the dilepton final state, the number of events with two tight lepton candidates, of which at least one
is fake/non-prompt, is:
 
\begin{equation}
\nTF =  \nTT - \nTTR = \\
F_{1}(\nLT-\nLTR) + F_{2}(\nTL-\nTLR) - F_{1}F_{2}(\nLL-\nLLR),
\label{eq:ff_2L}
\end{equation}
 
where \nTLgen is the number of events where the first lepton candidate is tight (loose) and the second lepton candidate is tight (loose), \nTLRgen is the contribution to \nTLgen from events where both lepton candidates are real leptons, and  $F_{1}$ and $F_{2}$ correspond to the fake factors associated with the first and second lepton candidate, respectively.
 
However, the algebraic simplification that leads to Eq.~(\ref{eq:ff_2L}), where the result depends simply on products of the fake factors and the observed tight and loose lepton
candidate yields, with a correction term that depends only on events where all the leptons are real, does not hold for all possible event selections nor all values of $m_\mathrm{f}$;
such a simplification is restricted to cases where the baseline and tight candidate lepton multiplicities are the same in all events, and where $m_\mathrm{f} = 1$.
 
\subsection{Use with weighted events}
 
In some cases, such as for self-consistency tests using simulated events, it may be advantageous to weight the events that are input to the fake/non-prompt background estimate. This is
straightforward for the fake-factor and asymptotic matrix methods, since the weight returned by the method for each event can be multiplied by the event weight $w_{\mathrm{evt}_i}$. For the
likelihood matrix method, this is handled by using the scaled Poisson distribution~\cite{Bohm:2013gla}, in which the values of $N_{\mathrm{tl},\mathrm{exp}_i}$ and $N_{\mathrm{tl},\mathrm{evt}_i}$ from
Eq.~(\ref{eq:lhood}) are scaled according to the event weights $w_{\mathrm{evt}_i}$:
 
\begin{equation*}
s_\mathrm{tl}  \equiv {\sum_{\mathrm{tl}, \mathrm{obs}} w_{\mathrm{evt}_i} \over \sum_{\mathrm{tl}, \mathrm{obs}} w_{\mathrm{evt}_i}^2},
\end{equation*}
so that the likelihood becomes
\begin{equation*}
L\left({\mathbf N}^\mathrm{t}_\mathrm{rf}\right)  = \prod_i P\left[N_{\mathrm{tl}_i}/s_\mathrm{tl} | N_{\mathrm{tl},\mathrm{exp}_i}\left({\mathbf N}^\mathrm{t}_\mathrm{rf}\right)/s_\mathrm{tl}\right]
\end{equation*}

\subsection{Performance studies}
\label{sec:perf}

MC simulations of experiments are utilised to assess the statistical performance of the methods.  These simulations consist of pseudoexperiments that mimic scenarios that might occur in an actual analysis.  Pseudoexperiments with sample sizes of \num{10} or \num{1000} dilepton events in the baseline sample are considered. The fraction of fake/non-prompt leptons in the baseline lepton sample is varied for each pseudoexperiment, with a uniform distribution between \num{0} and 100\%.  The values of $\varepsilon_\mathrm{r}$ and $\varepsilon_\mathrm{f}$ for each lepton are drawn from Gaussian distributions with specified means and widths~(limits are imposed such that the values are always between zero and one, and $\varepsilon_\mathrm{f}$ is always at least 10\% less than $\varepsilon_\mathrm{r}$).  Each lepton is randomly assigned as fake/non-prompt or real, in accord with the fraction of fake/non-prompt leptons assumed for the pseudoexperiment.  Then each lepton is judged to either meet or fail to meet the tight selection criteria, based on whether or not it is a real lepton and the values of $\varepsilon_\mathrm{r}$ and $\varepsilon_\mathrm{f}$ assigned to it.  The set of simulated leptons is input to the data-driven algorithms, and the estimated fake yield and its statistical uncertainty are determined for each pseudoexperiment.  These values are then compared with the expectation value for the number of fake/non-prompt lepton events in each pseudoexperiment, which is determined by the numbers of real and fake/non-prompt baseline leptons, the values of $\varepsilon_\mathrm{r}$ and $\varepsilon_\mathrm{f}$ for each lepton, and the value of $m_\mathrm{f}$ for the simulated analysis.  For example, in a dilepton sample with $m_\mathrm{f} \ge 1$, the expectation value is:
 
\begin{equation*}
\langle N_\mathrm{f}^\mathrm{t} \rangle = \sum_{\textrm{rf events}}\varepsilon_{\mathrm{r}_{1,i}}\varepsilon_{\mathrm{f}_{2,i}} + \sum_{\textrm{fr events}}\varepsilon_{\mathrm{f}_{1,i}}\varepsilon_{\mathrm{r}_{2,i}} + \sum_{\textrm{ff events}} \varepsilon_{\mathrm{f}_{1,i}}\varepsilon_{\mathrm{f}_{2,i}}.
\end{equation*}
 
For the Poisson likelihood matrix method,  the likelihood maximisation is implemented using the \textsc{minuit} function minimisation package~\cite{James:1975dr} to minimise the negative log likelihood. The interface to \textsc{minuit} is provided  by the TMinuit class in \textsc{root}~\cite{Brun:1997}.
 
The fake-factor method requires a two-step process, where the contribution from real-lepton events is subtracted in the second step.  In an actual physics analysis, this subtraction is done using MC simulation of the prompt lepton contribution.  For the simple MC simulation used here, the second step is modelled by running each pseudoexperiment a second time with the same parameters but a statistically independent sample of events, and running the fake-factor method only on the events that do not have fake/non-prompt leptons.  The second sample has ten times more events than the primary sample, to be consistent with the usual case where the MC simulation sample has a multiple of the number of events in the data sample. The result of this second run is then scaled down by a factor of ten and subtracted from the result when using the initial set of simulated events.

\begin{figure}[htbp]
\begin{center}
\subfloat[]{\label{fig:ratio_2lep_varrf_91_highstat}  \includegraphics[width=0.32\textwidth]{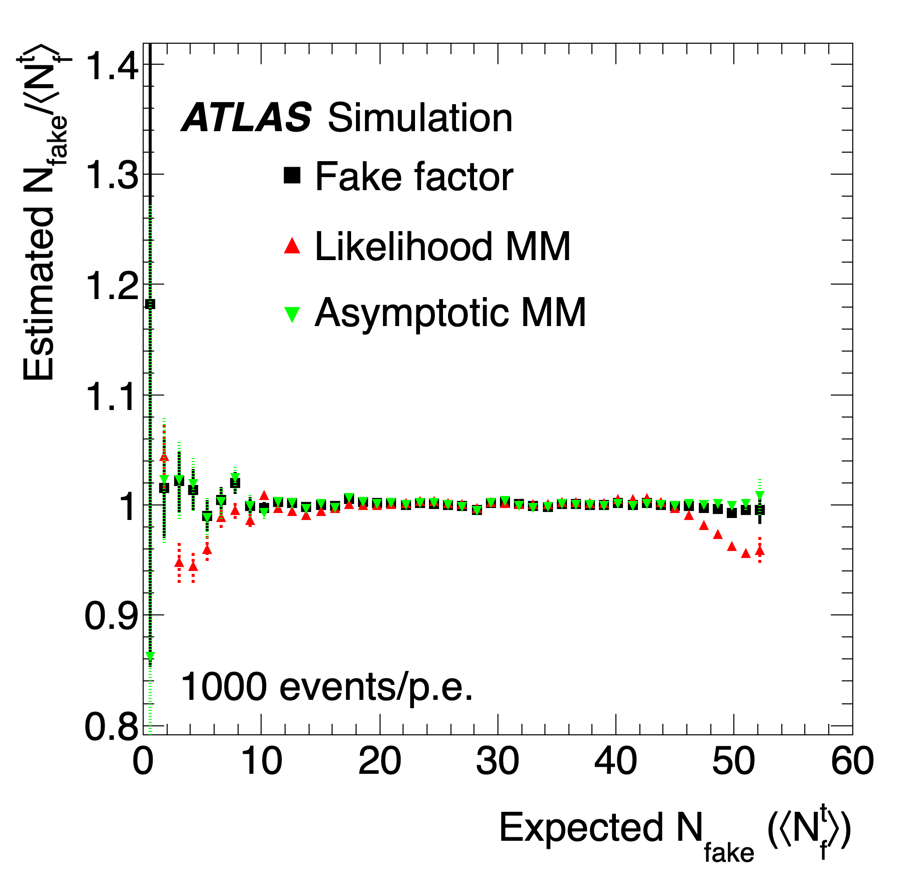} }
\subfloat[]{\label{fig:error_2lep_varrf_91_highstat}  \includegraphics[width=0.32\textwidth]{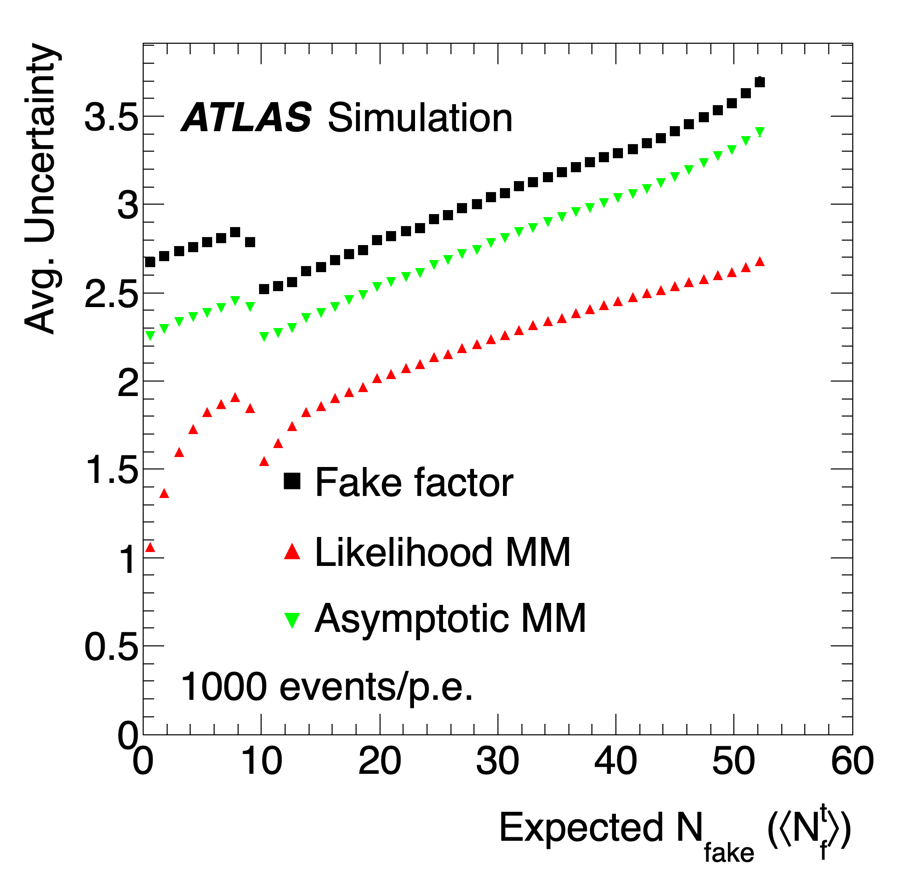} }
\subfloat[]{\label{fig:coverage_2lep_varrf_91_highstat}  \includegraphics[width=0.32\textwidth]{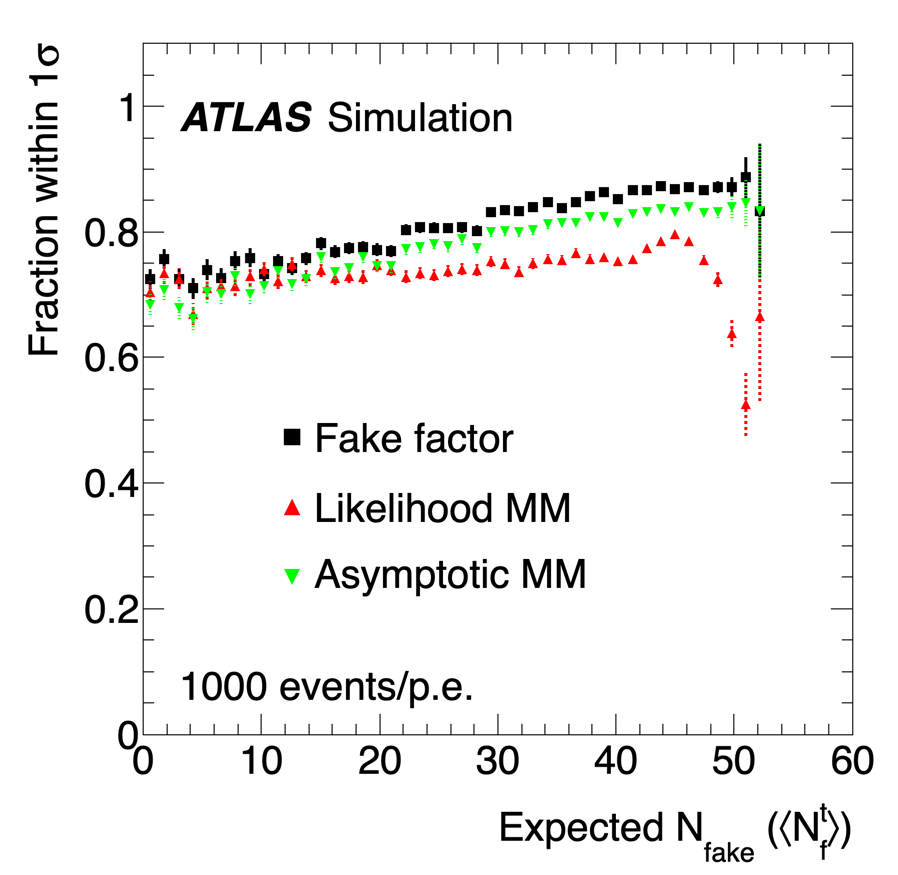} }
\caption{Performance of the three methods in pseudoexperiments with dilepton events where the leptons had an average $\varepsilon_\mathrm{r} = 0.9$ and $\varepsilon_\mathrm{f} = 0.1$, both values were varied according to a Gaussian distribution of width \num{0.1} when simulating each lepton, and there were \num{1000} events per pseudoexperiment. The quantity  $\langle N_\mathrm{f}^\mathrm{\
t} \rangle $ is the expectation value for the number of fake/non-prompt lepton events in each pseudoexperiment.  Plot (a)~shows the ratio of the estimated to expected fake-lepton yields, (b)~shows the absolute uncertainty estimate for each method~(for the Poisson likelihood method the average of the upward and downward uncertainties is taken), and (c)~shows the fraction of pseudoexperiments where the true fake yield lies within the reported one standard deviation ($1\sigma$) range.}
\label{fig:perf_2lep_varrf_91_highstat}
\end{center}
\end{figure}

\begin{figure}[htbp]
\begin{center}
\subfloat[]{ \includegraphics[width=0.32\textwidth]{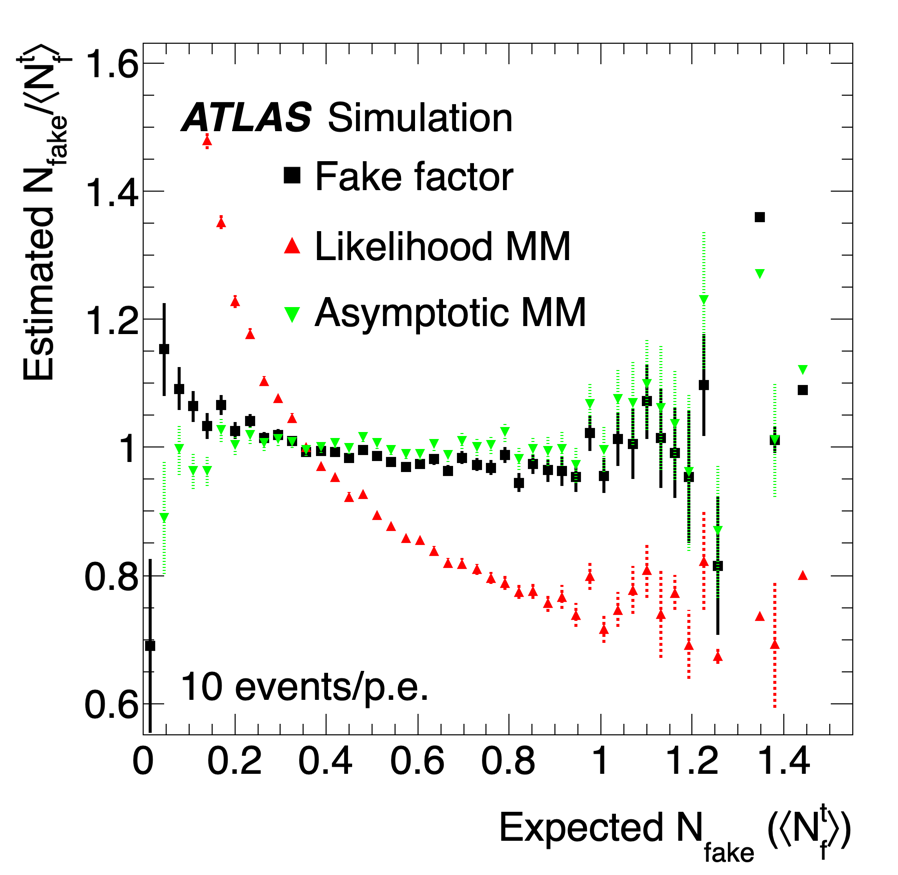} }
\subfloat[]{ \includegraphics[width=0.32\textwidth]{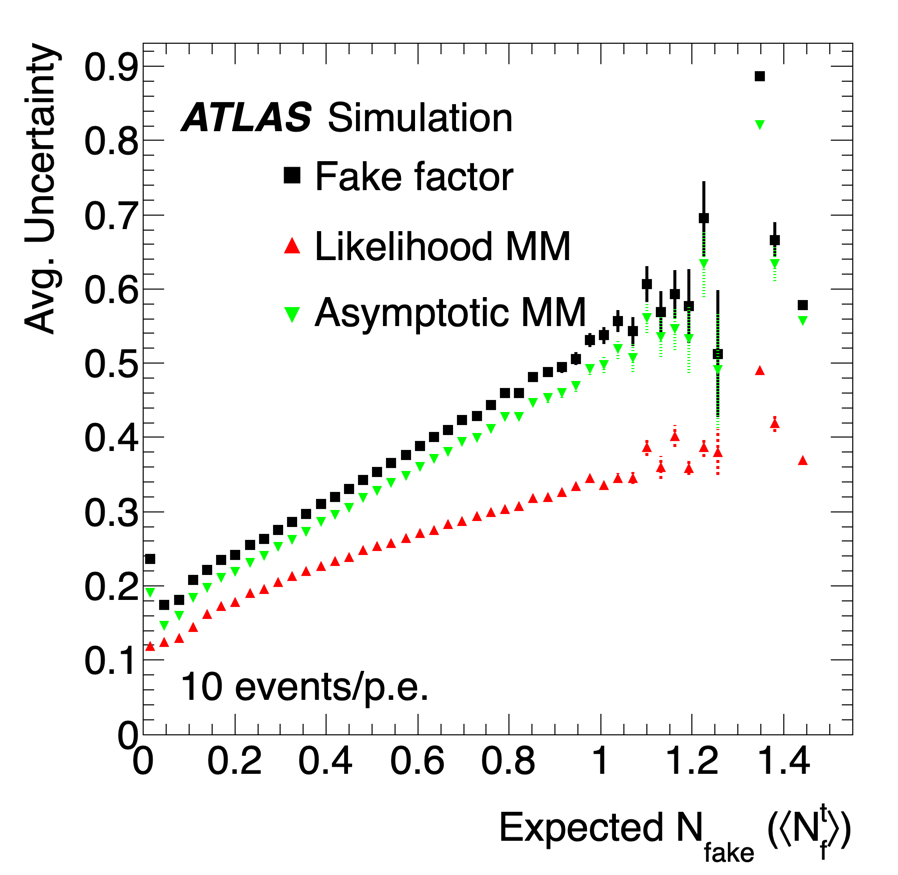} }
\subfloat[]{ \includegraphics[width=0.32\textwidth]{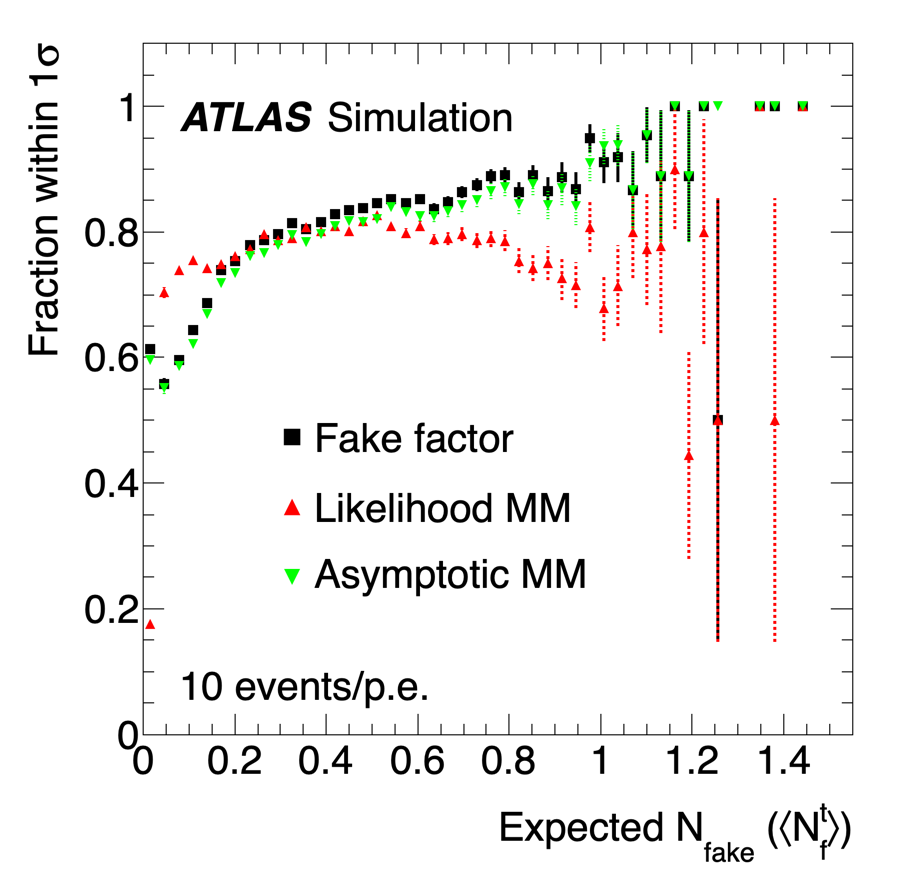} }
\caption{Performance of the three methods in pseudoexperiments with dilepton events where the leptons had an average $\varepsilon_\mathrm{r} = 0.9$ and $\varepsilon_\mathrm{f} = 0.1$, both values were varied according to a Gaussian distribution of width \num{0.1} when simulating each lepton, and there were \num{10} events per pseudoexperiment. The quantity  $\langle N_\mathrm{f}^\mathrm{t} \rangle $ is the expectation value for the number of fake/non-prompt lepton events in each pseudoexperiment. Plot (a)~shows the ratio of the estimated to expected fake-lepton yields, (b)~shows the absolute uncertainty estimate for each method~(for the Poisson likelihood method the average of the upward and downward uncertainties is taken), and (c)~shows the fraction of pseudoexperiments where the true fake yield lies within the reported one standard deviation ($1\sigma$) range.}
\label{fig:perf_2lep_varrf_91_lowstat}
\end{center}
\end{figure}

\begin{figure}[htbp]
\begin{center}
\subfloat[]{ \includegraphics[width=0.32\textwidth]{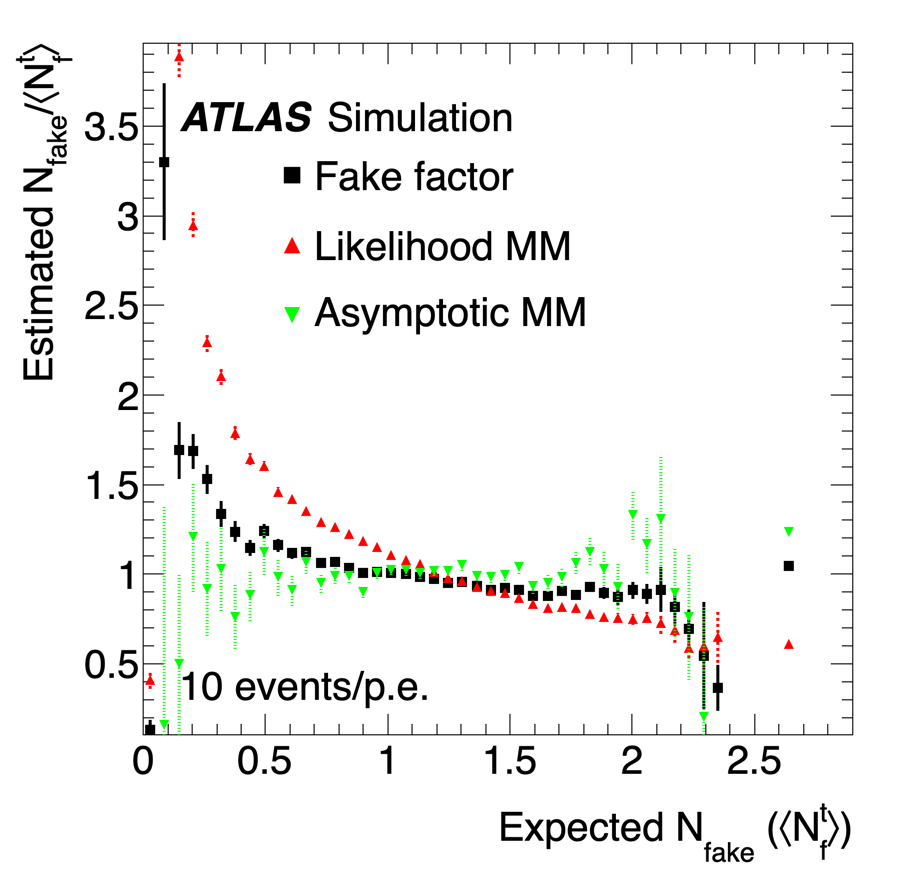} }
\subfloat[]{ \includegraphics[width=0.32\textwidth]{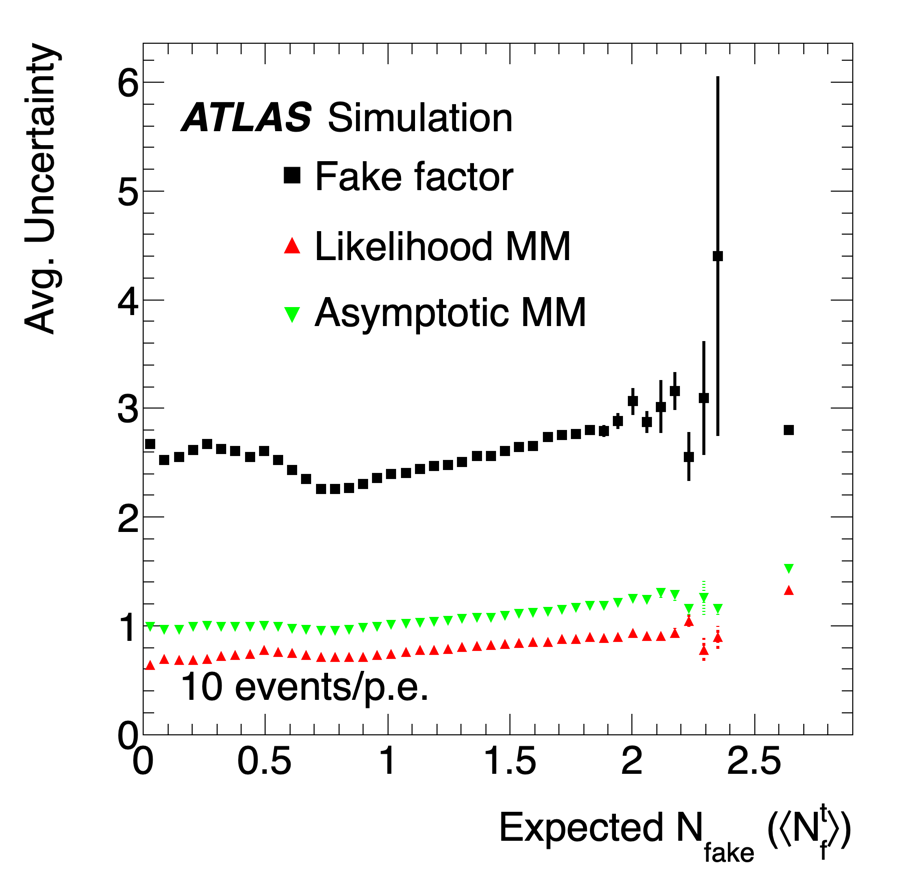} }
\subfloat[]{ \includegraphics[width=0.32\textwidth]{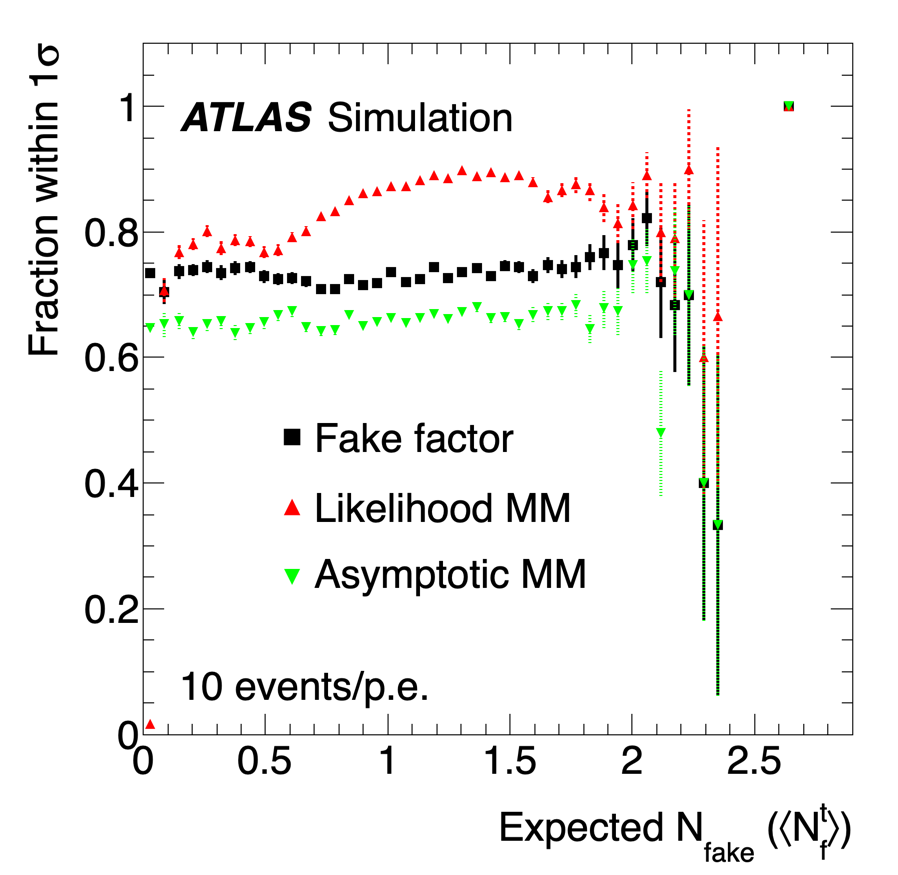} }
\caption{Performance of the three methods in pseudoexperiments with dilepton events where the leptons had an average $\varepsilon_\mathrm{r} = 0.7$ and $\varepsilon_\mathrm{f} = 0.3$, both values were varied according to a Gaussian distribution of width \num{0.1} when simulating each lepton, and there were \num{10} events per pseudoexperiment. The quantity  $\langle N_\mathrm{f}^\mathrm{t} \rangle $ is the expectation value for the number of fake/non-prompt lepton events in each pseudoexperiment. Plot (a)~shows the ratio of the estimated to expected fake-lepton yields, (b)~shows the absolute uncertainty estimate for each method~(for the Poisson likelihood method the average of the upward and downward uncertainties is taken), and (c)~shows the fraction of pseudoexperiments where the true fake yield lies within the reported one standard deviation ($1\sigma$) range.}
\label{fig:perf_2lep_varrf_73_lowstat}
\end{center}
\end{figure}

As an initial example, one can investigate the performance for dilepton events under conditions that are favourable for estimating the fake/non-prompt lepton background. This means that the samples are large~(\num{1000} events per pseudoexperiment), and the values of $\varepsilon_\mathrm{r}$ and $\varepsilon_\mathrm{f}$ are on average well-separated (here, $\langle\varepsilon_\mathrm{r}\rangle = 0.90$ and $\langle\varepsilon_\mathrm{f}\rangle = 0.10$).  The ratios of the estimated to true fake yields are shown versus $\langle N_\mathrm{f}^\mathrm{t} \rangle $ in Figure~\ref{fig:ratio_2lep_varrf_91_highstat}. The average statistical uncertainties in the estimates for each method are shown in Figure~\ref{fig:error_2lep_varrf_91_highstat}, and the fraction of pseudoexperiments in which the true fake yield lies within the uncertainty reported for each method is shown in Figure~\ref{fig:coverage_2lep_varrf_91_highstat}. The performance of the methods for dilepton analyses with low statistical precision~(\num{10} events per pseudoexperiment) are shown in Figure~\ref{fig:perf_2lep_varrf_91_lowstat}.
Finally, to represent a more challenging situation, the case where there is less separation between the values of $\varepsilon_\mathrm{r}$ and $\varepsilon_\mathrm{f}$~(due, for example,
to the application of stricter online lepton selection criteria that might be required when the LHC runs at higher luminosities) is also explored. Figure~\ref{fig:perf_2lep_varrf_73_lowstat}
shows the results when  $\langle\varepsilon_\mathrm{r}\rangle = 0.70$ and $\langle\varepsilon_\mathrm{f}\rangle = 0.30$, and there are \num{10} events per pseudoexperiment.

\begin{figure}[htbp]
\begin{center}
\includegraphics[width=0.50\textwidth]{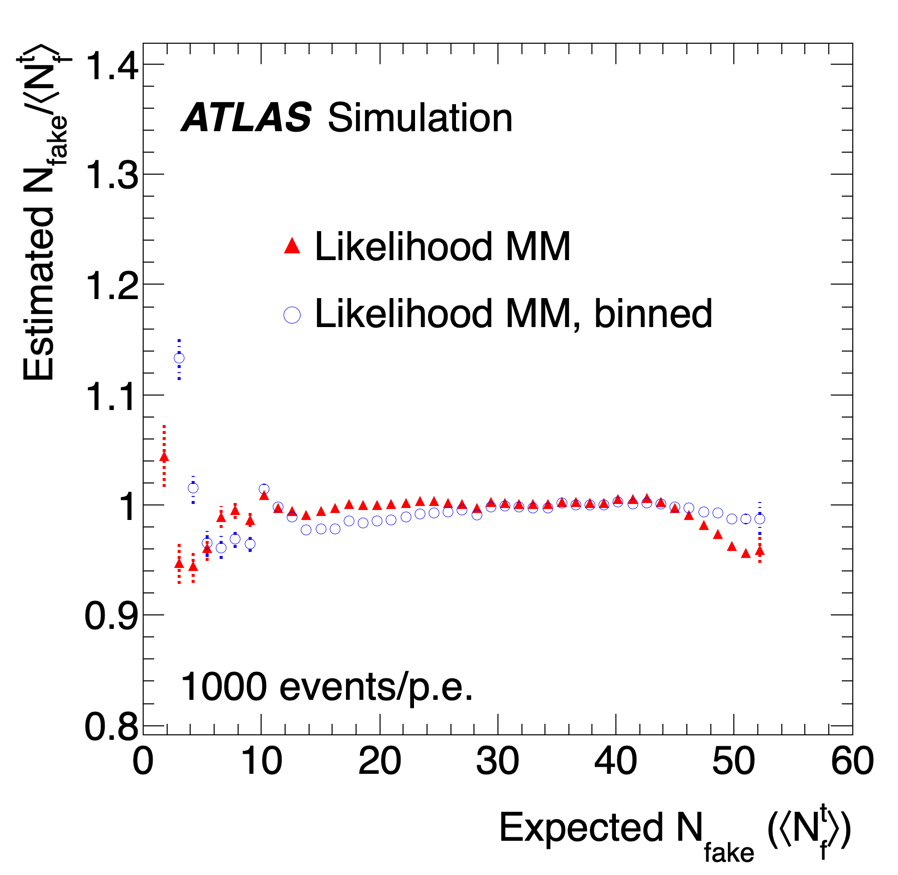}
\caption{Ratio of the estimated to expected fake-lepton yields for the Poisson likelihood matrix method in pseudoexperiments with dilepton events where the leptons had an average $\varepsilon_\mathrm{r} = 0.9$ and $\varepsilon_\mathrm{f} = 0.1$, both values were varied according to a Gaussian distribution of width \num{0.1} when simulating each lepton, and there were \num{1000} events per pseudoexperiment. The triangles show the  result with a single bin, while the circles show the result where the input events are binned according to the values of $\varepsilon_\mathrm{f}$ for each lepton.  Two bins are used for each of the two leptons, and the results from all four bins are summed to give the estimated $N_{\mathrm{fake}}$ value.}
\label{fig:perf_2lep_varrf_91_highstat_binned}
\end{center}
\end{figure}

These studies show that all three methods give accurate estimates, with nearly equivalent performance, in high-statistics samples with a large separation between $\varepsilon_\mathrm{r}$ and $\varepsilon_\mathrm{f}$ (as shown in Figure~\ref{fig:perf_2lep_varrf_91_highstat}).  One notable feature is the dip in the uncertainty values for all three methods near $\langle N_\mathrm{f}^\mathrm{t}\rangle = 10$ in Figure~\ref{fig:error_2lep_varrf_91_highstat}.  This occurs because there are two ways for the model to produce an expectation around \num{10}: either a very low fake fraction in the baseline to start with, or a very large fake fraction so that there are few real leptons and most of the background is from events with two fake/non-prompt leptons, which gives a minimum value of $\varepsilon_\mathrm{f}^{2} \cdot 1000 = 10$ when $\varepsilon_\mathrm{f} = 0.10$. These two processes will have very different uncertainties.   There is also a bias in the Poisson likelihood matrix method toward low values when the true number of fakes is large.  This bias arises due to the averaging of efficiencies over the entire baseline sample (see Eq.~(\ref{eq:avg_eff_matrix})) when in fact the efficiencies are on average different for real and fake leptons.  Such differences in the averages occur randomly in the `toy' MC tests, but can occur systematically in a physics analysis if, for example, the real and fake efficiencies have different kinematic distributions.  The biases can be mitigated by binning the baseline sample in the variables for which the real-lepton and fake-lepton distributions may differ, and performing the likelihood fit separately in each bin. As an example of the effect of such a binning, the pseudoexperiments can be run with the results binned according to the values of $\varepsilon_\mathrm{f}$.  The effect of using two such bins in the value of $\varepsilon_\mathrm{f}$ for each lepton is shown in Figure~\ref{fig:perf_2lep_varrf_91_highstat_binned}.
 
When the situation becomes more challenging, such as in Figures~\ref{fig:perf_2lep_varrf_91_lowstat} and~\ref{fig:perf_2lep_varrf_73_lowstat}, the characteristics of each method become more distinct.  For low-statistics samples, the Poisson likelihood approach tends to exhibit a bias toward high values when the true number of fakes is small, a natural consequence of the fact that it cannot return negative values. The coverage of the true value by the estimated uncertainty is, however, still reasonable.  A clear distinction between the precision of the methods also appears in Figures~\ref{fig:perf_2lep_varrf_91_lowstat} and~\ref{fig:perf_2lep_varrf_73_lowstat}, where the Poisson likelihood approach has the smallest statistical uncertainty, followed by the asymptotic matrix method and then the fake-factor method.  This is because the Poisson likelihood approach considers lepton efficiencies averaged over the entire sample and is therefore less susceptible to event-by-event fluctuations in the values of $\varepsilon_\mathrm{r}$ and $\varepsilon_\mathrm{f}$.
 
Despite the differences between them, all the approaches are valid, as shown by the statistical coverage plots: except in extreme cases, the confidence intervals built from the estimates and their statistical uncertainty do contain the true fake yield in at least 68\% of pseudoexperiments, and the visible overcoverage is due mostly to the fact that the uncertainties are computed under the assumption that all of the ${\mathbf N}_\mathrm{tl}$ values are independent, while the pseudoexperiments were generated with a fixed number of baseline events for each, which means there was some anticorrelation among the ${\mathbf N}_\mathrm{tl}$.
 
When selecting which method to use, the analyser needs to consider the relative benefits and complexities of implementing the methods, along with the size of the uncertainty in the fake/non-prompt lepton background yield relative to other uncertainties in the analysis.


\FloatBarrier

\section{The ATLAS detector}
\label{sec:ATLAS}

 
\newcommand{\AtlasCoordFootnote}{
ATLAS uses a right-handed coordinate system with its origin at the nominal interaction point~(IP) in the centre of the detector and the $z$-axis along the beam pipe.
The $x$-axis points from the IP to the centre of the LHC ring, and the $y$-axis points upwards. Cylindrical coordinates $(r,\phi)$ are used in the transverse plane,
$\phi$ being the azimuthal angle around the $z$-axis. The pseudorapidity is defined in terms of the polar angle $\theta$ as $\eta = -\ln \tan(\theta/2)$.
Angular distance is measured in units of $\DeltaR \equiv \sqrt{(\Delta\eta)^{2} + (\Delta\phi)^{2}}$.}
 
While the above description of the matrix and fake-factor methods is general, the remainder of this paper discusses the application of these methods to ATLAS
physics analyses, and therefore a brief description of the experimental apparatus follows.
 
The ATLAS detector~\cite{PERF-2007-01} at the LHC covers nearly the entire solid angle around the collision point.\footnote{\AtlasCoordFootnote}
It consists of an inner tracking detector surrounded by a thin superconducting solenoid, electromagnetic~(EM) and hadronic calorimeters, and a
muon spectrometer~(MS) incorporating three large superconducting toroidal magnets. The inner-detector system is immersed in a \SI{2}{\tesla} axial magnetic field
and provides charged-particle tracking in the range $\abseta < 2.5$.
 
The high-granularity silicon pixel detector covers the vertex region and typically provides four measurements per track, the first hit normally being in the
insertable B-layer~(IBL) installed before \RunTwo~\cite{ATLAS-TDR-2010-19,PIX-2018-001}. It is followed by the silicon microstrip tracker, which usually
provides eight measurements per track. These silicon detectors are complemented by the transition radiation tracker~(TRT), which enables radially extended track
reconstruction up to $\abseta = 2.0$. The TRT also provides electron identification information based on the fraction of hits~(typically \num{30} in total)
above a higher energy-deposit threshold corresponding to transition radiation.
 
The calorimeter system covers the pseudorapidity range $\abseta < 4.9$. Within the region $\abseta < 3.2$, EM calorimetry is provided by barrel and
endcap high-granularity lead/liquid-argon~(LAr) calorimeters, with an additional thin LAr presampler covering $\abseta < 1.8$ to correct for energy loss in
material upstream of the calorimeters. Hadronic calorimetry is provided by the steel/scintillator-tile calorimeter, segmented into three barrel structures within
$\abseta < 1.7$, and two copper/LAr hadronic endcap calorimeters. The solid angle coverage is completed with forward copper/LAr and tungsten/LAr calorimeter modules
optimised for EM and hadronic measurements respectively.
 
The MS comprises separate trigger and high-precision tracking chambers measuring the deflection of muons in a magnetic field generated by the superconducting air-core toroids.
The field integral of the toroids ranges between \num{2.0} and \SI{6.0}{Tm} across most of the detector. A set of precision chambers covers the region $\abseta < 2.7$
with three layers of monitored drift tubes, complemented by cathode-strip chambers in the forward region, where the background is highest. The muon trigger system covers the range
$\abseta < 2.4$ with resistive-plate chambers in the barrel, and thin-gap chambers in the endcap regions. Interesting events are selected to be recorded by the first-level trigger
system implemented in custom hardware, followed by selections made by algorithms implemented in software in the high-level trigger~\cite{TRIG-2016-01}. The first-level trigger accepts
events from the \SI{40}{\MHz} bunch crossings at a rate below \SI{100}{\kHz}, which the high-level trigger reduces in order to record events to disk at about \SI{1}{\kHz}.
 
An extensive software suite~\cite{ATL-SOFT-PUB-2021-001} is used in data simulation, in the reconstruction and analysis of real and simulated data, in detector operations, and in the
trigger and data acquisition systems of the experiment.
 

\section{Lepton selection criteria}
\label{sec:lepton_sel}

Full descriptions of the electron and muon reconstruction algorithms and available selection criteria used in ATLAS are provided in
Refs.~\cite{EGAM-2018-01} and~\cite{MUON-2018-03}, respectively. Here the features most relevant to the fake/non-prompt lepton background
estimation are summarised briefly.

\subsection{Electron reconstruction and identification}
\label{sec:elID}
Electron candidates are reconstructed within $\abseta <  2.47$ as tracks in the inner detector matched to energy clusters in the EM calorimeter.
 
In order to separate true electrons from other unwanted reconstructed candidates,
electron identification~(ID) algorithms are used.
These rely upon a set of variables that quantify the distribution of energy in the calorimeter, the quality of the spatial match
between the calorimeter deposit and the associated track, and the transition radiation signal in the TRT~(see Table~{1} of Ref.~\cite{EGAM-2018-01} for a complete list).
Rather than place individual requirements on these variables, they are combined into a likelihood discriminant based upon the probability density functions of
the variables measured for prompt electrons in \Zepem events and for background candidates reconstructed in inclusive collision events.
 
Since different analyses have different requirements for electron selection efficiency and background rejection, several \enquote{working points}~(WPs) are defined by different values of
the likelihood. The likelihood threshold values are varied according to the \pt and \abseta of the electron candidate, so that the selection efficiency varies smoothly with the electron \pt.
The most commonly used ID WPs~(and their average efficiencies\footnote{
Those efficiencies, as well as those quoted for muons in the next section,
do not include requirements for the leptons to be successfully identified in the context of the trigger decision,
which may include small additional inefficiencies~\cite{TRIG-2018-01,TRIG-2018-05}.
} measured for typical electroweak processes) are \enquote{Loose}~(93\%), \enquote{Medium}~(88\%) and \enquote{Tight}~(80\%).
In addition to the listed WPs, there is another~(\enquote{LooseAndBLayer}) WP that imposes the same requirement on the likelihood as the \enquote{Loose}~WP, but also requires that
the electron track have a hit in the IBL to suppress candidates originating from photon conversions.
 
Often, additional requirements are imposed on the impact parameter of the electron's track:
the impact parameter in the transverse plane, \dzero, with respect to the centre of the beamspot
must satisfy $|\dzero| < 5\sigma(\dzero)$, where $\sigma(\dzero)$ is its estimated uncertainty,
while the longitudinal separation $z_{0}$ between the point where \dzero is measured and the chosen primary vertex of the event,
multiplied by a moderating factor $\sin(\theta)$
which accounts for reduced $z_{0}$ accuracy for more forward tracks,
cannot exceed \SI{0.5}{\mm} in absolute value.

\subsection{Muon reconstruction and identification}
\label{sec:muID}
Muon candidates are reconstructed in the region $\abseta < 2.5$ by combining MS tracks with matching inner-detector tracks. The muon reconstruction efficiency is
approximately 98\% per muon in simulated \Zmpmm events. After reconstruction, high-quality muon candidates used for physics analyses are selected by a set of
requirements on the number of hits in the different inner subdetectors and different MS stations, on the track fit properties, and on variables that test the
compatibility of the individual measurements in the two detector systems, as detailed in Ref.~\cite{MUON-2018-03}. These criteria reduce the background from in-flight decays
of light-flavour hadrons, which often result in kinked tracks. The most commonly used muon ID WPs~(and their efficiencies measured in \ttbar MC events) are \enquote{Medium}~(98\%)
and \enquote{HighPt}~(80\%), the latter optimised to offer the best momentum resolution for $\pt > \SI{100}{GeV}$.
The same impact parameter requirements as defined for electrons
are also often applied to muon candidates,
with a tighter condition in the transverse plane: $|\dzero| < 3\sigma(\dzero)$.

\subsection{Lepton isolation}
 
In addition to the ID criteria mentioned above, most analyses place requirements on the isolation of the lepton
from other detector activity. This is especially helpful in reducing the contribution from leptons produced in heavy-flavour decays,
or muons from \pipm or $K^{\pm}$ decays, within jets\footnote{Jets are reconstructed from clusters of topologically connected calorimeter cells~(\topos), as described in Ref.~\cite{PERF-2013-05}. The \antikt algorithm~\cite{Cacciari:2008gp, Fastjet} is used to form jets from the \topos, with the radius parameter $R$ usually set to 0.4; when reconstructing jets formed from the merged decay products of boosted resonances, a larger $R$ value, typically 10, is used. Typically, jets with $\pt > \SI{20}{\GeV}$ and $\abseta < 4.5$ are considered in physics analyses.}
as there are often other components of the jet that are near the lepton in these cases.
In all of the methods described in Section~\ref{sec:methods},
the baseline lepton selection usually does not require isolation, while the tight lepton selection usually does.
However, many of the single-lepton triggers~\cite{TRIG-2018-05, TRIG-2018-01} used in ATLAS require isolation, so analyses that depend on such triggers cannot avoid applying isolation requirements for baseline leptons.
 
The calorimeter isolation~\cite{EGAM-2018-01, MUON-2018-03} is calculated from the sum of transverse energies of calorimeter energy clusters within $\DeltaR \equiv \sqrt{(\Delta \eta)^2 + (\Delta \phi)^2} = \text{XX}/100$ of the lepton candidate,
not including the contribution expected from the candidate itself.  Typical values of XX are \num{20}, \num{30} or \num{40}.
Expected residual contributions to the isolation from the lepton candidate,
as well as expected contributions from particles produced by additional proton--proton~(\pp) interactions,
are subtracted~\cite{Cacciari:2007fd},
resulting in the variable $E_\mathrm{T}^\text{coneXX}$.

The track-based isolation~\cite{EGAM-2018-01, MUON-2018-03}, denoted by $\pt^{\text{coneXX}}$, is based on tracks near the lepton candidate with either $\pt > 0.5$ or \SI{1}{\GeV} that satisfy basic track-quality requirements and are spatially consistent with the
primary vertex of the event. The scalar sum of the transverse momenta of such tracks, excluding tracks associated with the lepton candidate, is compared with the \pt of the candidate to assess the isolation. For muon candidates, only the single track associated with the candidate is excluded; for electron candidates, additional tracks consistent with pair-production from a bremsstrahlung photon are also excluded.  The track isolation
can also be defined with a variable cone size ($\pT^{\text{varconeXX}}$). In this case, the size of the cone around the lepton candidate within which tracks are considered is varied
as a function of the \pt of the candidate:
\begin{equation*}
\DeltaR = \min\left( \frac{ \SI{10}{\GeV} }{\pt \, [\GeV]},\, R_{\mathrm{max}}\right),
\end{equation*}
where $R_{\mathrm{max}}$ is the maximum cone size~(typically \num{0.2} or \num{0.3}).

Combining selections on track-based and calorimeter-based isolation provides even better fake/non-prompt lepton background rejection,
as the two isolation variables use complementary information. Track-based isolation was found to be less sensitive to detector noise and pile-up effects than
calorimeter-based isolation, and the inner detector provides a better \pt measurement than the calorimeters for individual soft hadrons. On the other hand,
calorimeter-based isolation includes neutral particles as well as some particles below the inner detector's track-\pt threshold, which are ignored when computing
track isolation.  However, track and calorimeter isolation variables measure hadronic activity in a redundant manner, since charged particles are measured by both the
calorimeters and the inner detector, and simple selection cuts applied independently to those two variables may not achieve optimal rejection power.
To avoid this, an analysis can use a \enquote{particle-flow} algorithm, which allows removal of overlapping contributions from the
track-based and calorimeter-based isolation, decreasing the correlation between the two variables. For the time being, particle-flow-based isolation variables are
defined only for muons, and discussed in Ref.~\cite{MUON-2018-03}.
 
For analyses where  the fake/non-prompt lepton background may be dominated by non-prompt electrons and muons from the decays of $b$- and $c$-hadrons~\cite{HIGG-2017-02,SUSY-2019-09}, isolation WPs using a
boosted decision tree~(BDT) discriminant based on isolation and secondary vertex information, referred to as the non-prompt lepton BDT, are also used.
 
Several isolation WPs based on tracking, a combination of calorimetry and tracking, particle-flow, or a non-prompt lepton BDT
are defined to allow for consistency across analyses that require different levels of lepton isolation.
They are described in Refs.~\cite{EGAM-2018-01} and~\cite{MUON-2018-03}.

\subsection{Removing overlaps between jets and leptons}
 
In some cases the same object can result in multiple signatures in the detector. For example, an electron will deposit energy in the calorimeter, and that energy
will generally also be clustered into a jet. In addition, sometimes objects will be spatially correlated due to the underlying physics, as when a muon is produced by
heavy-flavour decay within a jet. To avoid double-counting, and to select only the isolated leptons that are of interest in physics analyses, an overlap removal
procedure is applied to resolve these ambiguities. The procedure used for most physics analyses is as follows, where the lepton candidates are those that satisfy the baseline ID criteria
for the analysis:
 
\begin{itemize}
\item All electron candidates that are within $\DeltaR = 0.01$ of a muon candidate~(or share a track with a muon candidate) are removed.
\item All jets that are within $\DeltaR = 0.2$ of any remaining electron candidates are removed.
\item All electron candidates that are within $\DeltaR = 0.4$ of any remaining jet are removed.
\item Cases where a remaining jet is within $\DeltaR = 0.4$ of a muon candidate are examined to determine the number of tracks associated with the jet. If there are more
than two such tracks the muon candidate is removed; otherwise the jet is removed.
\end{itemize}
 
Variations of this procedure are also supported, primarily for analyses that focus on heavy-flavour jets or that select boosted massive particles using large-$R$ jets.

\FloatBarrier

\section{Monte Carlo simulation samples}
\label{sec:samples}

The studies of fake/non-prompt leptons that are presented in Sections~\ref{sec:fake_sources} and~\ref{sec:eff_meas}
make use of large MC samples of simulated events.
The production of \ttbar events at next-to-leading order~(NLO)
in the quantum chromodynamics~(QCD) coupling constant \alphas is described in Ref.~\cite{ATL-PHYS-PUB-2018-009}
and relies on the \POWHEGBOX[v2] event generator~\cite{Alioli:2010xd}
interfaced with \PYTHIA[8.230]~\cite{Sjostrand:2014zea} for parton showering and subsequent steps,
with the A14 set of tuned parameters~\cite{ATL-PHYS-PUB-2014-021}.
The parton distribution functions used for matrix element calculation
and parton showering are \NNPDF[3.0nlo]~\cite{Ball:2014uwa}
and \NNPDF[2.3lo]~\cite{Ball:2012cx} respectively.
Generated events were filtered such that
at least one of the top quarks decays semileptonically.
The \EVTGEN[1.2.0] program~\cite{Lange:2001uf} was used to model heavy-flavour hadron decays.
 
The production of Drell--Yan $Z/\gamma^{*}\to\ellell$ events ($\ell=e,\mu,\tau$)
at NLO in \alphas is described in Ref.~\cite{ATL-PHYS-PUB-2017-006}
and relies on the \SHERPA[2.2] event generator~\cite{Bothmann:2019yzt}
with the dedicated set of tuned parameters
and the \NNPDF[3.0nnlo]~\cite{Ball:2014uwa} parton distribution function.
Events were generated according to a partition of the phase space described in Ref.~\cite{ATL-PHYS-PUB-2017-006},
resulting in a set of orthogonal samples which were combined
with weights corresponding to the NLO cross-section calculated by the generator.
The \ellell invariant mass was required to be at least \SI{40}{\GeV}.
 
A full ATLAS detector simulation~\cite{SOFT-2010-01} based on \GEANT~\cite{Agostinelli:2002hh}
was then used to faithfully reproduce particle interactions with the detector and its response.
Additional \pp interactions in the same or neighbouring bunch crossings
were also simulated in order to reproduce the conditions of data-taking during
the LHC \RunTwo operation.
Simulated events are weighted to reproduce the measured performances of
various selection criteria, such as electron identification~\cite{EGAM-2018-01}.
In addition, reconstructed lepton and jet momenta are transformed~\cite{EGAM-2018-01,MUON-2022-01,JETM-2018-05}
to match the real detector resolution.
 
The source of a lepton candidate in simulation, discussed in Section~\ref{sec:fake_sources},
is identified by using preserved generator-level information
to match the track of each reconstructed lepton candidate
to a suitable stable particle produced by the generator or the~\textsc{Geant4} simulation,
among those preserved in the event record (Sections 3.6 and 5.3 in Ref.~\cite{SOFT-2010-01}).
The matching is performed by evaluating the contribution of each stable particle to the
simulated ID and (for muons) MS hits used to reconstruct the track,
and requiring that the matched particle accounts for at least half of those hits (Section 6.3 in Ref.~\cite{STDM-2015-02}).
While the latter is a general procedure used for all tracks,
for leptons specifically the result is discarded if the reconstructed track
has less than three combined hits in the pixel and SCT detectors,
or if the track and the stable particle do not satisfy $\Delta{R}<0.2$.
If the stable particle is a charged lepton
(or, in the case of electron candidates, may also be a photon),
its origin is checked so that
the different sources of non-prompt leptons can be distinguished.
 
In Section~\ref{sec:applications}, other simulated processes are used.
The corresponding MC samples are obtained with a similar workflow.
Complete information about the MC generators and their configurations,
as well as the cross-section calculations used to normalise the simulated samples,
is available in the references provided in that section.

\FloatBarrier

\section{Sources of fake/non-prompt leptons}
\label{sec:fake_sources}

\begin{figure}[t]
\centering
\subfloat[]{
\includegraphics[width=0.49\textwidth]{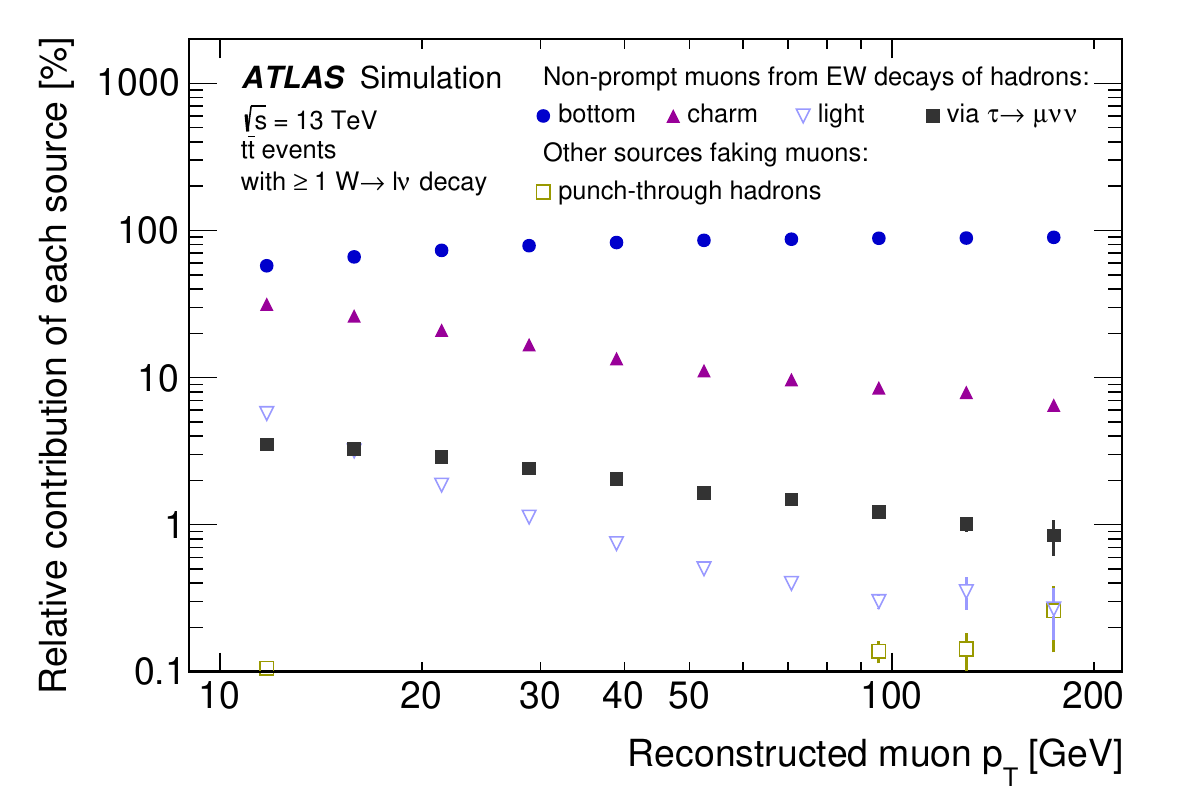}}
\subfloat[]{
\includegraphics[width=0.49\textwidth]{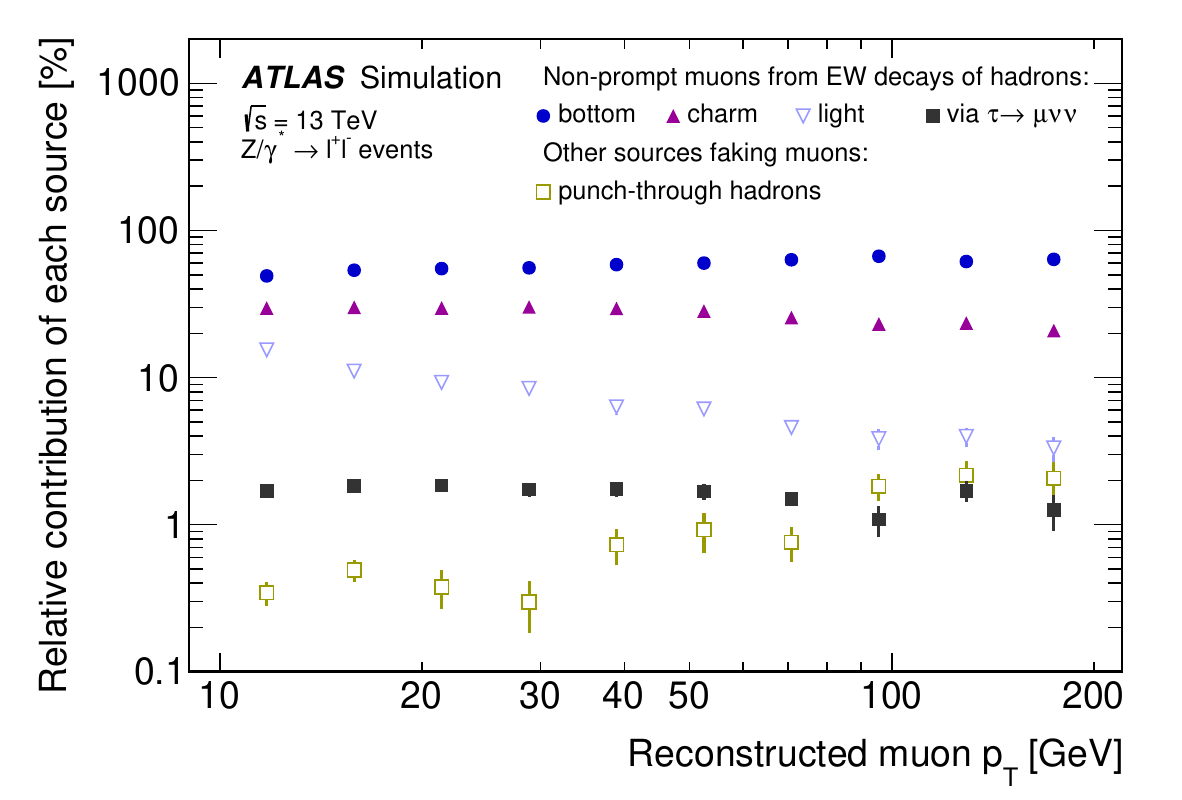}}
\caption{Relative contributions of fake or non-prompt muons from different sources as a function of \pt in simulated (a)~\ttbar and (b)~Drell--Yan processes.
Events are required to contain at least two leptons with identical charges.
Punch-through hadrons are charged hadrons reaching the MS and leading to the reconstruction of a fake muon candidate.
Some other minor sources are not displayed; the sum of their contributions is less than 2\% in every bin of (a),
and less than 4\% in bins of (b) up to $\pt = \SI{100}{\GeV}$.
The muons are required to satisfy minimal track-quality criteria specified in the text.
Error bars represent statistical uncertainties of the simulation.}
\label{fig:composition_muons}
\end{figure}
 
\begin{figure}[t]
\centering
\subfloat[]{
\includegraphics[width=0.49\textwidth]{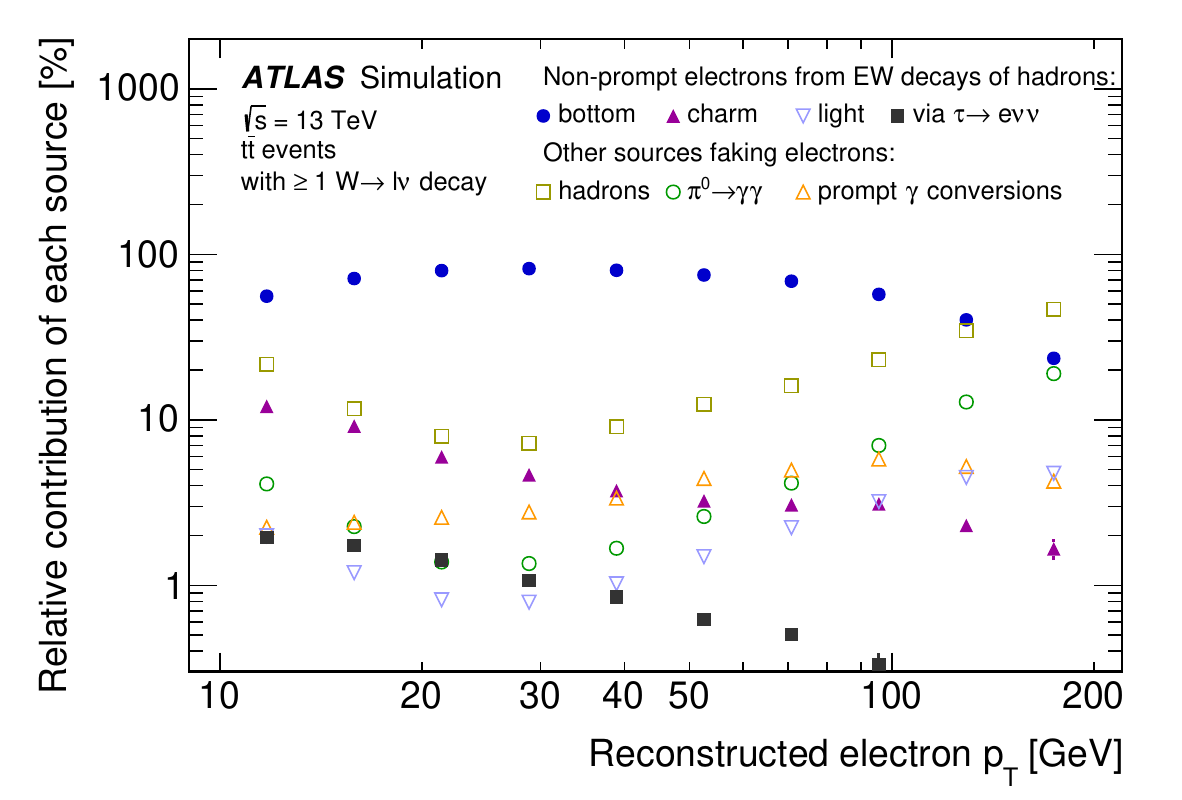}}
\subfloat[]{
\includegraphics[width=0.49\textwidth]{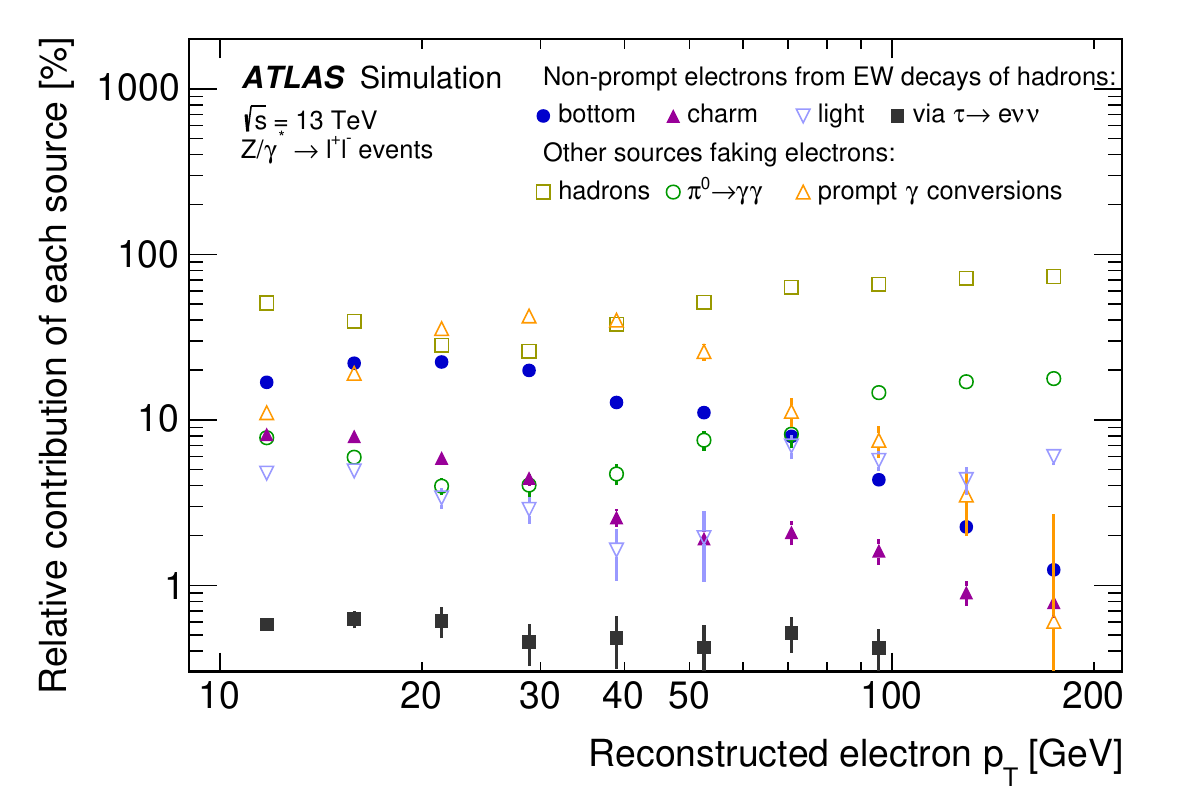}}
\caption{Relative contributions of fake or non-prompt electrons from different sources as a function of \pt in simulated (a)~\ttbar and (b)~Drell--Yan processes.
Events are required to contain at least two leptons with identical charges.
Some other minor sources are not displayed; the sum of their contributions is less than 0.5\% in every bin.
These do not include electron candidates corresponding to high-energy muon
ionisation signal in the calorimeters, or from conversions of final-state radiation~(FSR) photons with small angular separation from a genuine electron or muon.
The electrons are required to satisfy loose ID criteria specified in the text and Ref.~\cite{SUSY-2018-09}. Error bars represent statistical uncertainties of the simulation.}
\label{fig:composition_electrons}
\end{figure}

The relative contributions of fake/non-prompt leptons from different sources
to the sample of selected lepton candidates
depend on the energy and spatial location of the candidate in the detector, the identification, impact parameter
and isolation criteria applied, the overlap removal procedure,
and the nature of the selected final state~(e.g.\ the presence or absence of heavy-flavour hadrons).
 
Figures~\ref{fig:composition_muons} and~\ref{fig:composition_electrons} present for illustration the relative contributions of the different sources of
fake/non-prompt muons and electrons respectively, as a function of the transverse momentum of the candidate, as measured in MC simulated events. They are shown for two
different processes: \ttbar production, leading to final states rich in heavy-flavour hadrons, and Drell--Yan production of \ee or \mumu pairs;
for \ttbar events at least one of the two $W$ bosons produced in the \ttbar decay is required to decay leptonically ($W\to\ell\nu$).
For these particular figures, the reconstruction/ID of electron
and muons and the general event selection follow those described in Ref.~\cite{SUSY-2018-09} for baseline lepton candidates, which correspond to rather loose criteria~(in
particular, no isolation nor transverse impact parameter requirements are applied, and no overlap removal is done).
For both processes, events are considered only if they contain a pair of reconstructed baseline leptons with identical charges,
a signature for which fake/non-prompt leptons usually represent a non-negligible source of background.

The \enquote{prompt \gam-conversion} category only includes electron candidates where the photon is separated by $\DeltaR>0.1$ from any generator-level high-\pt electron from the hard-scatter
interaction~(photons emitted at smaller separation are generally reconstructed as part of the electron candidates).

Non-prompt leptons are those arising from electroweak decays of hadrons.
Heavy-flavour $b$- and $c$-hadrons decay close to the interaction point and the resulting leptons
are distinguishable from real leptons mostly by isolation and impact parameter.

Light-flavour hadrons can also be a major source of fake leptons via decay-in-flight in the tracker volume. This happens mainly in final states for which QCD multi-jet production
is a significant contributor. A charged hadron stopping early in the calorimeter, and generating a narrower-than-average shower, can mimic the experimental signature of an electron.
Electron ID criteria are particularly powerful in rejecting these candidates rather than those from other sources, based notably on three-dimensional profiles of the shower,
but since many orders of magnitude more hadrons than leptons are produced in collisions at the LHC, a substantial number of such fake electrons may be selected in a physics analysis.
With regard to muons, the depth of the ATLAS calorimeter is sufficient to stop most pions and kaons before they can reach the MS.
Muon candidates arising from kaons decaying semileptonically in flight before reaching the MS are a more important contribution to the fake/non-prompt lepton background.

Among hadrons faking electrons, one significant class is neutral pion decays into photons ($\pizero \to \gam\gam$); the collimated photons create a single energy deposit
in the EM calorimeter, while the associated track might be provided by the conversion of one of the photons in the upstream detector material. Due to the importance
of this phenomenon, the electron ID criteria specifically discriminate against these by attempting to identify a two-peak structure in the distribution of the cell
energies matched to the electron cluster~\cite{EGAM-2018-01}, which is not present for real electrons. Other contributions such as Dalitz decay of pions~\cite{Dalitz} also
create comparable experimental signatures.

The conversion of photons into electron--positron pairs represents the last important class of non-prompt electrons. These \gam-conversions must typically occur early on~(e.g.\ in the beam pipe)
and be largely asymmetric in the splitting of the momentum between the two electrons; otherwise, the conversion vertex can be reconstructed, or the candidate electron's track lacks
hits in the first layers of the inner detector, both leading to the proper classification of the reconstructed object as a photon instead of an electron~\cite{EGAM-2018-01}.
The origin of the photon itself influences the characterisation of the candidate as non-prompt or real: photons emitted close to a real electron, either due to bremsstrahlung or as
higher-order quantum electrodynamic~(QED) corrections to the production process, are typically considered part of the  electron candidate (the calorimeter energy deposits tend
to overlap to the extent that a single cluster is reconstructed);  furthermore, from the perspective of quantum field theory, well-defined electrons must include extra
radiation~(\enquote{dressed leptons}~\cite{ATL-PHYS-PUB-2015-013}).  The electron reconstruction procedure~\cite{EGAM-2018-01} accounts for bremsstrahlung, in particular
by allowing kinks in the track consistent with bremsstrahlung emission in dense material regions. In contrast, photons from other origins, such as initial-state radiation,
QED processes not involving leptons or where photons are sufficiently separated from leptons, or hadronic jet fragmentation, may be considered as sources of fake electrons.

It can be seen in Figure~\ref{fig:composition_muons} that non-prompt muons constitute
the only substantial contribution to the fake/non-prompt muon background,
while for electrons Figure~\ref{fig:composition_electrons} shows more variety:
in general, non-prompt electrons are particularly represented in the lower \pt range,
especially in processes involving the production of heavy-flavour hadrons,
while hadron fakes and converted photons populate the higher \pt range.\footnote{However,
for the range $\pt < \SI{10}{\GeV}$, not shown in the figure, hadron fakes are also the dominant contribution.}

\begin{figure}[t]
\centering
\subfloat[]{
\includegraphics[width=0.49\textwidth]{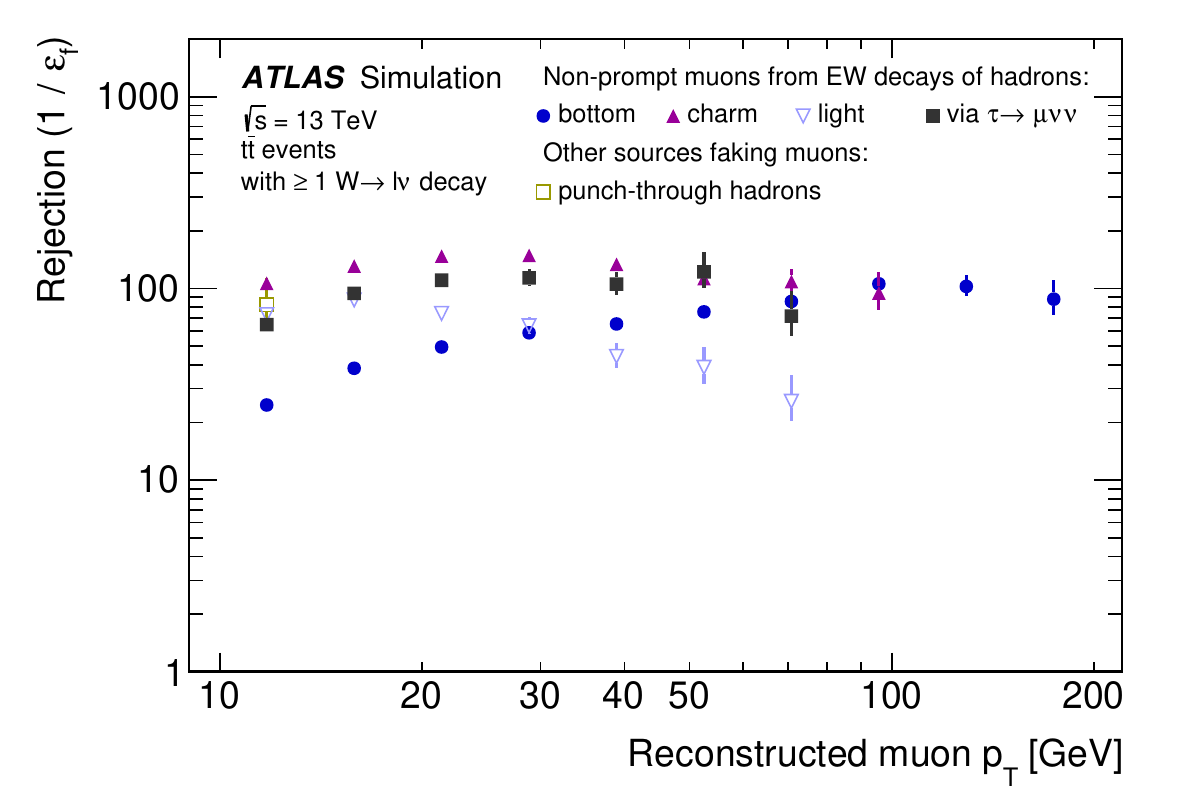}}
\subfloat[]{
\includegraphics[width=0.49\textwidth]{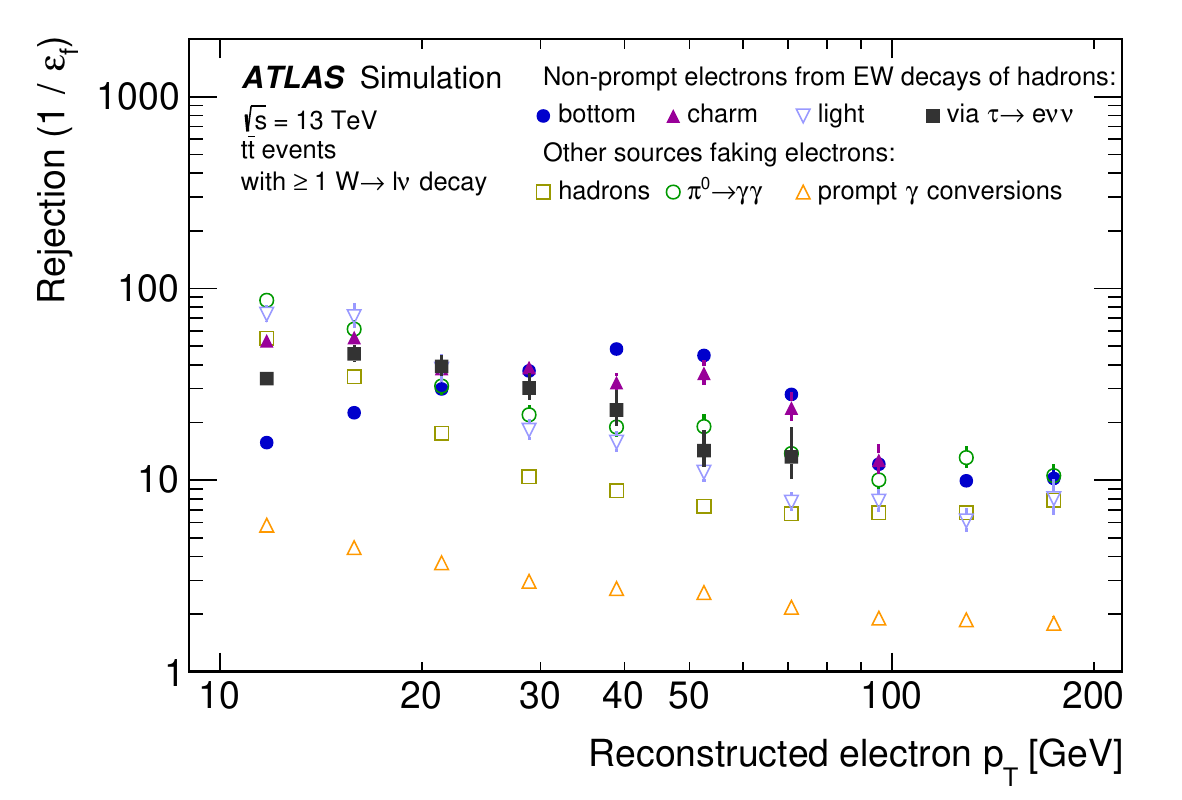}}
\caption{
Rejection (defined as 1/$\varepsilon_\mathrm{f}$) as a function of \pt
for different sources of fake/non-prompt (a) muons and (b) electrons in simulated \ttbar events
containing two leptons with identical charges.
The efficiency $\varepsilon_\mathrm{f}$ is the fraction of lepton candidates selected
using the same baseline criteria as in Figures~\ref{fig:composition_muons} and~\ref{fig:composition_electrons}
that also pass the tight selection criteria of Ref.~\cite{SUSY-2018-09}.
Error bars represent statistical uncertainties of the simulation,
and for purposes of clarity, values with relative uncertainties greater than 30\% are not shown.
}
\label{fig:fake_efficiency_per_source}
\end{figure}
 
The different sources of fake/non-prompt leptons have distinct probabilities to satisfy the
tight lepton selection criteria described in Section~\ref{sec:lepton_sel}.
Figure~\ref{fig:fake_efficiency_per_source} illustrates those differences
for the particular example of signal leptons definitions used in Ref.~\cite{SUSY-2018-09} (including lepton--jet overlap removal)
and simulated \ttbar events containing two leptons with identical charges.
Such variability between sources is unwelcome,
as the fake efficiencies required for the application of the methods described in Section~\ref{sec:methods}
then depend upon the relative contributions of each source to the regions of interest,
which may not be easy to assess.
It is therefore desirable to measure the efficiencies in regions similar in composition
to the regions where the background estimate is needed,
otherwise large extrapolation uncertainties may apply.
 
Precise simulation of these various sources of background, including their relative contributions, is indeed very challenging, as it relies heavily on the modelling
of the soft-QCD regime by event generators, including modelling of fragmentation and hadronisation processes, hadron decay modelling, and soft emissions and detector modelling.
Another issue is that only a small fraction of fake or non-prompt lepton candidates survive the ID and isolation requirements,
so the simulation of a very large number of events is needed to obtain a statistically accurate prediction.
For inclusive processes with large cross-sections~(e.g.\ multi-jet production), this is often impractical.
 
For these reasons, many of the fake/non-prompt lepton background predictions used in ATLAS publications are based on methods using the data, such as the ones described in
this paper. They rely on common properties shared to some extent by the different sources of fake/non-prompt leptons that differentiate them from real leptons,
such as a high likelihood to not meet the combination of ID and isolation criteria.

\FloatBarrier

\section{Measurement of real and fake/non-prompt lepton efficiencies}
\label{sec:eff_meas}

The methods described in Section~\ref{sec:methods} both rely on knowledge of the efficiency for leptons that pass the baseline selection to also pass the
tight selection. For the fake-factor method, only the efficiency for fake/non-prompt leptons is used explicitly in the calculation,
while for the other methods the efficiency for real leptons must also be measured.
In many cases these efficiencies depend on the properties of the lepton (such as its
\pT or angular distance from a jet) or on the event in which it is found (such as the overall activity of the event, as measured for example by
the number of reconstructed primary vertices).
 
\subsection{Real-lepton efficiencies}
\label{sec:eff_prompt}
 
The efficiencies of specific working points of lepton selections
are calibrated precisely in ATLAS for general purposes,
primarily with \enquote{tag-and-probe methods} based e.g.\ on \Zll events.
By performing these measurements on both data and MC-simulated events,
\enquote{scale factors}~(SFs) that account for differences between the efficiencies observed in data and simulation are derived.
These SFs can then be applied to simulated events using the selection criteria that are relevant to a given analysis to determine the appropriate real-lepton efficiencies.
This MC-based approach is valid as long as both the baseline and tight lepton selection criteria are taken from the set for which SFs have been measured; only in
extraordinary cases would an analysis utilise different selection criteria.
Since the efficiencies depend more strongly on the environment than the SFs do,
the main advantage of this approach over purely data-based measurements
is to allow efficiencies to be obtained directly in the desired environment (i.e.\ the region in which the fake/non-prompt background estimate is needed),
rather than being extrapolated from a more distant region which would be needed for reliable measurements in data.
 
Details of the real-electron and real-muon efficiency measurements can be found in Refs.~\cite{EGAM-2018-01} and~\cite{MUON-2018-03}, respectively.
The real efficiencies are often parameterised with respect to the \pT and \abseta of the leptons, and measured separately for electrons and muons.

\subsection{Fake/non-prompt lepton efficiencies}
\label{sec:eff_fake}
 
The fake/non-prompt lepton efficiencies are specific to each analysis,
primarily since there are several sources for such leptons~(see Section~\ref{sec:fake_sources}),
which will contribute with different weights depending on the chosen selection criteria.
In general, though, the first step in the efficiency measurement is to identify a region that has a large contribution from fake leptons. Two approaches are commonly
used. In the first, events with a pair of leptons with the same electric charge are selected. Since such lepton pairs are only rarely produced at the LHC~(via processes
such as \WZjets, \ZZjets, and $\ttbar + X$~($X = W/Z$) production) it is likely that one of the two leptons is fake/non-prompt. By placing stringent quality criteria on one lepton
in these events, the probability that the remaining lepton is fake/non-prompt is enhanced.  The second approach is to use single-lepton events, where criteria are imposed to
suppress the contribution from real leptons. Examples of such criteria are requiring the missing transverse momentum\footnote{
\METvec is defined as the negative vector sum of the \pt of the reconstructed
and calibrated objects in the event, with a correction applied for inner detector tracks
that originate from the primary collision vertex and are not associated with any other objects, and \MET is defined as the magnitude of \METvec ~\cite{PERF-2016-07}.}~\MET,
or transverse mass\footnote{Here $\mT \equiv \sqrt{2 \, \pt^{\ell} \, \MET (1 - \cos \Delta \phi_{\ell, \MET})}$, where $\Delta \phi_{\ell, \MET}$
is the azimuthal angle between the lepton and \METvec directions.}~\mT, to be below specific thresholds, thereby reducing the contribution from \Wjets or \ttbar events, or requiring
via the track impact parameters that the lepton originate from a position inconsistent with the primary event vertex, thereby enhancing the contribution of leptons from
heavy-flavour decay.
 
In either approach, there will be a residual contribution from events with only real leptons in the selected sample. This contribution is typically estimated using MC simulation
and subtracted separately from both the tight and baseline samples before the ratio of these samples is taken to measure the efficiency.
 
As with the real-lepton efficiencies, the fake/non-prompt lepton efficiencies depend on properties of the lepton candidates or of the event in which they are found.
Therefore, it is generally helpful to bin the efficiencies in the lepton \pt and \abseta, and possibly in terms of other quantities as well. The optimal binning to be used
is chosen in the context of each physics analysis, considering the given numbers of events and potential changes in the relative contributions from the different sources of fake/non-prompt leptons;
illustrative examples are provided in Section~\ref{sec:applications}.
 
\begin{figure}[t!]
\subfloat[]{
\includegraphics[width=0.49\textwidth]{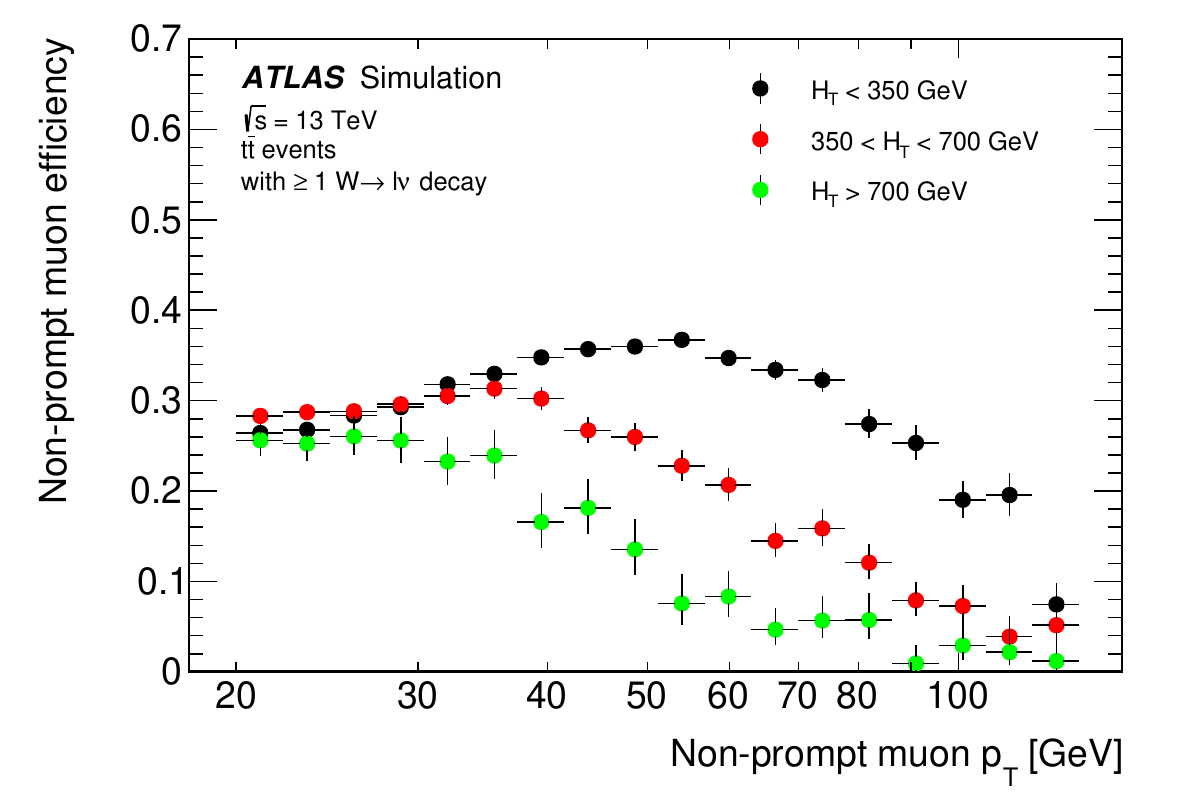}}
\subfloat[]{\label{fig:fake_rates_ps_vs_pthadr_b}\includegraphics[width=0.49\textwidth]{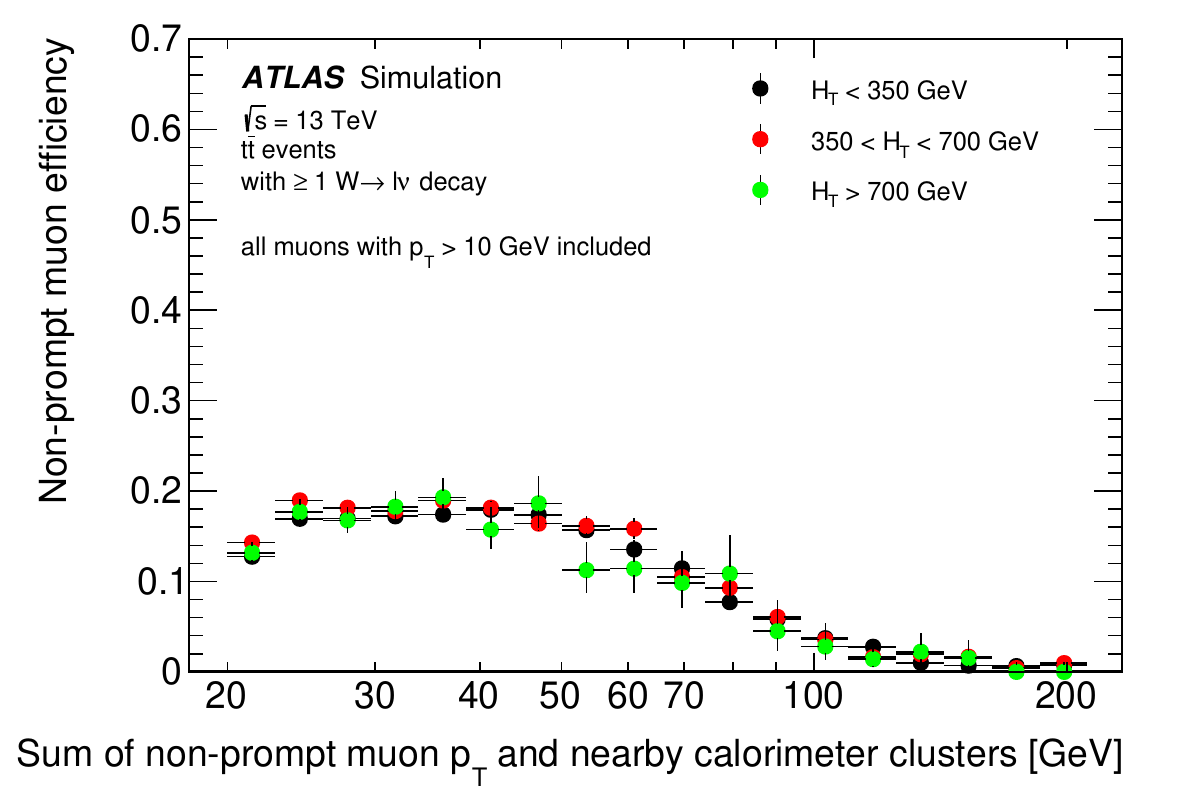}}
\caption{Probability for a muon produced in the decay of a $b$-hadron to satisfy a track-based isolation requirement,
in simulated \ttbar events, shown as a function of (a)~the muon \pt; and (b)~the sum $\pt + \et^\text{cone40}$ of the muon \pt and nearby calorimeter energy deposits.
In each case, the universality of these probabilities is gauged  by changing the kinematic properties of the parent hadron~(markers of various colours), achieved
indirectly by selecting events in different ranges of global transverse energy \HT. Precise definitions of the isolation criteria, the $\pt + \et^\text{cone40}$ and \HT variables,
as well as descriptions of the muon and jets selections, are given in Section~\ref{sec:eff_fake}.
}
\label{fig:fake_rates_ps_vs_pthadr}
\end{figure}
 
Adopting a parameterisation of the efficiency with respect to variables other than \pt and $\eta$ can sometimes be beneficial. For example, the probability that a fake or non-prompt lepton satisfies the isolation criteria is correlated with the fraction of the jet’s visible momentum carried by the lepton.
Therefore parametrising the fake efficiency as a function of the lepton \pt amounts to assuming that, for a particular lepton \pt, the distribution of the parent jet momentum is similar in
the regions where the efficiencies are measured and the regions where the background estimates are needed. If this assumption does not hold, it can be useful to adopt
instead a parameterisation as a function of the parent jet momentum. Since this quantity is not easily accessed experimentally~(unlike the lepton \pt), proxy observables
are used in practice. An example of successful application is the analysis in Ref.~\cite{STDM-2014-05}, which employed the sum of the lepton's \pt and the transverse energy
in a cone around the lepton as a proxy.

The preceding discussion is also illustrated in Figure~\ref{fig:fake_rates_ps_vs_pthadr} for non-prompt muons produced in the decay of $b$-hadrons in simulated \ttbar events.
The probabilities for such muons to satisfy a track-based isolation requirement, as defined in Section~\ref{sec:lepton_sel}, are shown for two alternative parameterisations:
one based on the muon \pt, and the other on the scalar sum of the muon \pt and the transverse energy deposited in calorimeter-cell clusters within a cone of size $\DeltaR = 0.4$
around the muon~(referred to as $\et^\text{cone40}$ in Section~\ref{sec:lepton_sel}). This scalar sum serves as a proxy for the parent jet's transverse momentum.
 
In the second parameterisation, the fake efficiency is the fraction of jets with a non-prompt muon and visible momentum $\pt + \et^\text{cone40}$ in which the muon is
mostly isolated, i.e.\ $\pt \gg \et^\text{cone40}$. While one might consider jets with arbitrarily soft muons in the denominator of this fraction, for practical reasons
Figure~\ref{fig:fake_rates_ps_vs_pthadr_b} only includes events where muons satisfy $\pt > \SI{10}{\GeV}$.
To study the dependence of these efficiencies on the momentum distribution of the underlying jet,
different regions of jet momentum are emphasised by imposing different requirements on the global transverse energy \HT in the event, defined for
this purpose as the scalar \pt sum of all jets with $\pt > \SI{25}{\GeV}$ and $\abseta < 2.8$ that are a distance $\DeltaR > 0.6$ from the muon. This quantity is indeed partially correlated
with the kinematics of the muon's parent jet, via the momentum of the top quarks producing all these jets. The reconstruction, calibration and selection of jets and muons
for this figure are otherwise those detailed in Ref.~\cite{SUSY-2018-09}.
 
It can be observed that for the case of a \pt-dependent parameterisation, the efficiencies vary strongly with \HT, although large differences occur mostly for $\pt > \SI{40}{\GeV}$.
Since most non-prompt muons are produced at low \pt, the overall impact of this non-captured dependency might be small, unless regions of interest in the analysis specifically
select high-\pt leptons. In contrast, the parameterisation as a function of $\pt + \et^\text{cone40}$ is much less influenced by \HT, making the measured efficiencies
less dependent on the event topology. In practice, a compromise has to be found between this observation  and other elements evoked above justifying a \pt-dependent parameterisation,
especially for electrons because ID criteria are usually employed in addition to isolation. The direct dependency of efficiencies on other variables that are also
correlated with the event topology~(e.g.\ \MET) may also be reduced by such a parameterisation.
 
As detailed in the next section, suitable uncertainties must be assigned
to the use of the fake efficiencies
in different regions than those in which they are measured,
in particular to account for potential differences in relative contributions of the different sources of fake/non-prompt leptons.
To minimize these uncertainties,
it has sometimes been found beneficial to use an approach closer to that of section~\ref{sec:eff_prompt},
in which efficiencies are evaluated in the simulation and supplemented by data-driven correction factors that depend on the source of the fake/non-prompt lepton. These correction factors are derived using dedicated control regions that are enriched in a particular source of fake/non-prompt leptons.
The main assumptions are then the universality of the correction factors across different processes,
and the ability of the simulation to adequately predict the relative contributions of each source in the regions of interest.
Such an approach has for example been used in Ref.~\cite{SUSY-2018-02}.

\FloatBarrier

\section{Systematic uncertainties}
\label{sec:systematics}

Systematic uncertainties in the fake/non-prompt lepton background estimates from the matrix method and the fake-factor method arise from uncertainties in the values of
$\varepsilon_\mathrm{r}$ and $\varepsilon_\mathrm{f}$.
These uncertainties can be traced to statistical uncertainties from the samples used to measure the efficiencies,
to potential biases that may cause the efficiencies in the signal region for a particular analysis to differ from the values obtained from control samples
(such as differences in the origin of fake/non-prompt leptons between these regions),
and to uncertainties in the modelling of contamination from real-lepton processes in the samples used to measure the fake efficiency.
Details of these uncertainties and their estimation are provided below.
The overall impact of variations in $\varepsilon_\mathrm{r}$ and $\varepsilon_\mathrm{f}$
depends on the characteristics of the analysis.  For example, in the matrix method Equations~\ref{eq:nfakes_b} -- ~\ref{eq:nreal_t_1lep} imply that for analyses with
\begin{equation*}
N_\mathrm{r}^\mathrm{t} \times \frac{\Delta\varepsilon_\mathrm{r}}{\varepsilon_\mathrm{r}^2}  > N_\mathrm{f}^\mathrm{t} \times \frac{\Delta\varepsilon_\mathrm{f}}{\varepsilon_\mathrm{f}^2}
\end{equation*}
the uncertainty in $\varepsilon_\mathrm{r}$ will dominate, and vice versa.

\subsection{Statistical uncertainties in the measured efficiencies}
 
Statistical uncertainties in the real-lepton and fake-lepton efficiencies can be accounted for either by analytically propagating the uncertainties through to the estimated event
yields or by varying the efficiencies input to the nominal yield calculation by their statistical uncertainties and using the resulting difference in the estimated fake/non-prompt
background yield as the resulting uncertainty. The real- and fake-efficiency uncertainties are generally uncorrelated since the efficiencies are measured in statistically independent samples.
In the usual case where the efficiencies are measured in bins of one or more quantities, the bins are uncorrelated, so the variations in each bin are
applied separately. The total statistical uncertainty in the fake/non-prompt lepton background estimate is given by the sum in quadrature of the uncertainties from all variations.
This source of systematic uncertainty is usually not dominant.

\subsection{Systematic uncertainties in the measured efficiencies}
 
Systematic uncertainties in $\varepsilon_\mathrm{f}$ are generally larger and more challenging to assess than for $\varepsilon_\mathrm{r}$:
 
\begin{enumerate}
\item{Real-lepton efficiencies have only slight variations due to event environment effects since real leptons have small contributions to their measurements from underlying
event and jet activity. This also means there are a wide variety of samples with which they can be calibrated in great detail.}
 
\item{There are several sources of fake/non-prompt leptons, and the efficiencies may differ between these sources. Therefore, any differences in the
fake/non-prompt lepton composition between the sample used to measure the efficiencies and the signal region for an analysis may lead to a bias in the efficiencies.}
\end{enumerate}
 
Several methods are used to estimate systematic uncertainties in the fake-lepton efficiencies. One is simply to vary the selection criteria for events in the control
region used to measure the fake-lepton efficiencies, since the composition of fake/non-prompt leptons in the standard and alternative control regions may differ. A more
sophisticated approach is to use MC simulation to estimate the fake/non-prompt lepton compositions in both the control and analysis regions. This information, combined
with the MC-estimated selection efficiencies for each source of fake/non-prompt leptons, can be used to provide an estimate of the uncertainty.

\subsection{Uncertainties in the modelling of real-lepton processes}
 
When measuring the fake-lepton efficiencies, a correction must be applied to account for contamination from processes with real leptons in the control sample used for the measurement:
 
\begin{equation*}
\varepsilon_\mathrm{f} = \frac{ N^\mathrm{t} - N_\mathrm{r}^\mathrm{t}}{ (N^\mathrm{t} + N^\mathrm{l}) - (N_\mathrm{r}^\mathrm{t} + N_\mathrm{r}^\mathrm{l}) }
\end{equation*}
where $N_\mathrm{r}^\mathrm{t}$ and $N_\mathrm{r}^\mathrm{l}$ are the numbers of real leptons in the selected tight and loose samples, respectively. MC simulation of real-lepton processes is generally used to
estimate $N_\mathrm{r}^\mathrm{t}$ and $N_\mathrm{r}^\mathrm{l}$, with corrections applied to account for known differences in object selection efficiencies between simulation and data.
Nonetheless, several sources of systematic uncertainty in the real-lepton contamination remain: uncertainties in the cross-sections of the real-lepton processes,
uncertainties in the correction factors, and uncertainties in the parameters, e.g.\ parton distribution functions~(PDFs) and factorisation/renormalisation scales~(\muF, \muR), used in
the simulation.
 
\subsection{Uncertainties due to biases in the Poisson likelihood matrix method}
 
The biases that occur in some situations for the Poisson likelihood matrix method (see Section~\ref{sec:perf}) may also be considered as a source of systematic uncertainty, especially when the signal region contains many bins with few events in each.  The magnitude of the bias can be estimated either by repeating the analysis with coarser bins, or by constraining the total fake/non-prompt lepton yield estimate to be the value returned by applying the Poisson likelihood matrix method to the entire unbinned sample.  Any resulting differences in the binned estimates can be taken as a systematic uncertainty.
 

\FloatBarrier

\section{Examples of application in ATLAS analyses}
\label{sec:applications}
 
In this section the application of the fake/non-prompt lepton background estimation methods is described using two example ATLAS analyses. The first is a measurement of
the \ttZ differential cross-section using events that contain three or four lepton candidates, and the second is a model-independent search for \enquote{beyond the Standard Model}~(BSM)
phenomena in events with three or more lepton candidates. In both cases, the high lepton multiplicity suppresses the Standard Model~(SM) backgrounds, which makes the relative
contribution of the fake/non-prompt lepton background larger.
 
Data from $\rts = \SI{13}{\TeV}$ \pp collisions recorded by the ATLAS detector between 2015 and 2018 are used to perform these analyses. In this period, the
LHC delivered colliding beams with a peak instantaneous luminosity of $L = \SI{2.1e34}{cm^{-2}s^{-1}}$, achieved in 2018, and an average number of \pp
interactions per bunch crossing of \num{33.7}. After applying beam, detector and data-quality criteria, the total integrated luminosity of the dataset is
\SI{139}{\ifb}~\cite{DAPR-2018-01}. The uncertainty in the combined 2015--2018 integrated luminosity is 1.7\%~\cite{ATLAS-CONF-2019-021}, obtained using
the LUCID-2 detector~\cite{LUCID2} for the primary luminosity measurements.
 
\subsection{Measurement of the \texorpdfstring{\ttZ}{ttZ} cross-section in final states with three or four leptons}

This analysis measured the differential production cross-section of the \ttZ process in final states with a total of three or four electron and muon candidates. The tools described
above are used for the lepton efficiency measurements and the application of the Poisson likelihood matrix method~(described in Section~\ref{subsubsec:lhMM}). In addition to the fake/non-prompt
lepton estimation in the signal regions of the analysis, the results are checked in several validation regions enriched in fake/non-prompt leptons. The Poisson likelihood matrix method
was chosen for this analysis since the number of fake/non-prompt leptons in the signal regions with three or four leptons
is expected to be very low, and the Poisson likelihood matrix method provides more stable results for binned estimations, which are necessary for differential background predictions
in these low-statistics signal regions. The fraction of events with more than one fake/non-prompt lepton in the signal or validation regions has been checked and found to
be negligible. More details about the analysis and the fake/non-prompt lepton background estimation can be found in Ref.~\cite{TOPQ-2018-08}.
 
\subsubsection{Real-lepton efficiencies}
 
\begin{figure}[htbp]
\centering
\subfloat[]{\includegraphics[width=0.8\textwidth]{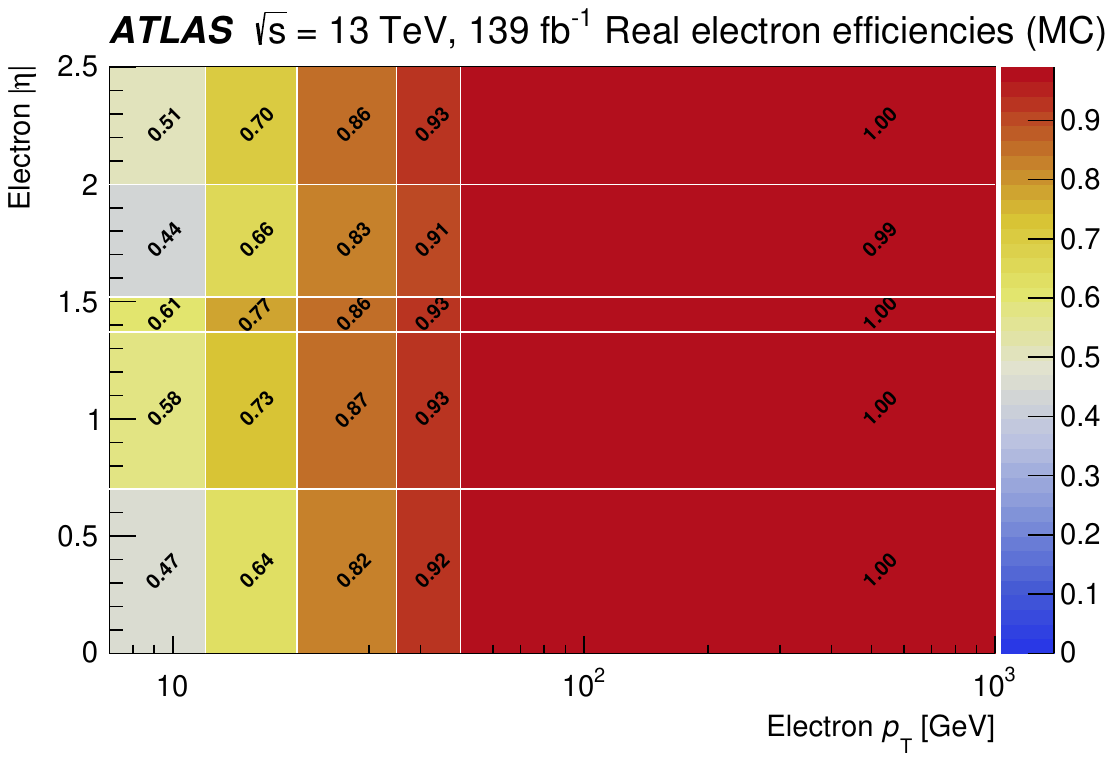}} \\
\subfloat[]{\includegraphics[width=0.8\textwidth]{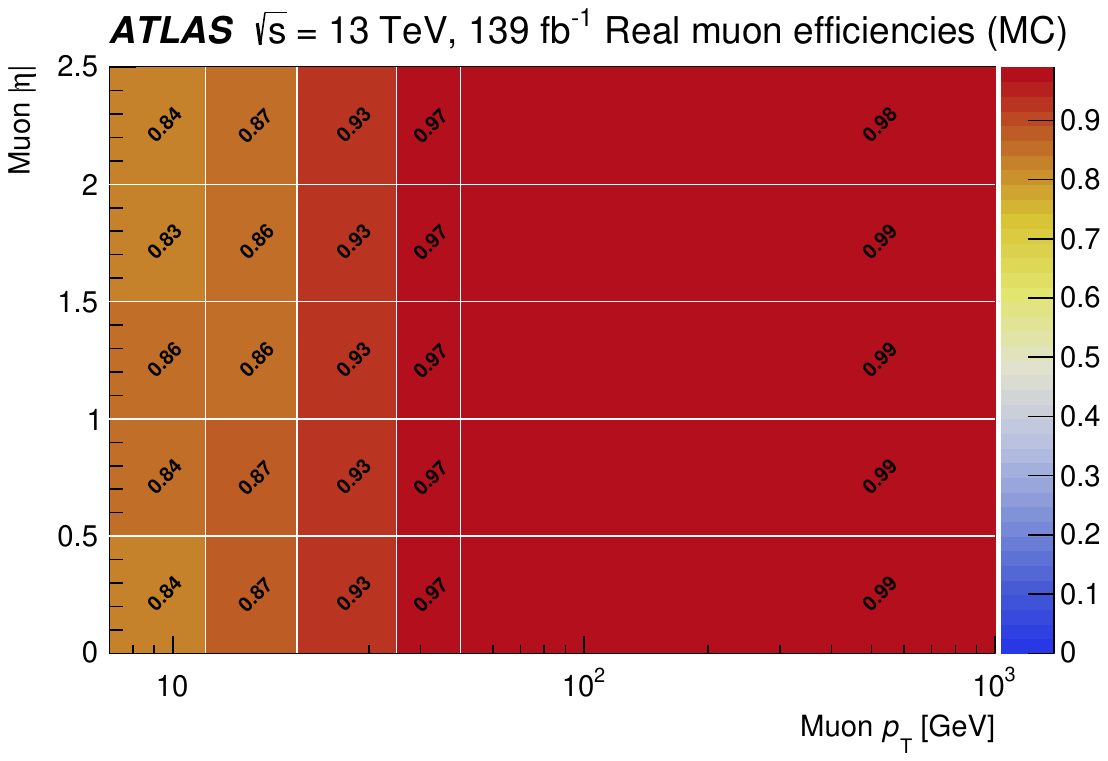}}
\caption{The two-dimensional real-lepton efficiencies obtained for (a)~electrons and (b)~muons, in bins of \pT and \abseta of the leptons. The last \pt bin is inclusive.
The real-lepton efficiencies are obtained using MC simulation, but corrected to match the performance seen in data control samples.}
\label{fig:realrates_elmu2D}
\end{figure}
 
The first step in the application of the matrix method is to measure the efficiencies for real and fake/non-prompt leptons that satisfy the baseline criteria to also
satisfy the tight criteria. In the \ttZ cross-section measurement, baseline electrons are required to satisfy the \enquote{LooseAndBLayer} ID WP, whereas tight electrons are
required to satisfy the stricter \enquote{Medium} ID criteria and to be isolated from nearby tracks and calorimeter energy deposits. Both the baseline and tight muons are required to satisfy
the \enquote{Medium} WP, and tight muons are in addition required to be isolated from nearby tracks.
 
As discussed in Section~\ref{sec:eff_prompt}, the real-lepton efficiencies
are obtained using MC simulation, corrected to match the performance seen in data control samples.
Those efficiencies are shown in Figure~\ref{fig:realrates_elmu2D}, binned in lepton \pT and \abseta. To check for potential dependencies on
the number of additional jets in the events used for the measurements, the efficiencies are derived for different jet multiplicities. No significant differences between the
real-lepton efficiencies are observed.

\subsubsection{Fake/non-prompt lepton efficiencies}
 
The fake/non-prompt lepton efficiencies are measured with same-charge electron--muon~($e\mu$) or muon--muon~($\mu\mu$) data events using a tag-and-probe method, where one \enquote{tag
lepton} with very stringent requirements on momentum and isolation~($\pT > \SI{40}{\GeV}$, $\textrm{max}(\pT^{\textrm{cone}20}, \, \et^{\textrm{cone}20})/ \pT < 0.01$)
is selected and the remaining \enquote{probe lepton} is used for the efficiency measurement. Events with more than two leptons are not considered. Only the aforementioned baseline electron or muon requirements are used to select
the probe lepton.
The samples are dominated by events containing at least one fake/non-prompt lepton,
and are orthogonal to the signal regions,
which require a minimum of three tight lepton candidates.
In addition, \ttZ production is negligible in these samples.
 
\begin{table}[!htb]
\footnotesize
\caption{\label{tab:fake_control_region} Definition of the fake-lepton control regions used for the electron~($e$-fakes-CR) and muon~($\mu$-fakes-CR) fake-efficiency
measurements. $N_{\ell}$ is the number of leptons ($\ell$), while $N_{\textrm{jets}}$ ($N_{b\textrm{-jets}}$) is the number of jets~($b$-tagged jets, see text),
respectively. For additional details of the definitions of physics objects, see Ref.~\cite{TOPQ-2018-08}.}
\def\arraystretch{1.3}
\begin{center}
\begin{tabular}{l| c c c c c}
\toprule
Region         &  $N_{\ell}$~$(\ell=e,\mu)$   & $\pt$~$(\ell)$   & $\pt$~(jets)      & $N_{\textrm{jets}}$  &  $N_{b\textrm{-jets}}$  \\
\midrule
$e$-fakes-CR   &  $=2$~($e^{\pm}\mu^{\pm}$)   & $> \SI{7}{\GeV}$ & $> \SI{25}{\GeV}$ & $\geq 1$             &  $\geq 1$ \\
$\mu$-fakes-CR &  $=2$~($\mu^{\pm}\mu^{\pm}$) & $> \SI{7}{\GeV}$ & $> \SI{25}{\GeV}$ & $\geq 1$             &  $\geq 1$ \\
\bottomrule
\end{tabular}
\end{center}
\end{table}
The definition of the regions used for the fake-efficiency measurements is summarised in Table~\ref{tab:fake_control_region}. For the electron fake efficiencies an $e\mu$
signature is used, with the muon being used as the tag lepton.
For the muon fake efficiencies, the $\mu\mu$ region is used,\footnote{In $\mu\mu$ events with both muons satisfying the tag-selection, the one with the higher \pt is
chosen as the tag lepton.} since the $e\mu$ region also contains unwanted events where an electron with an incorrectly measured charge is selected as the tag lepton and paired with a real muon.
 
\begin{figure}[htbp]
\centering
\subfloat[]{\includegraphics[width=0.8\textwidth]{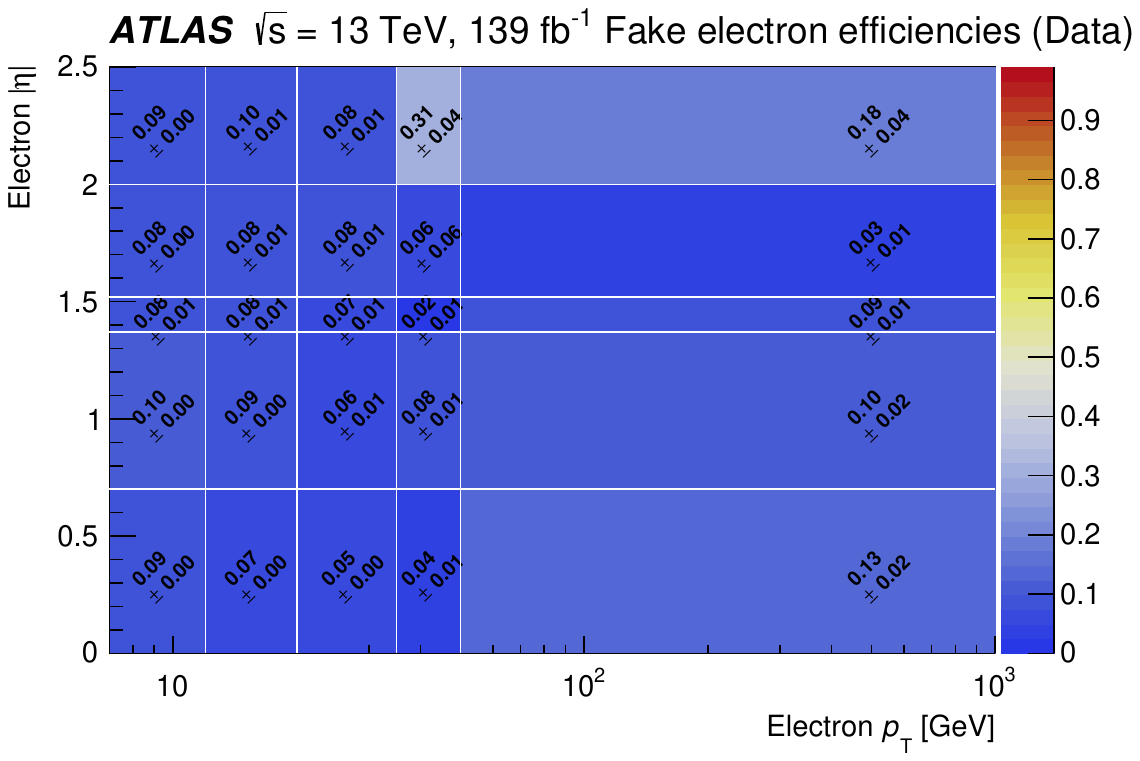}} \\
\subfloat[]{\includegraphics[width=0.8\textwidth]{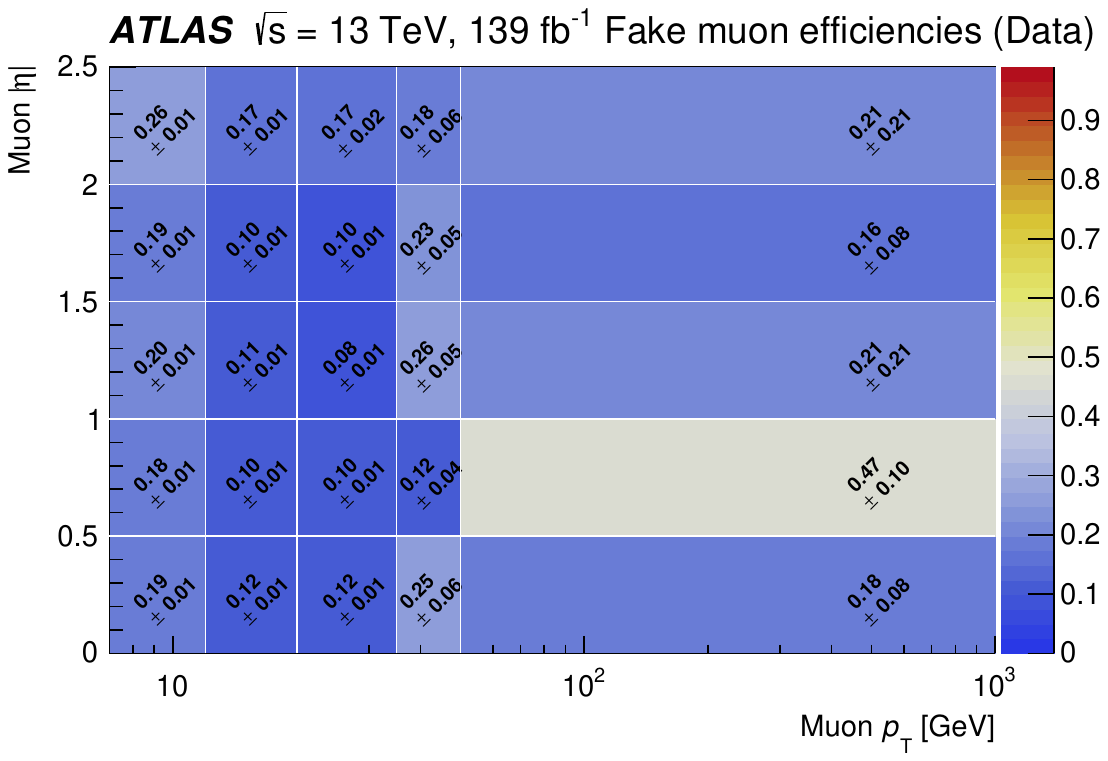}}
\caption{The two-dimensional fake/non-prompt lepton efficiencies measured for (a)~electrons and (b)~muons, in bins of \pT and \abseta of the leptons. The last \pt bin is inclusive.
The indicated uncertainties show only the statistical errors in the given bins. Compared to the real-lepton efficiencies shown in Figure~\ref{fig:realrates_elmu2D}, the fake/non-prompt lepton efficiencies depend
much more on the specifications of the analysis.}
\label{fig:fakerates_elmu2D}
\end{figure}
 
Real-lepton background processes (primarily diboson and $t\bar{t}w$ production) leading to the same-charge dilepton signature are estimated using MC-simulated events and are subtracted from data to obtain an unbiased
efficiency measurement. The contribution from electrons with misassigned charge in the same-charge $e\mu$ region is also subtracted using estimates from MC-simulated events.
 
After this subtraction, the dominant source of fake/non-prompt leptons is found to be heavy-flavour hadron decays. The fake efficiencies, binned in lepton \pT and \abseta,
are shown in Figure~\ref{fig:fakerates_elmu2D}.
 
It is assumed that since the loose and tight lepton selection criteria depend on quantities related to the lepton itself or to its immediate surroundings, the chosen parameterisation
captures the main variations in the fake efficiencies, and residual dependencies on the event environment can be covered by systematic uncertainties~(see Section~\ref{sec:systematics}).
Indeed, in the simulation, the dependence of the fake efficiencies on the number of light-flavour jets or $b$-tagged\footnote{Jets containing $b$-hadrons are identified~(tagged) by the
MV2c10 $b$-tagging algorithm~\cite{FTAG-2018-01}. The algorithm uses a multivariate discriminant with quantities such as the impact parameters of associated tracks, and well-reconstructed
secondary vertices.} jets is mild. The uncertainties are evaluated by comparing, in the simulation, the $\varepsilon_f$ values in the control regions described in Table~\ref{tab:fake_control_region} to the values in the signal regions.
These differences, evaluated as a function of \pt
and \abseta, are applied as a systematic uncertainty of the fake-efficiency measurement~(as discussed in Section~\ref{sec:systematics}), and are of the order of 10\%--20\%, except for muons
with $\pt > \SI{50}{\GeV}$, for which they reach 40\%. Furthermore, normalisation uncertainties are considered for the real-lepton background processes, which are subtracted in the fake-efficiency
measurement. They are evaluated by scaling the real-lepton background processes upwards and downwards within their cross-section uncertainties before the subtraction
and using the differences between the modified and nominal efficiencies as uncertainties, which are added in quadrature to the aforementioned uncertainties.

\subsubsection{Results in the fake/non-prompt lepton validation regions}

To validate the performance of the method, predictions are obtained and compared with data in two dedicated validation regions called \enquote{VR-$3\ell$-$1b3j$}
and \enquote{VR-$3\ell$-$1b3j$-no$Z$}, which have a larger proportion of fake/non-prompt leptons than is expected in the signal regions. The definitions of these
two validation regions are summarised in Table~\ref{tab:fake_val_regions}. No charge requirements are placed on the reconstructed lepton candidates in these regions.
 
\begin{table}[!htb]
\footnotesize
\caption{\label{tab:fake_val_regions} Definition of the fake/non-prompt lepton enriched validation regions, VR-$3\ell$-$1b3j$ and VR-$3\ell$-$1b3j$-no\Zboson.
$N_{\ell}$ stands for the number of leptons~(${\ell}$), while $N_{\textrm{jets}}$~($N_{b\textrm{-jets}}$) is the number of jets~($b$-tagged jets), respectively.
The leptons are ordered by decreasing \pt~($\ell_{1,2,3}$). The mass~($m$) variables are discussed in the text. For additional details of the object definitions, see
Ref.~\cite{TOPQ-2018-08}.}
\def\arraystretch{1.3}
\begin{center}
\begin{tabular}{l| c c c c c c}
\toprule
Region         &  $N_{\ell}$~$(\ell=e,\mu)$      &  \pT~$(\ell_{1,2,3})$  & $\pt$~(jets) &  $|m_{\ell\ell}^{\textrm{SF}}-\mZ|$   & $N_{\textrm{jets}}$  &  $N_{b\textrm{-jets}}$  \\
\midrule
VR-$3\ell$-$1b3j$  &  $=3$        &  $> 27,20,\SI{20}{\GeV}$    & $> \SI{25}{\GeV}$ &   --       & $=3$                  &  $=1$  \\
& (2 tight, 1 loose) &                   &                   &            &                       &     \\
VR-$3\ell$-$1b3j$-no\Zboson  &  $=3$  &  $> 27,20,\SI{20}{\GeV}$  &   $> \SI{25}{\GeV}$ &   $> \SI{10}{\GeV}$        & $=3$                  &  $=1$  \\
& (all tight) &             &           &                       &      &  \\
\bottomrule
\end{tabular}
\end{center}
\end{table}
 
The variable $m_{\ell\ell}^{\textrm{SF}}$ refers to the invariant mass of the same-flavour opposite-charge (SFOC) lepton pair with the invariant mass closest to the \Zboson boson mass. VR-$3\ell$-$1b3j$ is
a region similar to the actual signal regions defined in Ref.~\cite{TOPQ-2018-08}, but without a requirement on the \Zboson mass for the (SFOC) lepton
candidate pair. Therefore, it contains a higher fraction of fake/non-prompt leptons after the selection. To further enhance the fake/non-prompt lepton contribution, the third-highest-\pt lepton
that satisfies the baseline selection criteria must not satisfy the tight criteria. An additional validation region, VR-$3\ell$-$1b3j$-no\Zboson, is defined by requiring
all three leptons to satisfy the tight selection criteria, but placing a veto on SFOC lepton pairs that have an invariant mass consistent with the $Z$ boson,
thereby enhancing the fake/non-prompt lepton fraction in this region.
Both regions are orthogonal to the analysis signal regions and intended to validate the predictions of the matrix method for different levels of fake/non-prompt lepton contamination.
 
\begin{figure}[htbp]
\centering
\subfloat[]{
\includegraphics[width=0.49\textwidth]{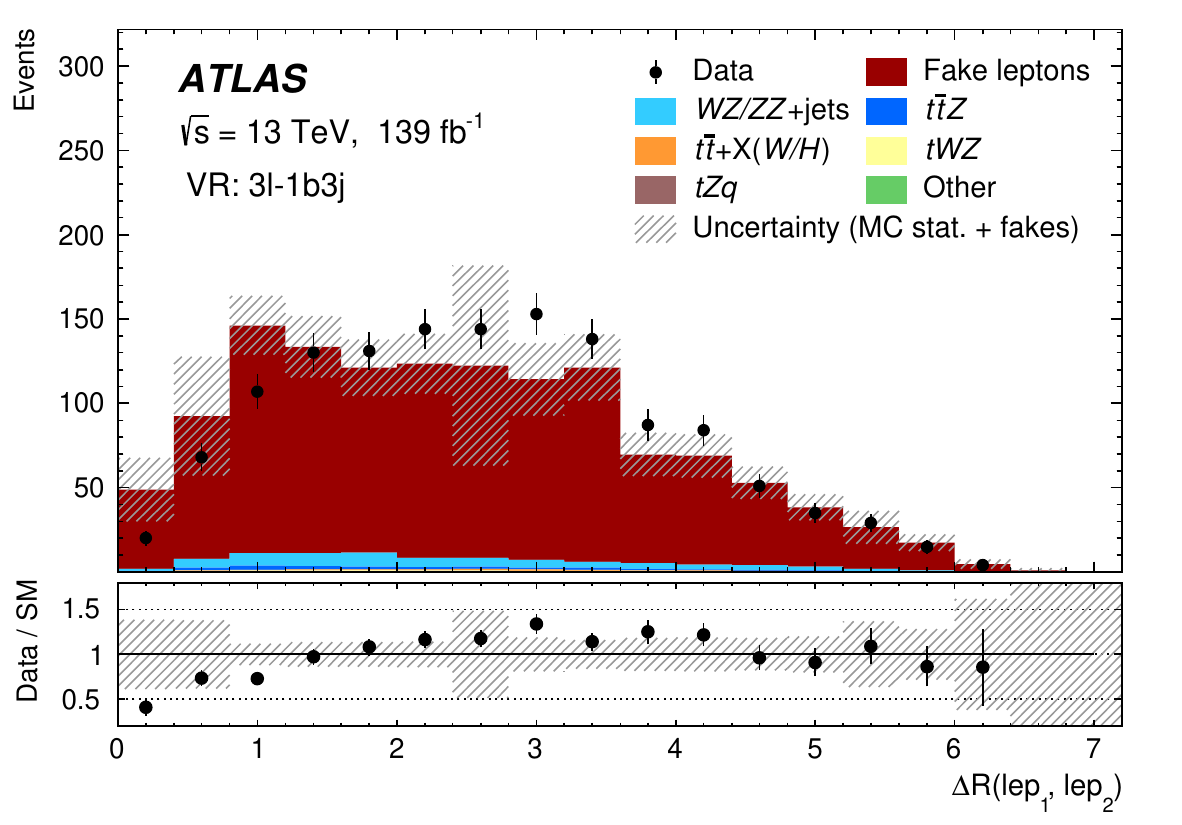}}
\subfloat[]{
\includegraphics[width=0.49\textwidth]{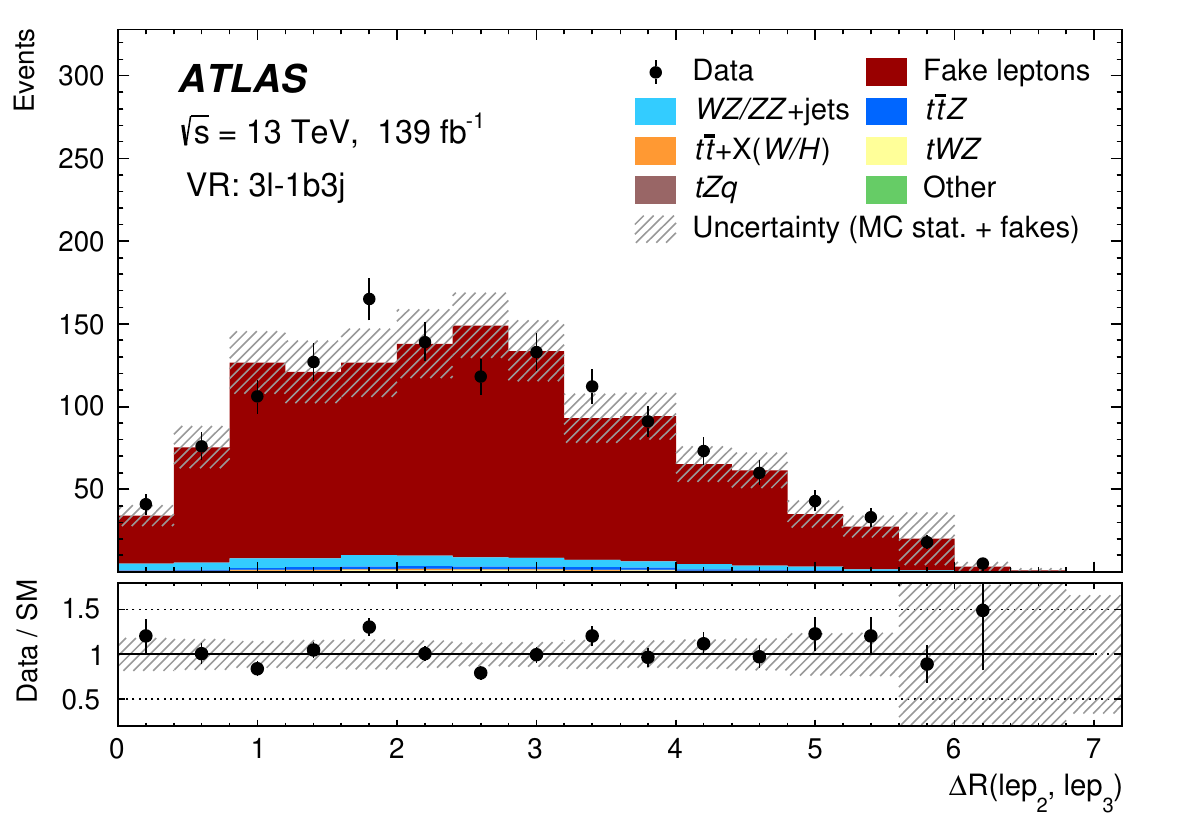}}
\\
\subfloat[]{
\includegraphics[width=0.49\textwidth]{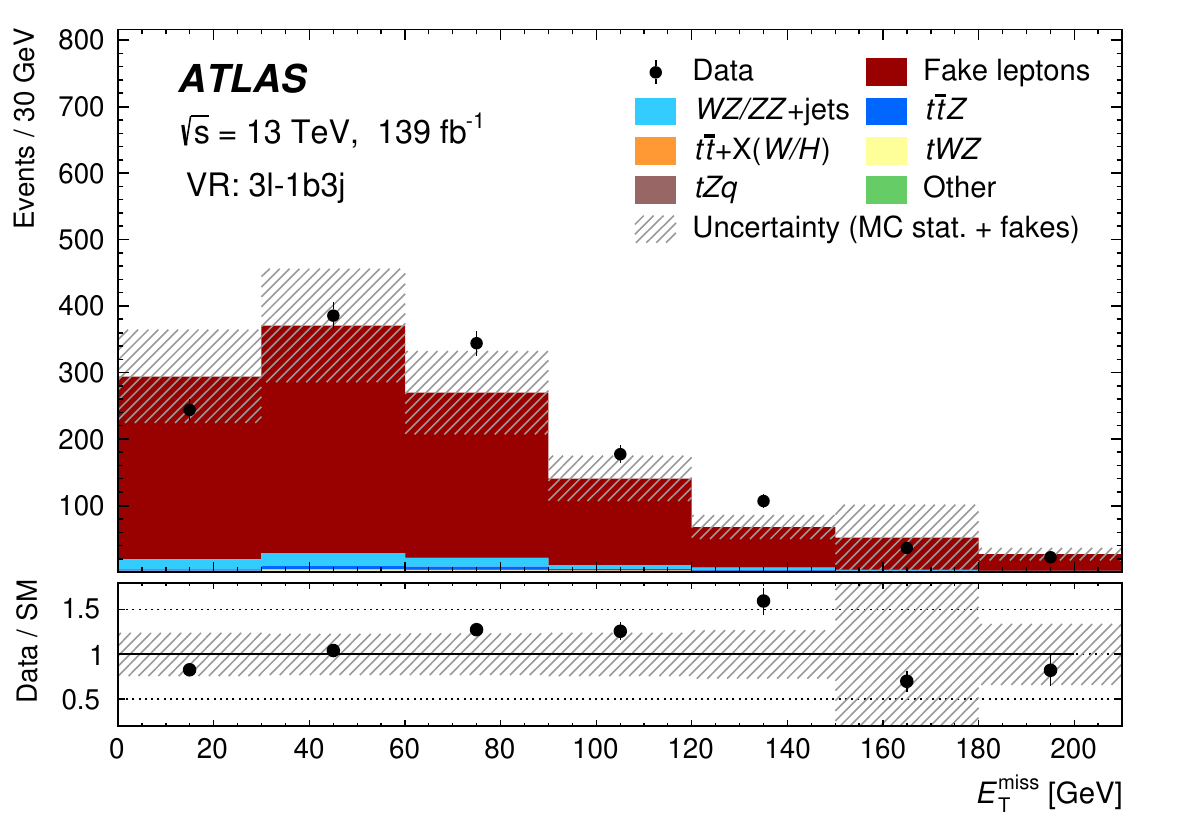}}
\subfloat[]{\label{fig:vr-3l-1b3j_d}
\includegraphics[width=0.49\textwidth]{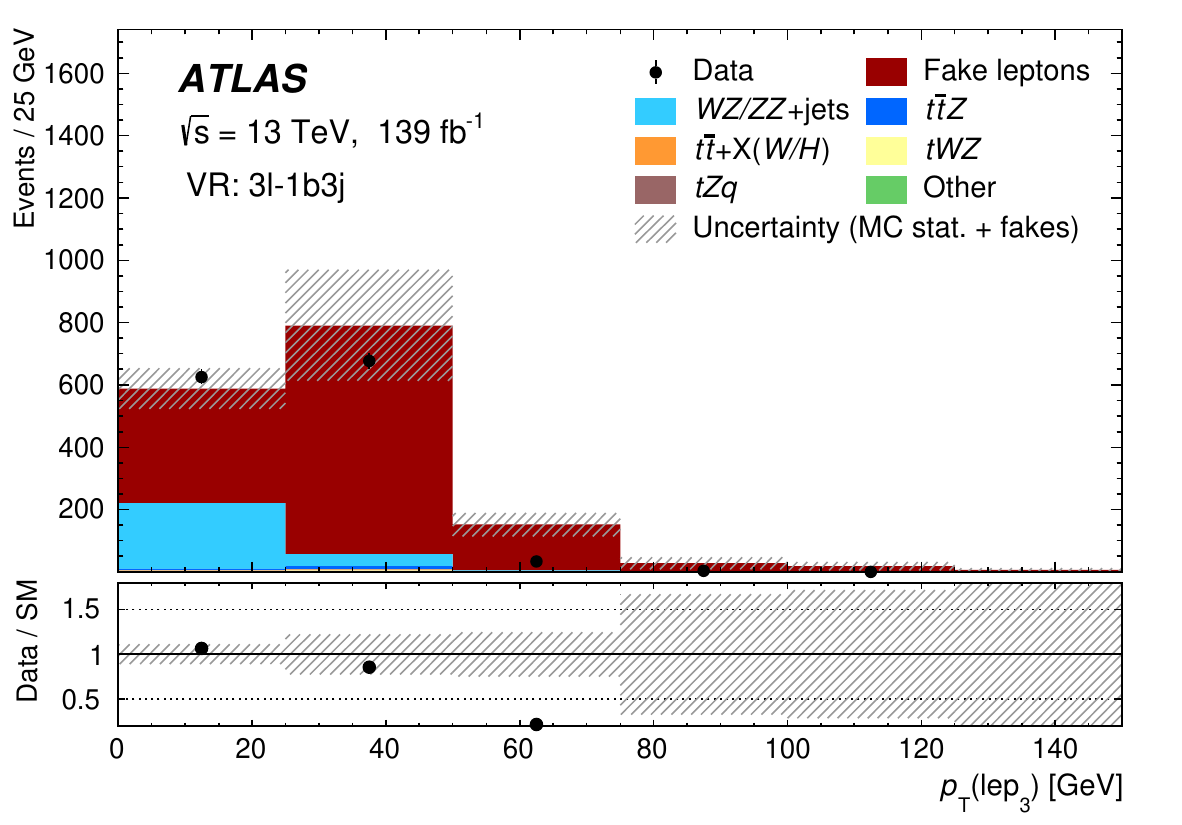}}
\caption{(a)~Angular separation between the leading and second leading (in \pt) lepton candidates, $\DeltaR(\ell_{1}, \ell_{2})$, (b)~angular separation between the second- and third-leading lepton candidates,
$\DeltaR(\ell_{2}, \ell_{3})$, (c)~missing transverse momentum in the event, \met, and (d)~the \pt of the third-leading lepton in VR-$3\ell$-$1b3j$. The processes with three real
leptons are modelled with MC simulation, while the contribution from fake/non-prompt leptons~(dark red) comes from the Poisson likelihood matrix method as described above.  The hatched band shows the uncertainty from the MC statistics and the fake/non-prompt background estimate.
The rightmost bins are inclusive and contain all events above the $x$-axis ranges. The lower panel shows the ratio of data to the total SM prediction~(sum of the real-lepton background contributions estimated with MC samples and the fake/non-prompt lepton contribution estimated with the matrix method).  Further details are available in Ref.~\cite{TOPQ-2018-08}.}
\label{fig:vr-3l-1b3j}
\end{figure}

\begin{figure}[htbp]
\subfloat[]{
\includegraphics[width=0.49\textwidth]{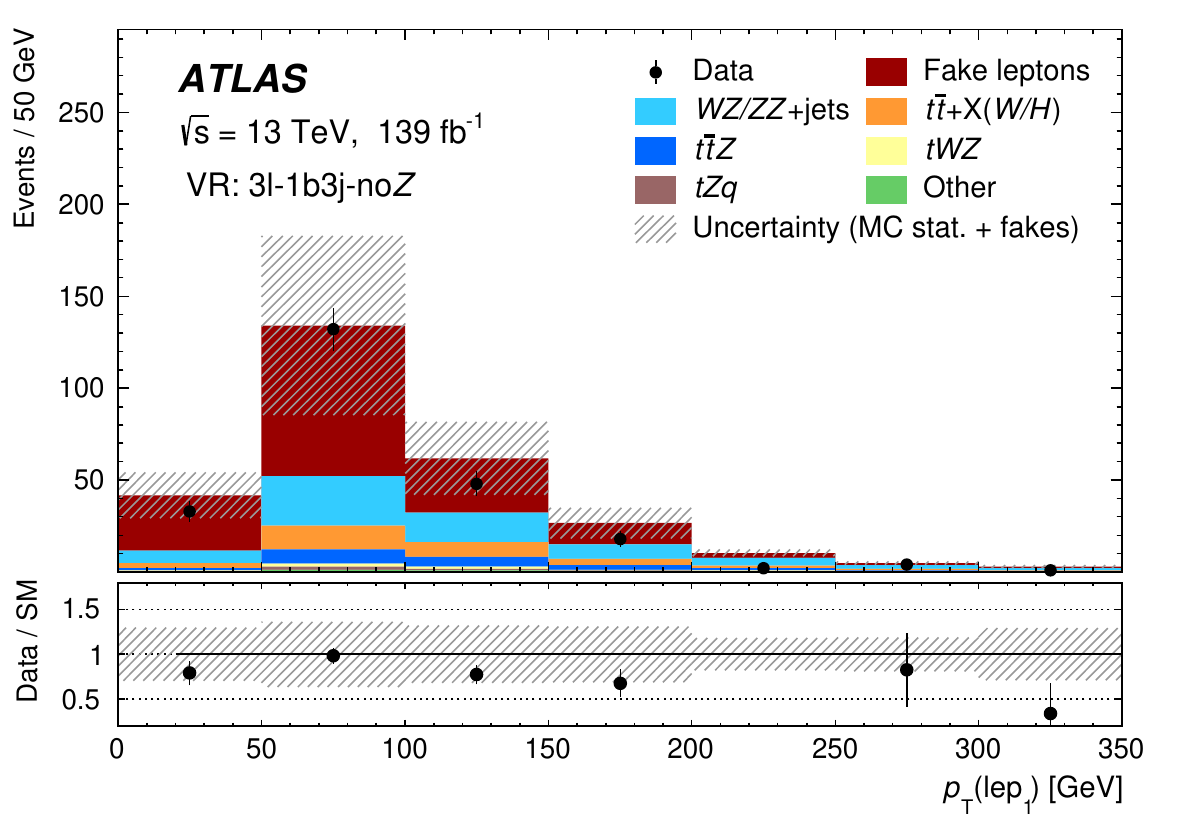}}
\subfloat[]{
\includegraphics[width=0.49\textwidth]{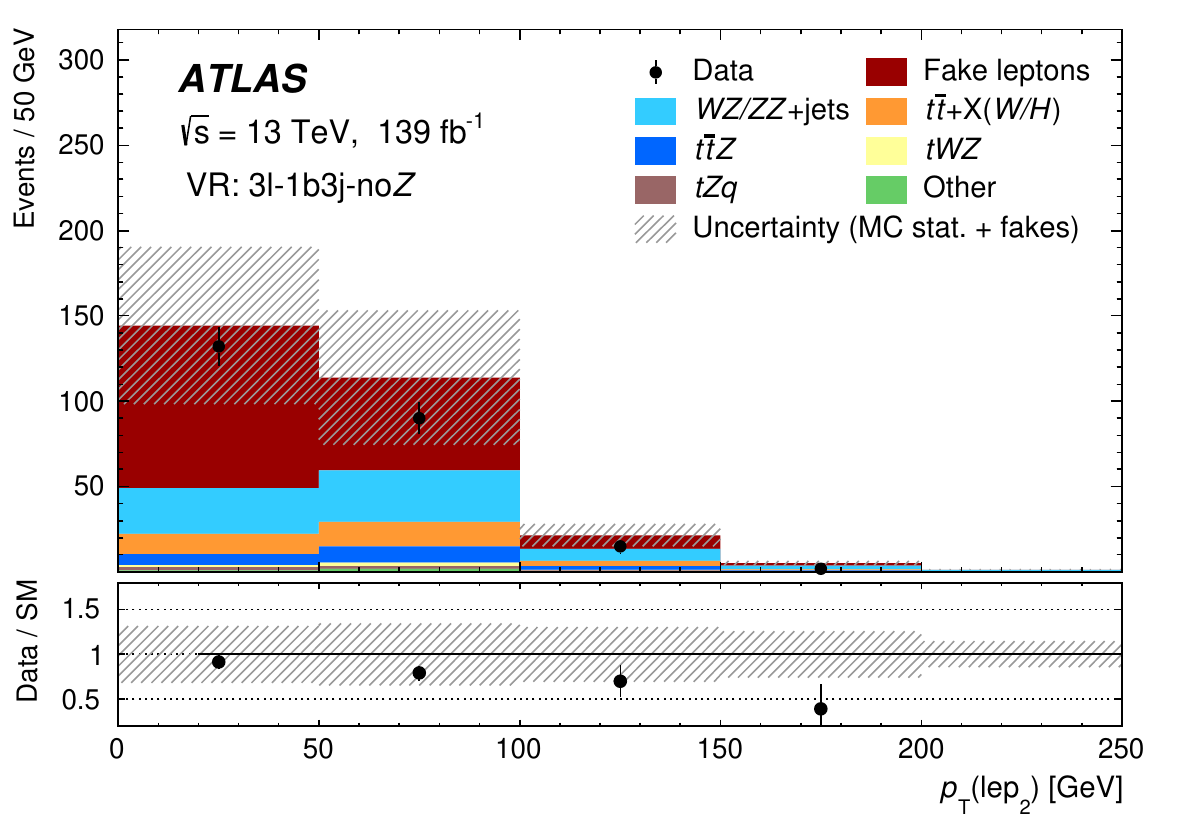}}
\\
\subfloat[]{
\includegraphics[width=0.49\textwidth]{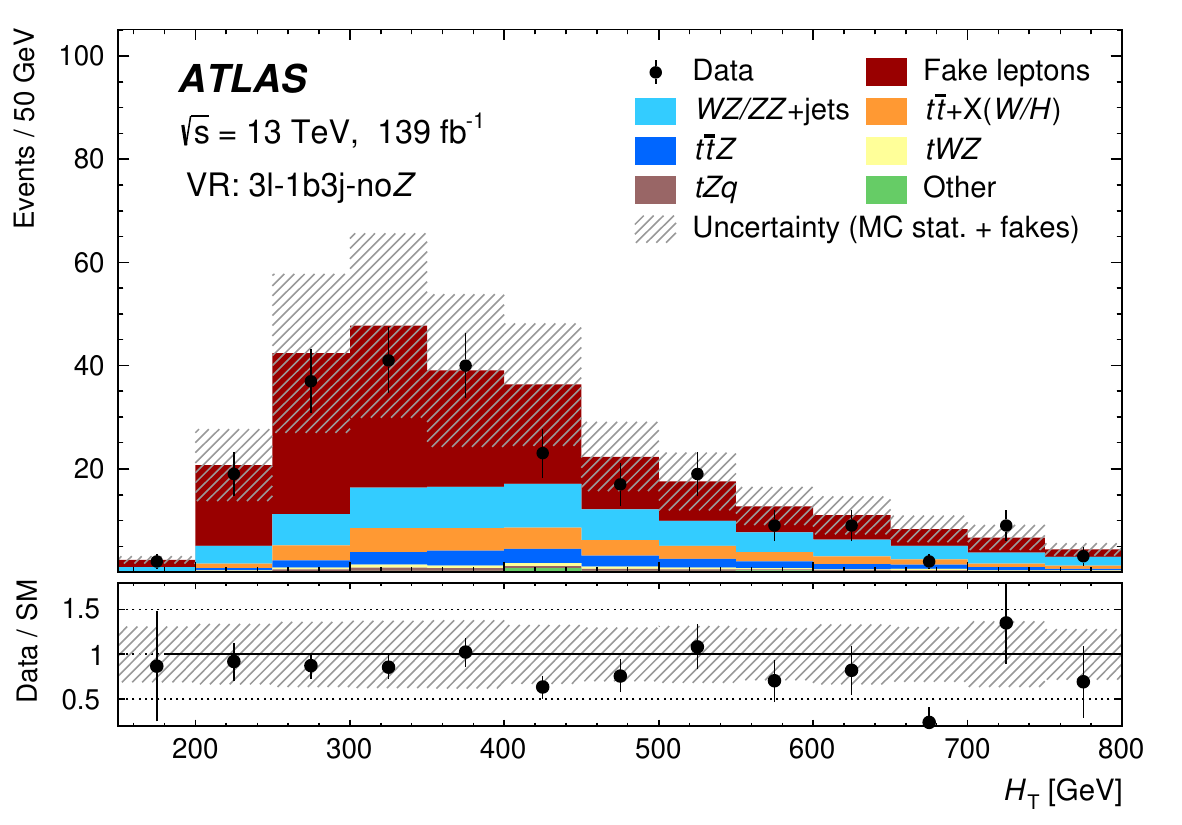}}
\subfloat[]{
\includegraphics[width=0.49\textwidth]{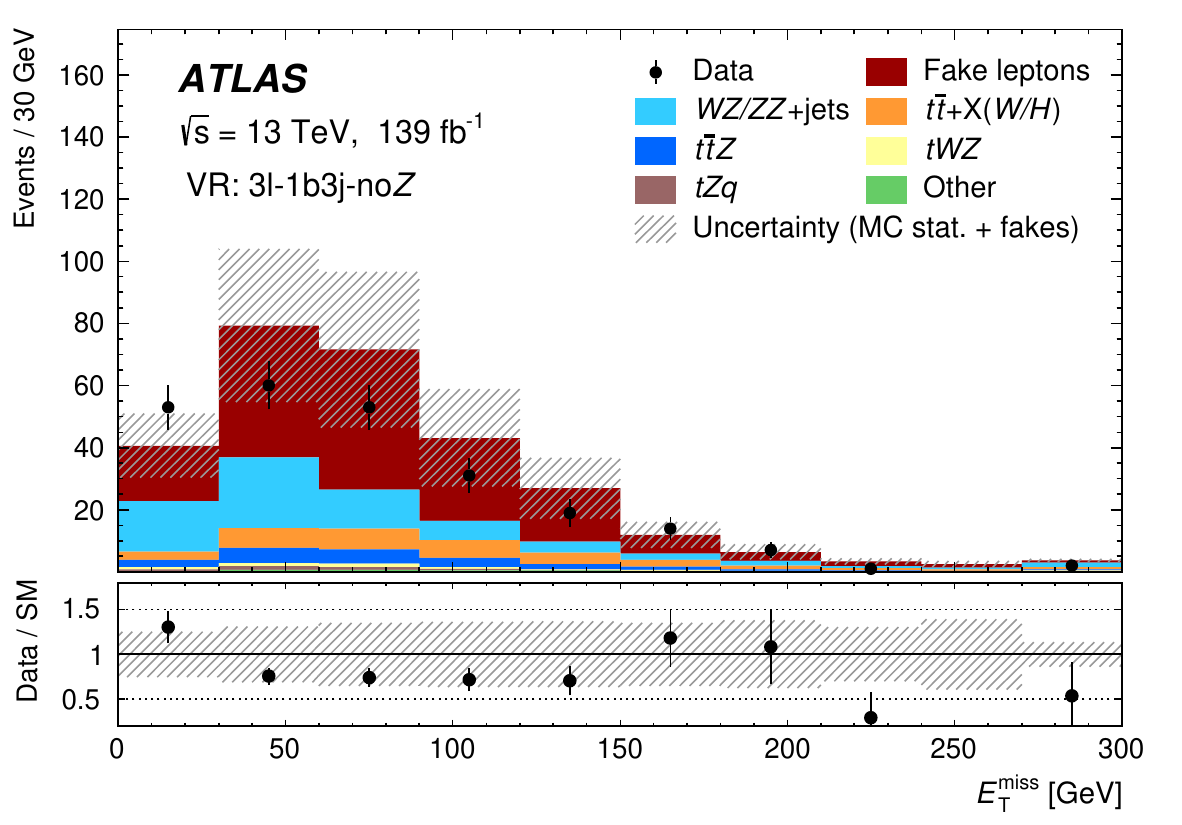}}
 
\caption{Comparisons of the predicted and observed yields in VR-$3\ell$-$1b3j$-no\Zboson, with respect to the \pT of the (a)~leading and (b)~subleading lepton candidates, (c) the scalar sum of the lepton and jet transverse momenta, \HT, and (d)~the missing transverse momentum in the event, \met. The processes with three real leptons are modelled with
MC simulation, while the contribution from fake/non-prompt leptons~(dark red) comes from the Poisson likelihood matrix method as described above.  The hatched band shows the uncertainty from MC statistics and the fake/non-prompt background estimate.
The rightmost (leftmost) bins are inclusive and contain all events above (below) the $x$-axis ranges. The lower panel shows the ratio of data to the total SM prediction~(sum of the real-lepton background contributions estimated with MC samples and the fake/non-prompt lepton contribution estimated with the matrix method).  Further details are available in Ref.~\cite{TOPQ-2018-08}.}
\label{fig:vr-3l-1b3j-noZ}
\end{figure}

Some example distributions are shown in Figure~\ref{fig:vr-3l-1b3j} for VR-$3\ell$-$1b3j$ and Figure~\ref{fig:vr-3l-1b3j-noZ} for VR-$3\ell$-$1b3j$-no\Zboson. The processes with three
real leptons~(modelled with MC simulations) plus the prediction from the matrix method can be compared with data in these regions.
 
The hatched bands in Figures~\ref{fig:vr-3l-1b3j} and~\ref{fig:vr-3l-1b3j-noZ} show only the statistical uncertainties of the MC prediction and the uncertainties
associated with the fake/non-prompt lepton estimates~(i.e.\ no theoretical or detector-related systematic uncertainties are included). The total uncertainty associated with
the fake/non-prompt lepton estimate itself contains a systematic component, which is evaluated from variations of the input fake/real efficiencies~(described in the previous section),
and the statistical uncertainty of the data sample to which the Poisson likelihood matrix method is applied.\footnote{The total uncertainties of the fake/non-prompt lepton estimates may be different from
the uncertainties reported in Ref.~\cite{TOPQ-2018-08}, as the number of loose leptons in these validation regions is larger than in the signal region of the \ttZ analysis and,
therefore, the statistical uncertainties are smaller.}
 
There is generally good agreement between the data and the total background estimate, except that the background is overestimated at low $\DeltaR(\ell_1, \ell_2)$. One contribution to that difference is that the two leading (two highest-\pt) lepton candidates are likely to be real, yet when they are near each other they have a lower
efficiency for satisfying the isolation criteria, and thus are misinterpreted as fake/non-prompt leptons by the matrix method. Analyses that are sensitive to such issues may benefit from
imposing a minimum \DeltaR requirement between leptons. A discrepancy is also observed in the higher \pt bins of Figure~\ref{fig:vr-3l-1b3j_d}. Events in these bins have three leptons with
\pt above \SI{50}{\GeV}. As shown in Figures~\ref{fig:realrates_elmu2D} and~\ref{fig:fakerates_elmu2D}, only a single \pt bin above \SI{50}{\GeV} is available for measuring $\varepsilon_\mathrm{r}$
and $\varepsilon_\mathrm{f}$, due to the limited number of events in the control regions, so variations above \SI{50}{\GeV} may be missed. This point was not investigated thoroughly since only a small
fraction of events in the analysis were impacted.

\FloatBarrier

\subsection{Model-independent search for new phenomena in multi-lepton final states}

Many interesting new models for BSM physics predict final states with three or more leptons. The general multi-lepton search for new phenomena~\cite{EXOT-2019-36}
agnostically considers such final states. Its aim is to be sensitive to BSM phenomena in often-overlooked corners of phase space.
The background estimation for the multi-lepton search uses MC predictions to account for events that contain only real leptons, and the fake-factor
method for events containing at least one fake/non-prompt lepton. The dominant sources of fake/non-prompt leptons are semileptonic heavy-flavour
decays~(primarily of $b$-hadrons), light-hadron decays, and misidentification of light hadrons as leptons. These mainly arise in \Zjets and \ttbar events.
 
\subsubsection{Fake/non-prompt lepton selection}
 
The fake factors are measured using events with a single lepton candidate, where the selection targets QCD dijet events. In order to prevent a bias in the fake factor
due to trigger selection criteria~\cite{TRIG-2018-01,TRIG-2018-05}, the selected events are required to have fired a loose single-lepton trigger where isolation requirements are not imposed. However, due to the high rate of events that
pass such triggers, a prescale factor is applied, which reduces the number of events available for measuring the fake factors (the luminosity of the prescaled samples is $\sim 0.5$ fb$^{-1}$ for electrons and  $\sim 3.7$ fb$^{-1}$ for muons). Baseline lepton candidates must pass a common object selection, as detailed in Ref.~\cite{EXOT-2019-36}. Electron candidates are required to pass either the \enquote{Loose} ID WP with calorimeter- and track-based isolation requirements, or the \enquote{Tight} ID WP with no isolation requirement.  Baseline muon candidates are required to pass the \enquote{Medium} ID WP~(\enquote{HighPt} for $\pt > \SI{300}{\GeV}$). Only leptons with $\pt > \SI{25}{\GeV}$
and satisfying the longitudinal and transverse track impact parameter requirements (see Sections~\ref{sec:elID} and~\ref{sec:muID}) are considered.
 
More stringent selection criteria are imposed for tight lepton candidates.  For muons, where fake/non-prompt candidates are
mainly muons from semileptonic heavy-flavour decays, only the isolation criteria is modified: the tight selection requires that muons satisfy track-based isolation criteria. For electrons, where both light- and heavy-flavour hadrons are a non-negligible source of fake/non-prompt candidates, both the ID and isolation criteria are modified. Tight electrons must satisfy both the \enquote{Tight} ID WP and calorimeter- and track-based isolation
requirements.
 
Additional selection requirements are imposed on the single-lepton sample to ensure a high purity of fake/non-prompt leptons, which reduces the statistical
uncertainty of the computed fake factor: the \MET is required to be~$< \SI{25}{\GeV}$ ($< \SI{40}{\GeV}$) for events with electron (muon) candidates, and
the number of jets in the event is required to be $\geq 1~(2)$ for events with electron~(muon) candidates. For muon-candidate events, there must also be at least
one jet with $\pt > \SI{35}{\GeV}$ (the \enquote{tag jet}), and the azimuthal angle between the muon candidate and this jet is required to be $> 2.7$~radians.
 
The fake factor is binned in both the \pt and \abseta of the lepton candidate. The bins are defined to have tolerable statistical uncertainties
while preventing sizeable differences between the fake-factor values in adjacent bins. The highest muon-\pt bin includes all muon candidates with  $\pt > \SI{80}{\GeV}$, as
there are very few fake/non-prompt muons at higher transverse momenta.

\subsubsection{Fake factors}
 
\begin{figure}[htbp]
\centering
\subfloat[]{\includegraphics[width=0.45\textwidth]{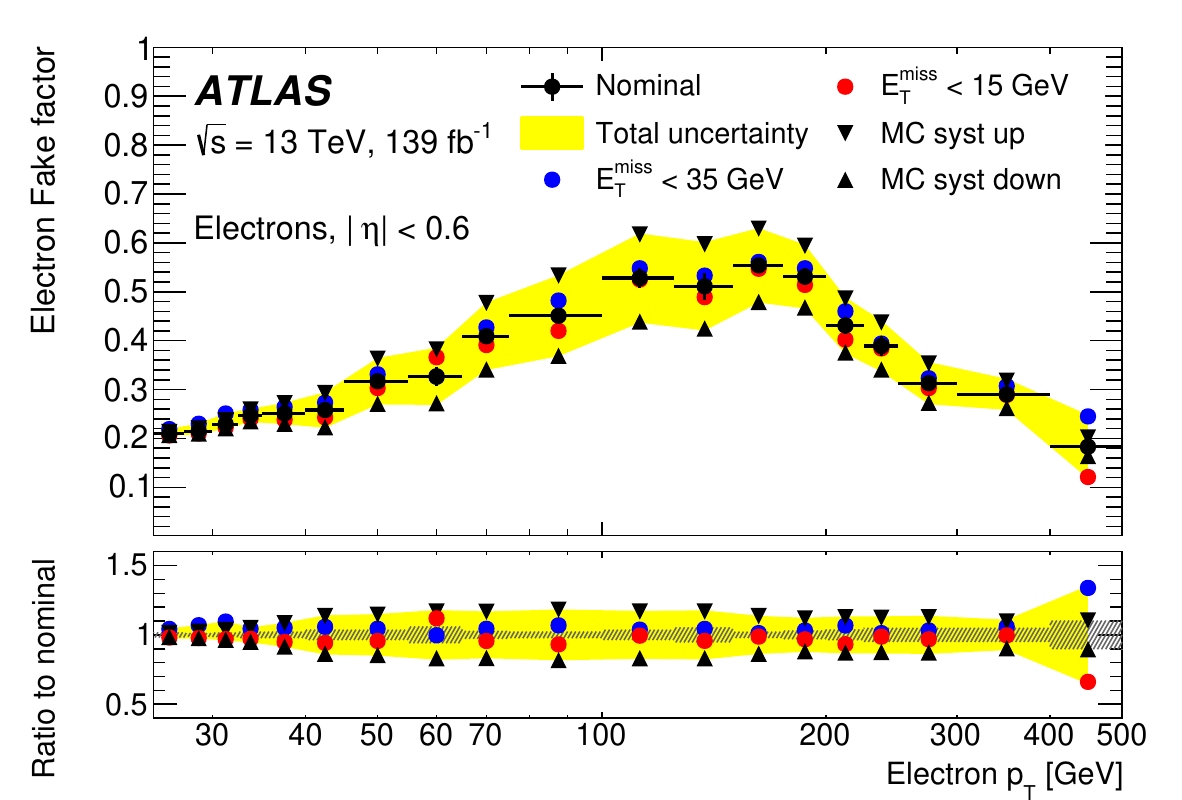}}
\subfloat[]{\includegraphics[width=0.45\textwidth]{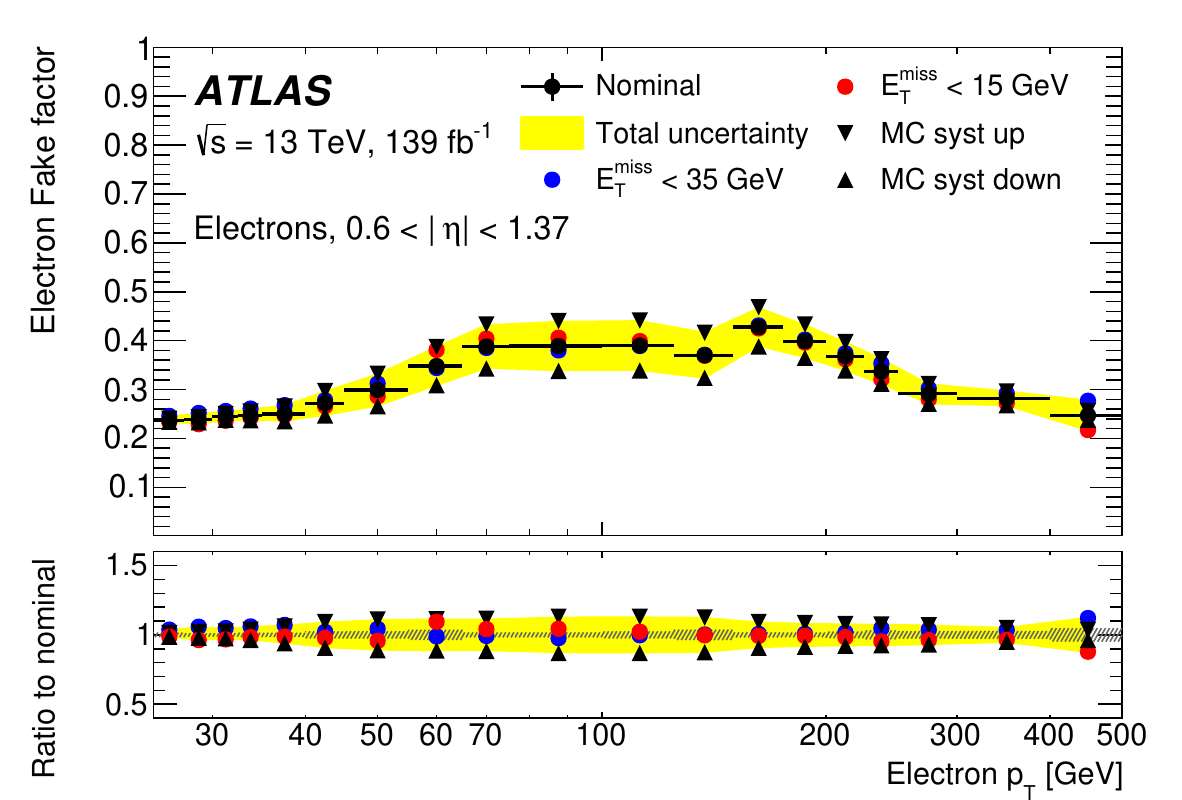}}\\
\subfloat[]{\includegraphics[width=0.45\textwidth]{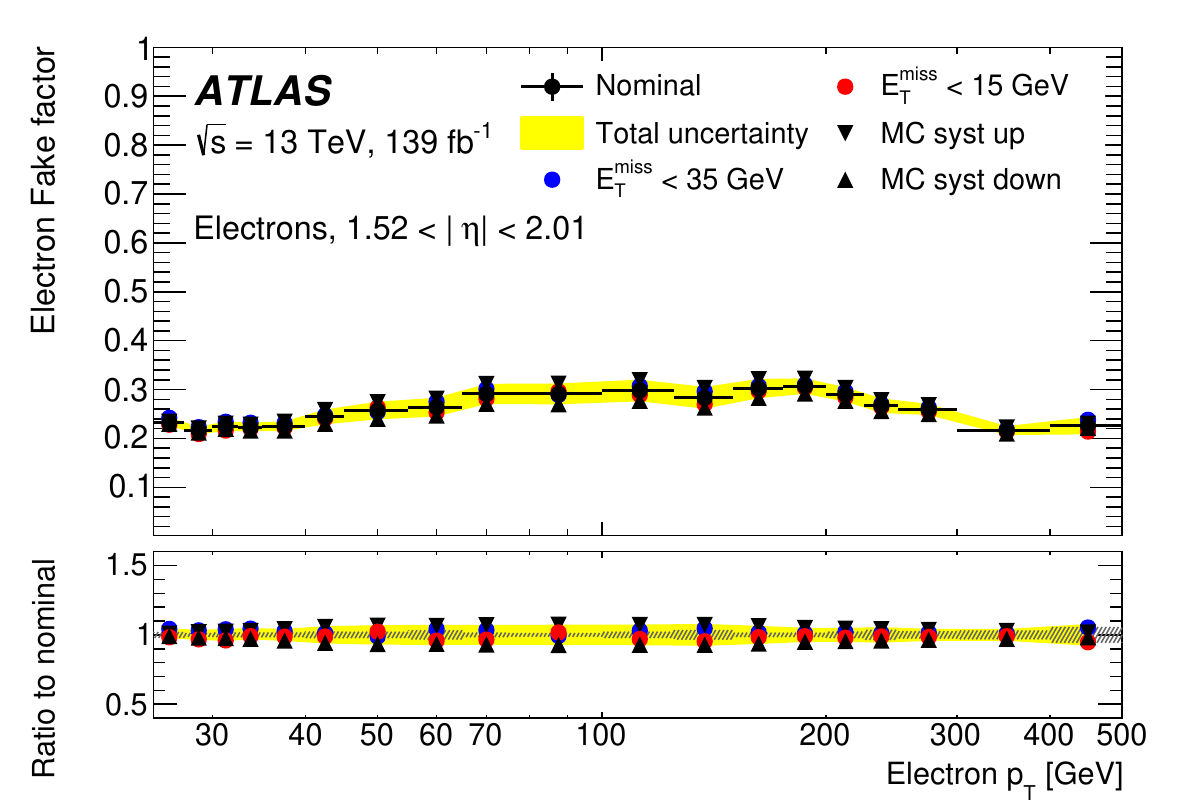}}
\subfloat[]{\includegraphics[width=0.45\textwidth]{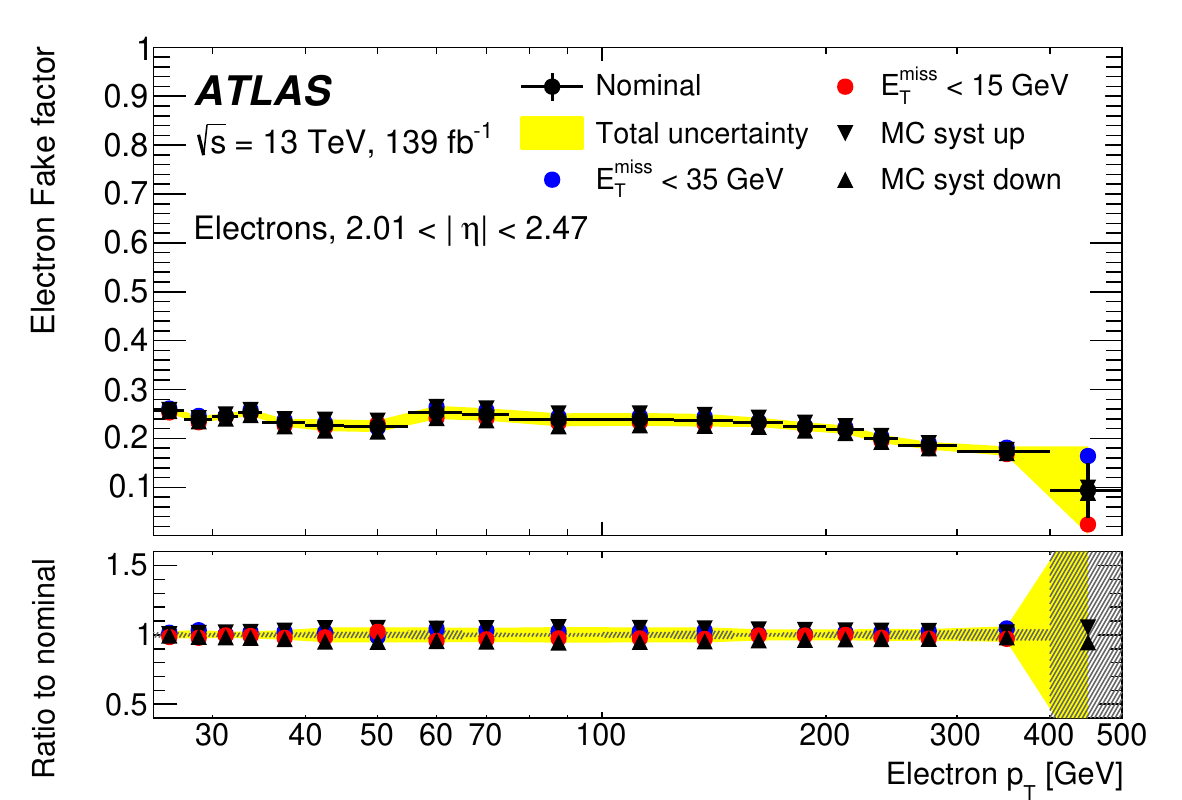}}\\
\caption{Measured fake factors for electrons as a function of \pt for different \abseta ranges,
and their dependence on individual variations of parameters of the measurement,
used to determine systematic uncertainties.
The \enquote{MC syst} uncertainty
covers variations in the model used to subtract the real-lepton contribution in the control regions used to measure $\varepsilon_{\text{f}}$. The combined impact
of statistical and systematic uncertainties added in quadrature is indicated by the shaded yellow area~(the grey area represents only the statistical uncertainty).}
\label{fig:ml_fakesyst_el}
\end{figure}

\begin{figure}[htbp]
\centering
\subfloat[]{\includegraphics[width=0.45\textwidth]{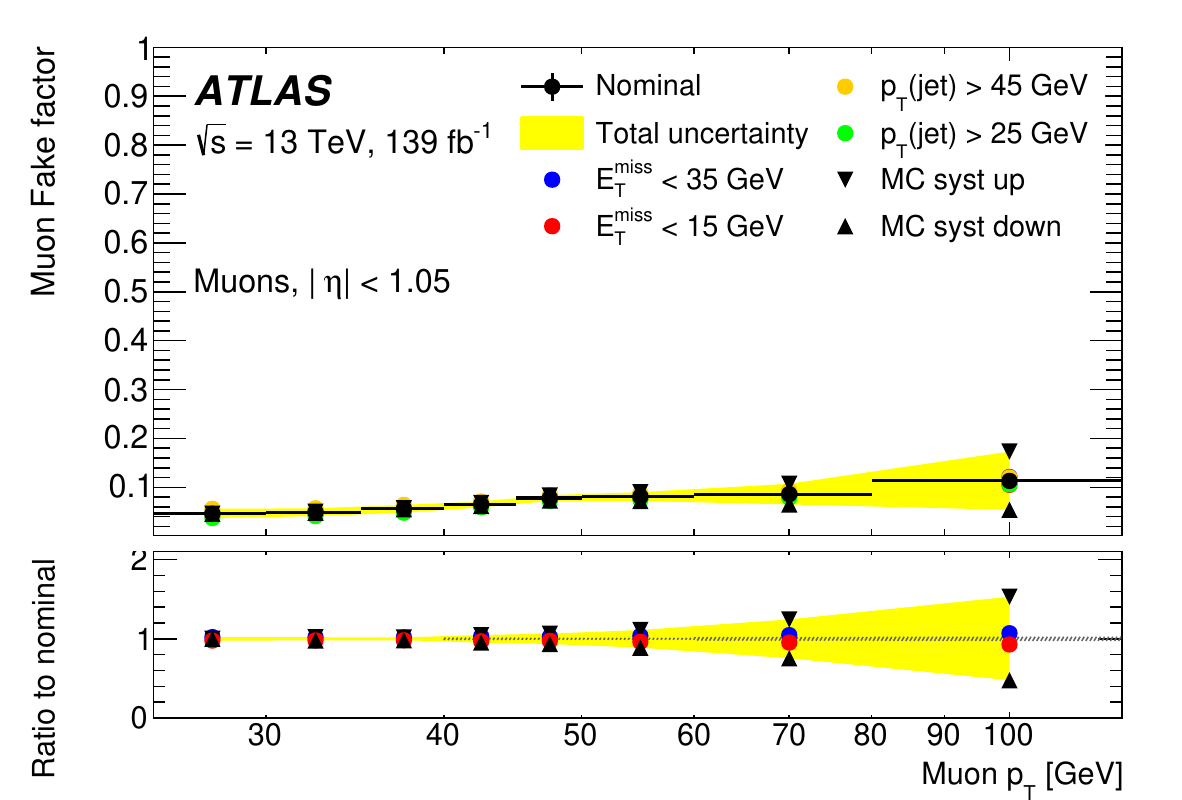}}
\subfloat[]{\includegraphics[width=0.45\textwidth]{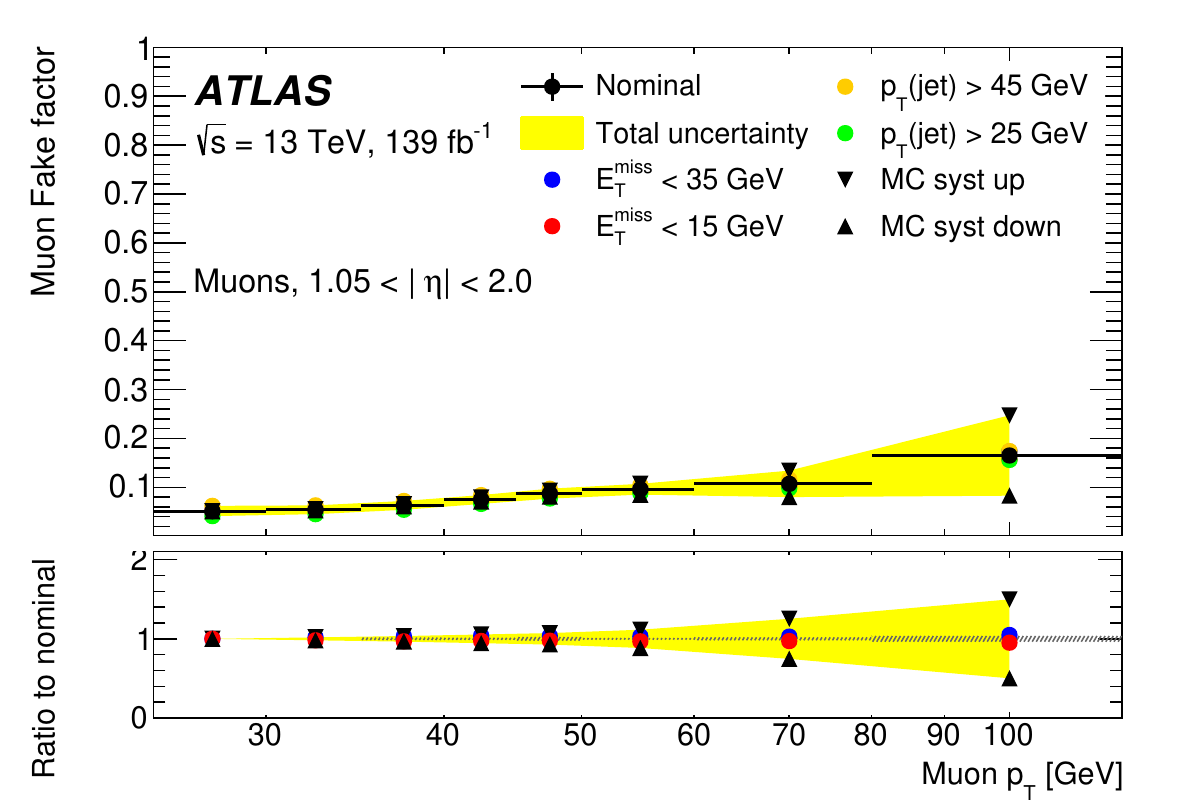}}\\
\subfloat[]{\includegraphics[width=0.45\textwidth]{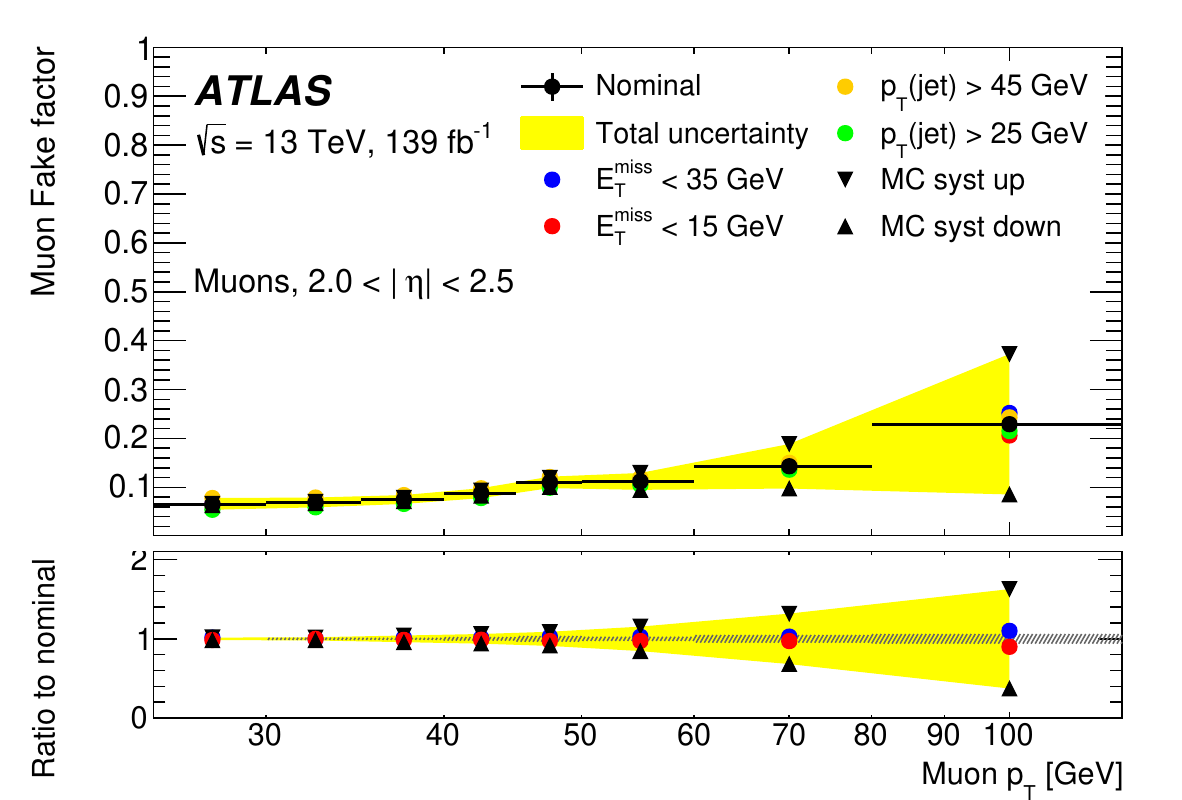}}
\caption{Measured fake factors for muons as a function of \pt for different \abseta ranges,
and their dependence on individual variations of parameters of the measurement,
used to determine systematic uncertainties.
The combined impact of statistical and
systematic uncertainties added in quadrature is indicated by the shaded yellow area~(the grey area represents only the statistical uncertainty).}
\label{fig:ml_fakesyst_mu}
\end{figure}

The fake factors calculated for the multi-lepton analysis are shown in Figures~\ref{fig:ml_fakesyst_el} and~\ref{fig:ml_fakesyst_mu} for electrons and muons.
The uncertainties in the fake factors are categorised and evaluated as described below.
 
There are statistical uncertainties due to the data and MC sample sizes in each fake-factor bin, which due to their small size are summed in
quadrature into a single uncertainty.
 
In the regions enriched in fake/non-prompt leptons, MC predictions are used to subtract the real-lepton contribution from the data. The uncertainties in the MC contributions
are propagated to the fake factors. The main contributions from real leptons in the single-lepton regions are from \Zjets, \Wjets and \ttbar events. For the $Z/\Wjets$ processes, an
uncertainty of 5\% is applied to the cross-section. For the \ttbar process,  uncertainties on the cross section due to renormalization and factorization scales ($\sim 3$\%) and PDF ($\sim 4$\%) are applied.
 
Extrapolating the fake factors from the single-lepton sample to the multi-lepton samples used in the analysis introduces an uncertainty because these samples differ in the kinematic distributions of fake/non-prompt leptons, and possibly also in the fake/non-prompt lepton composition.
Two uncertainties are included to address the bias caused by imposing a \met upper bound in the fake/non-prompt lepton estimation sample, and by imposing a \pt
requirement on the tag jet in the fake/non-prompt muon estimation sample. These uncertainties are estimated by varying the requirements on these variables upwards
and downwards by \SI{10}{\GeV}.\footnote{Only the \met variation is displayed separately in the plots, as the variation of the jet \pt results in an uncertainty that is too small to be visible.}
Plots showing the impact of these systematic effects on the fake factors are given in Figures~\ref{fig:ml_fakesyst_el} and~\ref{fig:ml_fakesyst_mu}.
 
\begin{figure}[htbp]
\centering
\subfloat[]{\includegraphics[width=0.43\textwidth]{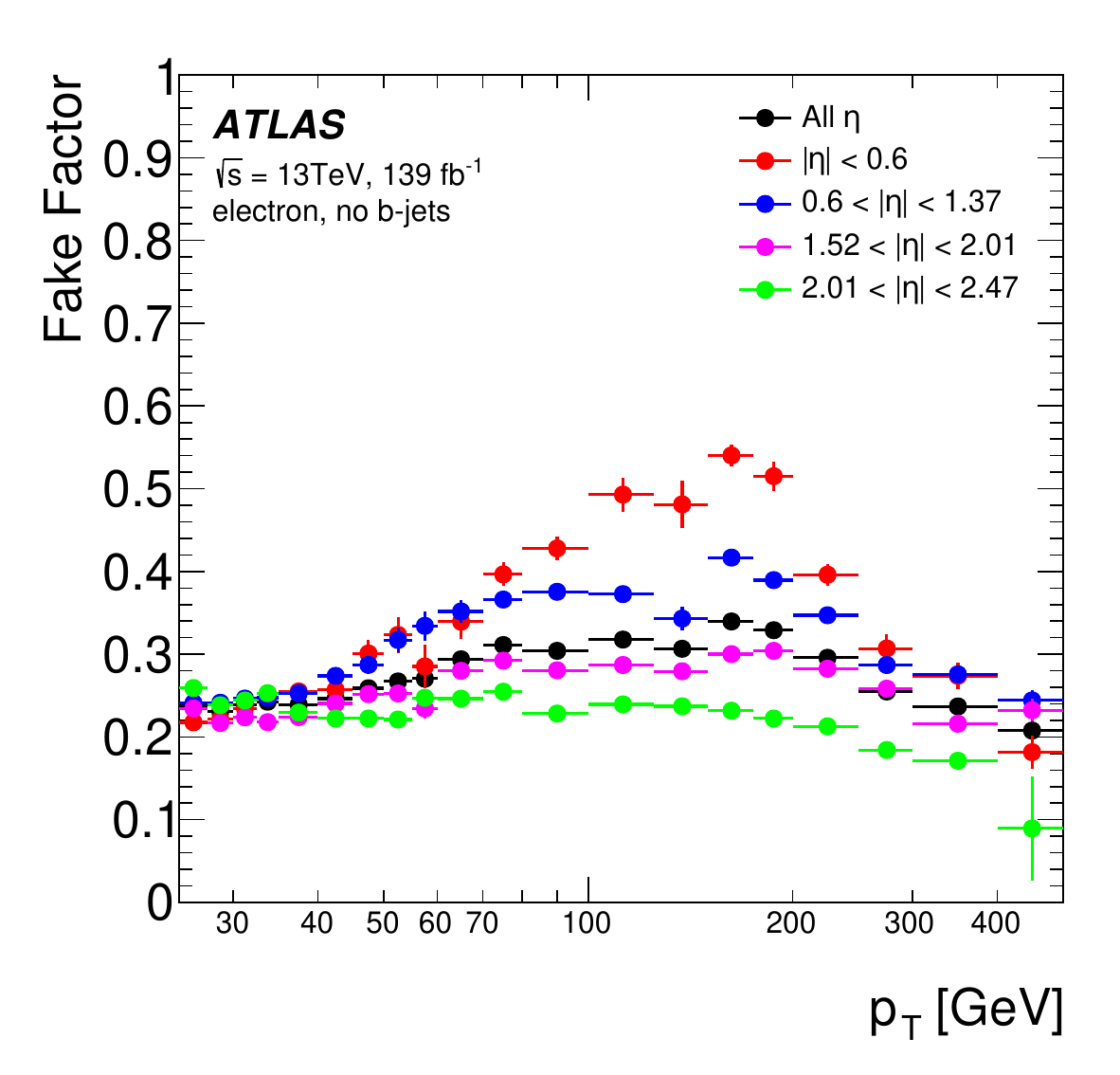}}
\subfloat[]{\includegraphics[width=0.43\textwidth]{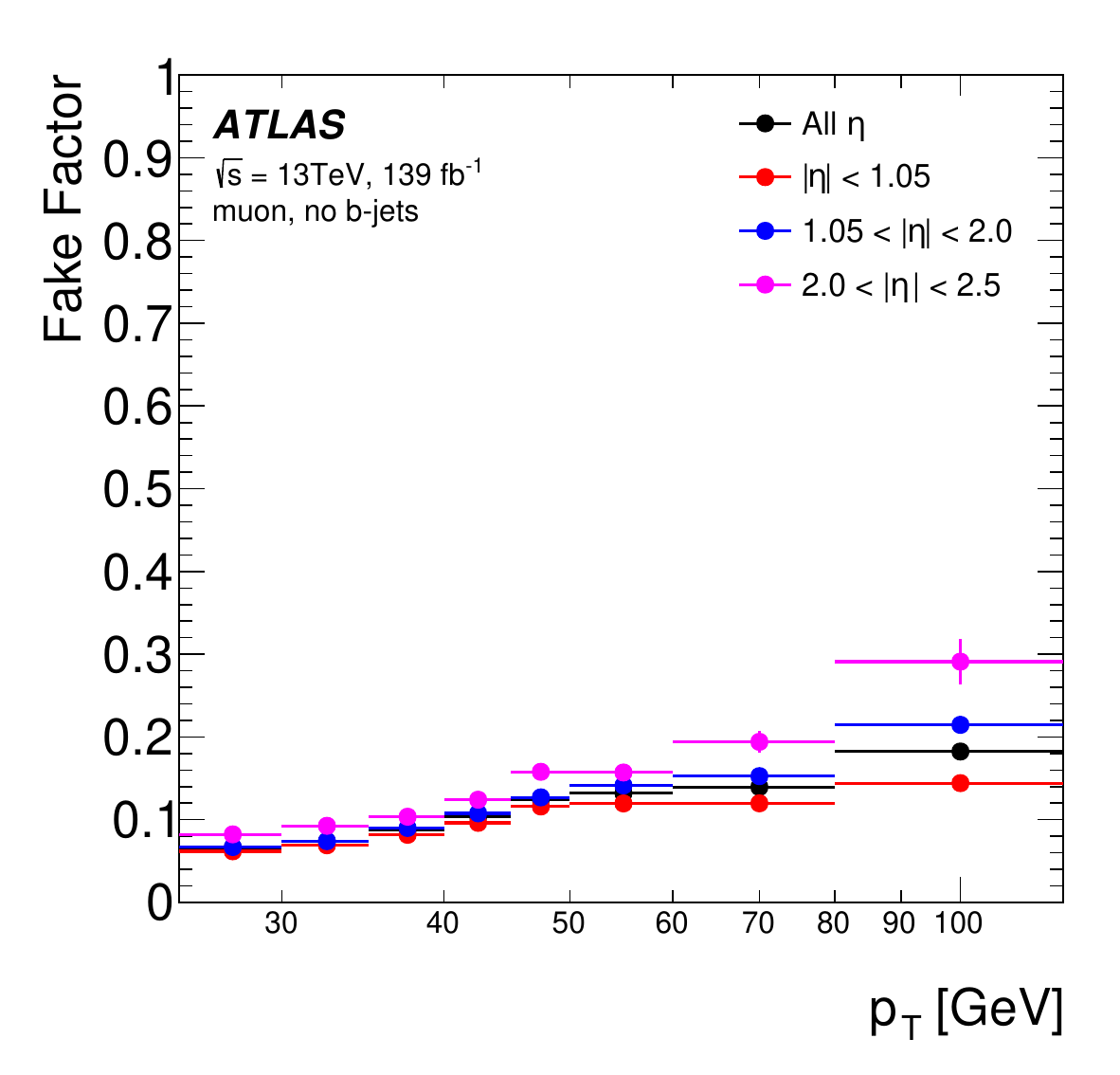}}\\
\subfloat[]{\includegraphics[width=0.43\textwidth]{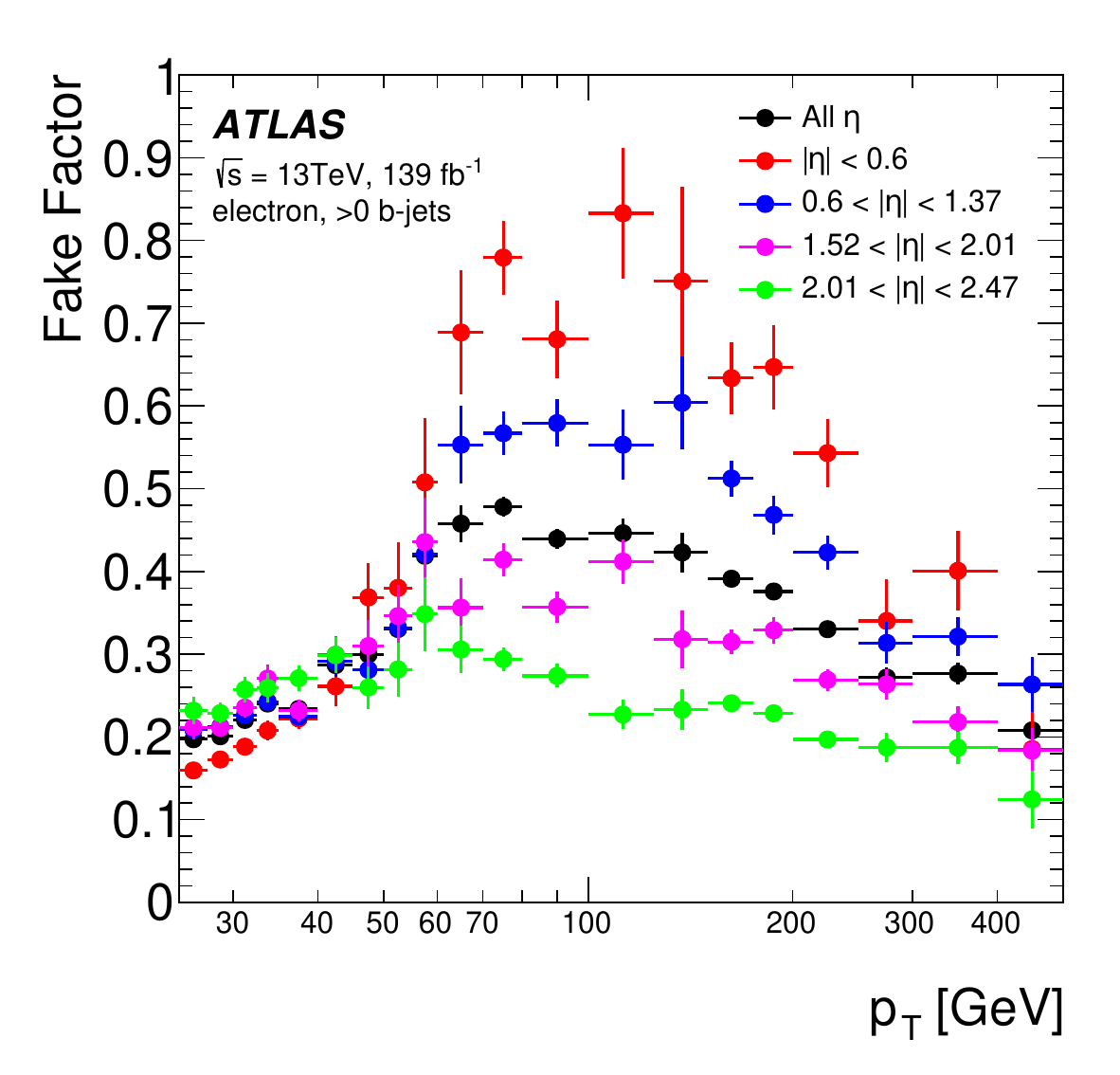}}
\subfloat[]{\includegraphics[width=0.43\textwidth]{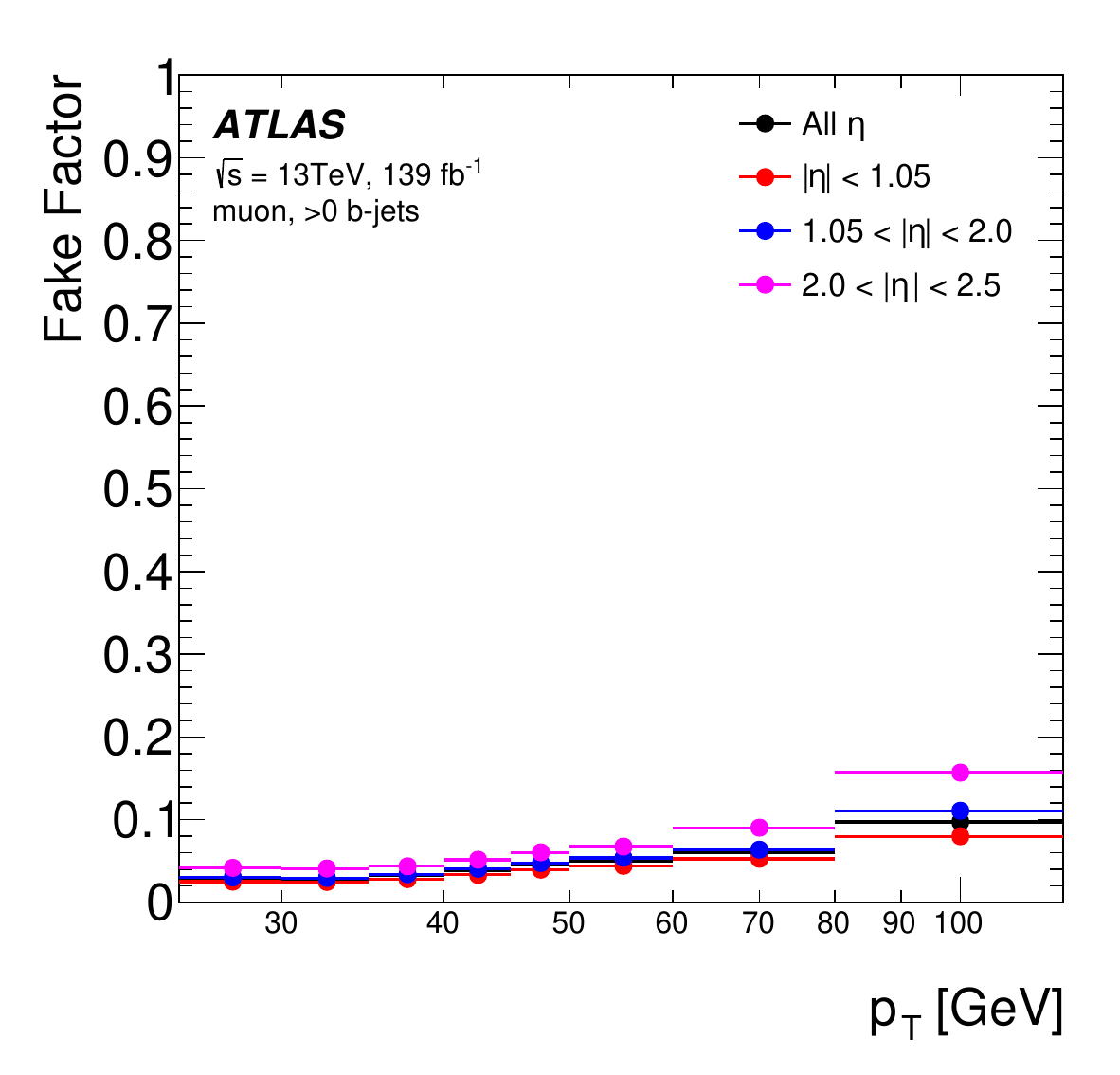}}
\caption{Fake factors measured in bins of the lepton candidate's \pt and \abseta for events with (a,c)~electron candidates and (b,d)~muon candidates in events (a,b)~without and (c,d)~with $b$-tagged jets.}
\label{fig:ml_ff_result_bb}
\end{figure}

Finally, a direct assessment of the uncertainty in the composition of the fake/non-prompt lepton background in the multi-lepton sample is made. Since fake/non-prompt leptons can come from
both light- and heavy-flavour sources, it is possible that the relative abundances from these sources can vary between samples. This possibility is addressed through
an additional uncertainty which leverages the different event signatures produced by light- and heavy-flavour hadrons, the latter consisting primarily of $b$-hadrons. To evaluate
this uncertainty, an alternative set of fake factors is computed which, in addition to the binning in \pt and \abseta, are binned according to the presence or absence
of $b$-tagged jets in the event. This alternative set of fake factors is shown in Figure~\ref{fig:ml_ff_result_bb}.
 
The composition uncertainty in the total event yields is then derived from the difference between estimates calculated with either
the nominal fake factors oblivious to the presence of $b$-tagged jets in the events, or the alternative set that requires such a jet. This uncertainty is found to have a negligible
impact on the analysis, since most of the jets in the signal region and in the sample used to measure the fake factors are not $b$-tagged.
 
\FloatBarrier

\subsubsection{Validation regions}
 
Validation regions are defined using appropriate sub-selections of $ee\mu$ and  $e\mu\mu$ events. These are used to check that the computed fake factors extrapolate correctly
from the regions where they are calculated to the regions in which they are applied. The \enquote{on-$Z$} validation region requires an SFOC lepton pair
with a dilepton mass within \SI{10}{\GeV} of the \Zboson boson mass. The \enquote{off-$Z$} validation region also requires a SFOC pair of leptons, but requires the dilepton
mass to fall outside of the \Zboson-mass window. Only mixed-flavour final states are selected for these validation regions so that which lepton to choose as the third one,
assumed to be the fake/non-prompt lepton, is unambiguous. The sources of fake/non-prompt leptons contribute in different ratios to the on-\Zboson and off-\Zboson validation regions:
the on-\Zboson validation region is more sensitive to \Zjets events than the off-\Zboson region, while the inverse is true for \ttbar events, although in absolute terms,
\Zjets events are more numerous than \ttbar events in both cases. Both validation regions target, through a \mT requirement of $\mT(\ell,\met) < \SI{40}{\GeV}$, a third lepton
that is likely to be fake/non-prompt. The union of the on-\Zboson and off-\Zboson validation regions is called the \enquote{fakes validation region}.
 
\begin{figure}[htbp]
\centering
\subfloat[]{\includegraphics[width=0.48\textwidth,keepaspectratio]{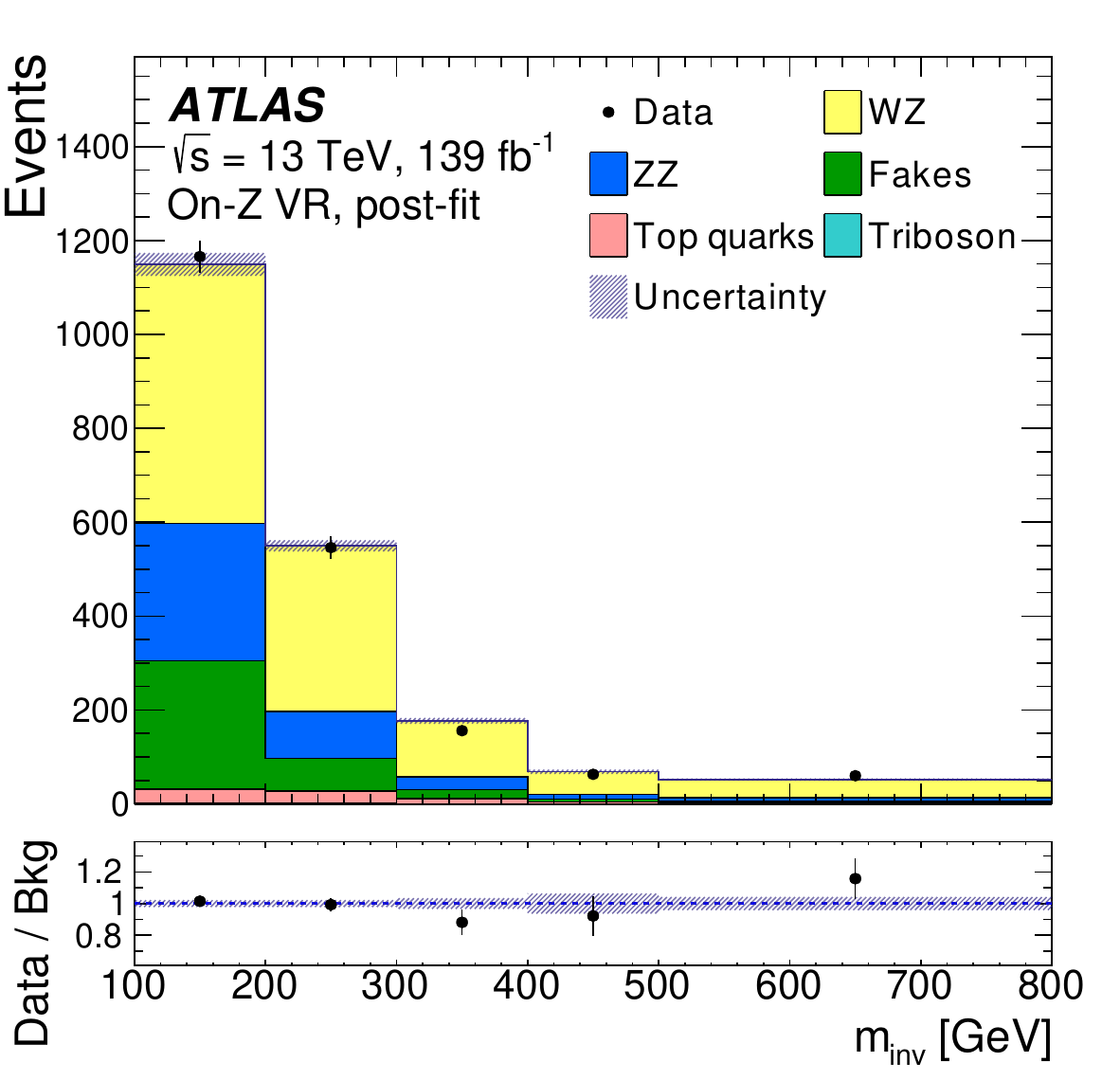}}
\subfloat[]{\includegraphics[width=0.48\textwidth,keepaspectratio]{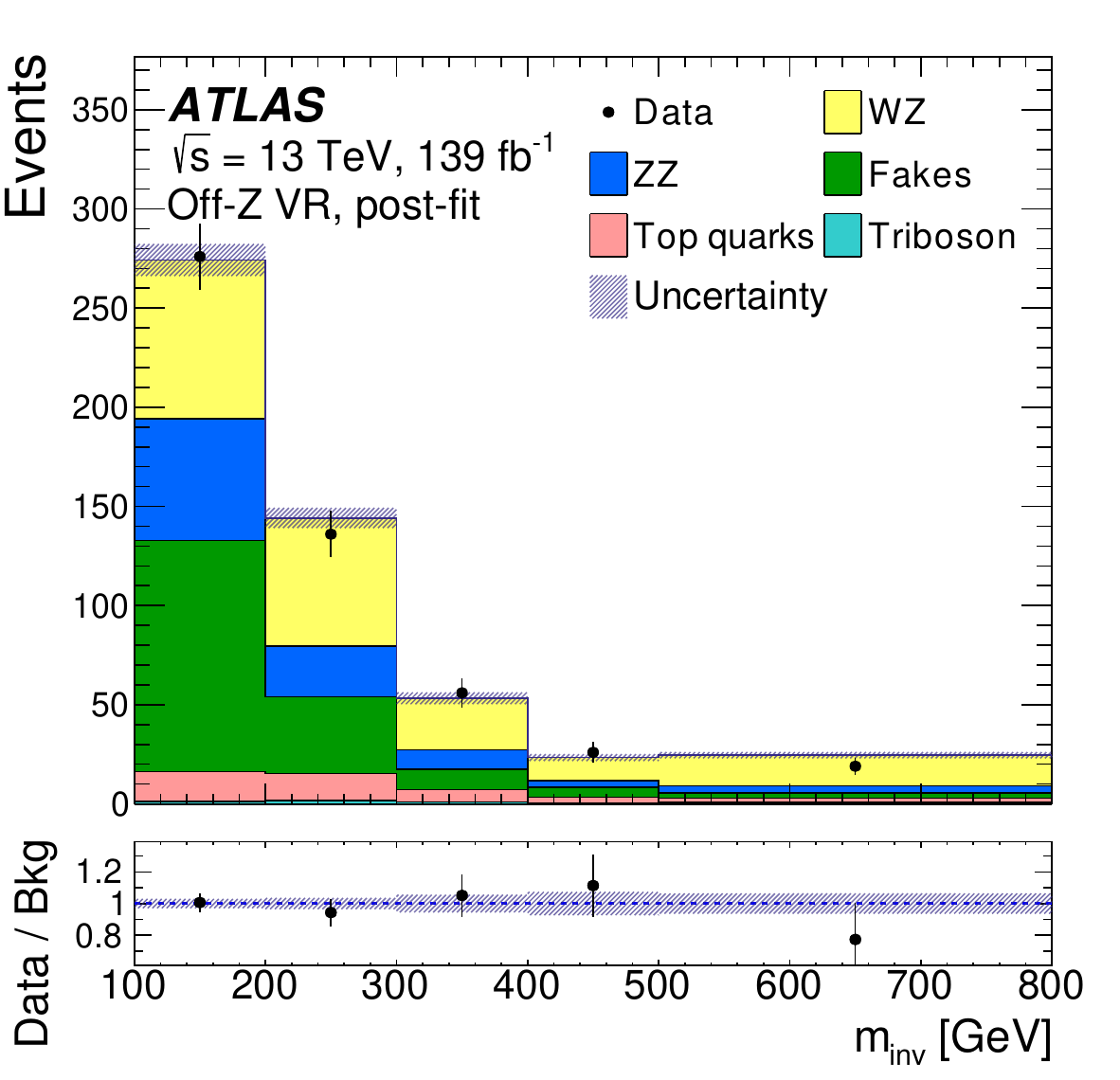}}
\caption{Comparison between data and prediction for the \minv distribution in the~(a)~on-\Zboson and~(b)~off-\Zboson validation regions, after fitting the
normalisation factors for the \WZjets and \ZZjets backgrounds and systematic uncertainties~\cite{EXOT-2019-36}. All uncertainties, systematic and statistical, are included.
The leftmost (rightmost) bin is inclusive and contains all events with $\minv < \SI{200}{\GeV }\hbox{ } (> \SI{500}{\GeV})$. The hatched grey area in these figures shows the total uncertainty. Further details are available in Ref.~\cite{EXOT-2019-36}.}
\label{fig:ml_fit_vr}
\end{figure}

\begin{figure}[htbp]
\centering
\subfloat[]{\includegraphics[width=0.48\textwidth,keepaspectratio]{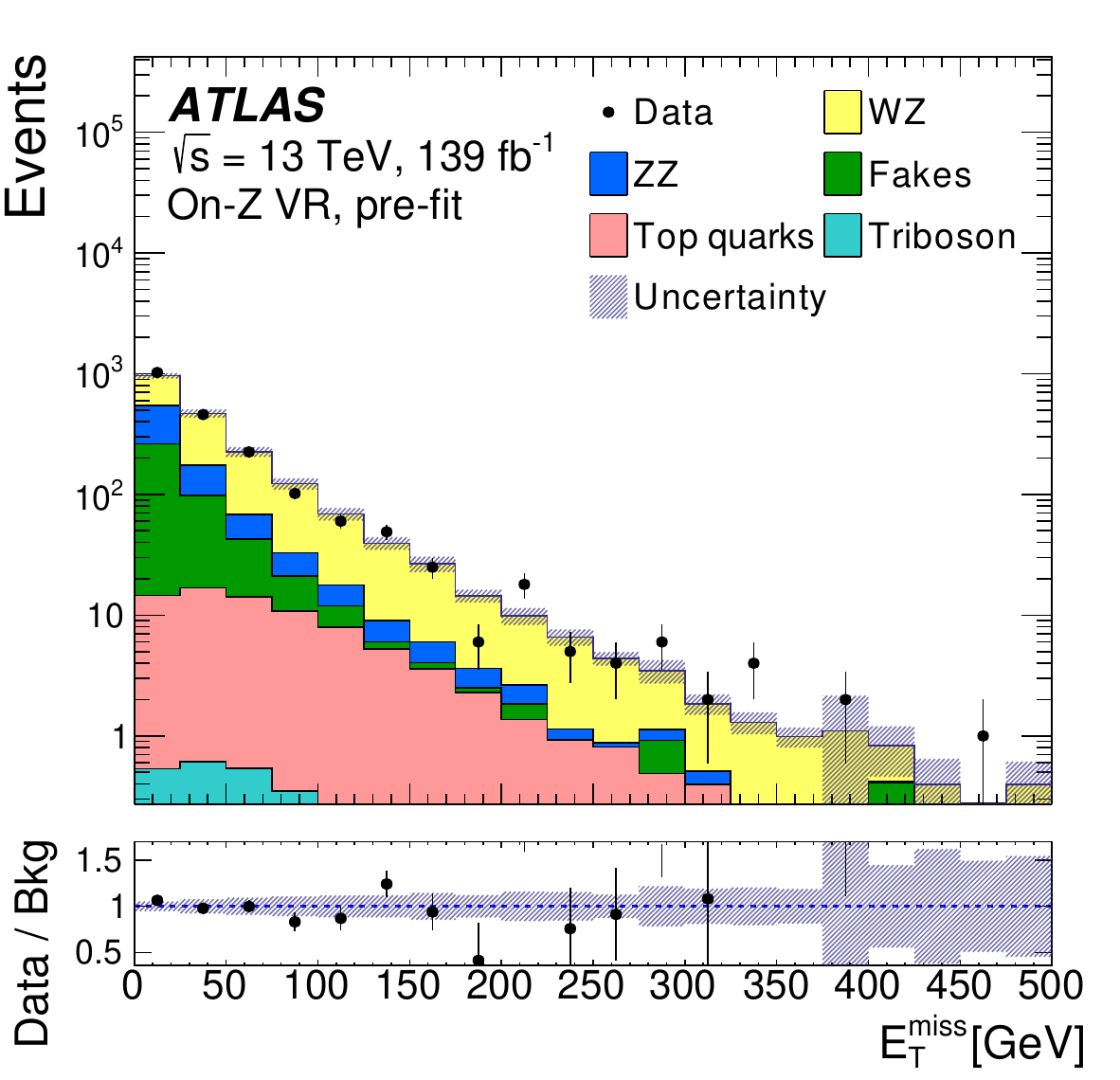}}
\subfloat[]{\includegraphics[width=0.48\textwidth,keepaspectratio]{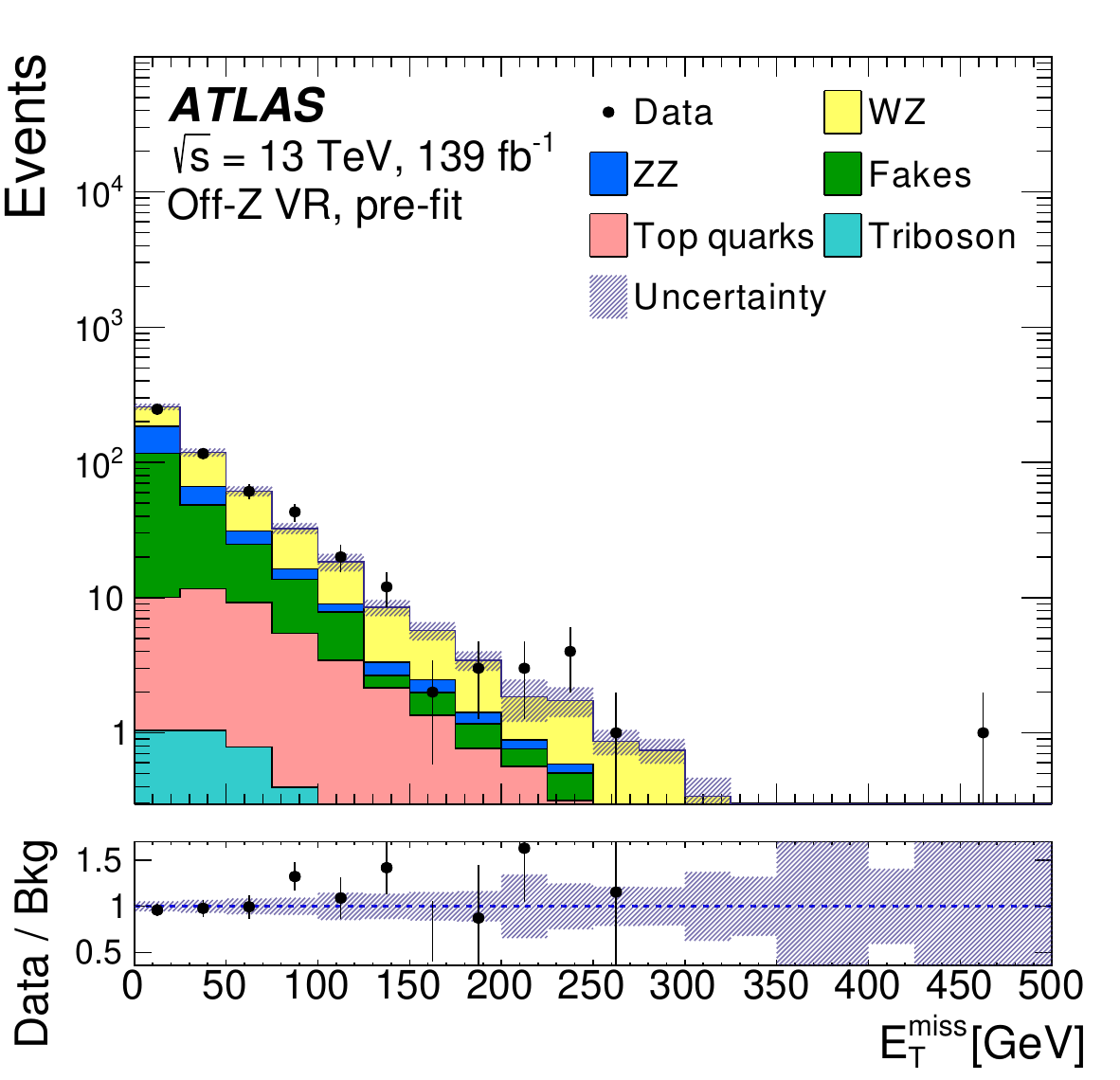}}
\caption{Comparison between data and prediction for the \met distribution in the~(a)~on-\Zboson and~(b)~off-\Zboson validation regions. All uncertainties,
systematic and statistical, are included. The rightmost bin is inclusive and contains all events with $\met > \SI{475}{\GeV}$. The hatched grey area in these figures shows the total uncertainty.  Further details are available in Ref.~\cite{EXOT-2019-36}.}
\label{fig:ml_fit_vr_met}
\end{figure}

\begin{figure}[htbp]
\centering
\subfloat[]{\includegraphics[width=0.48\textwidth,keepaspectratio]{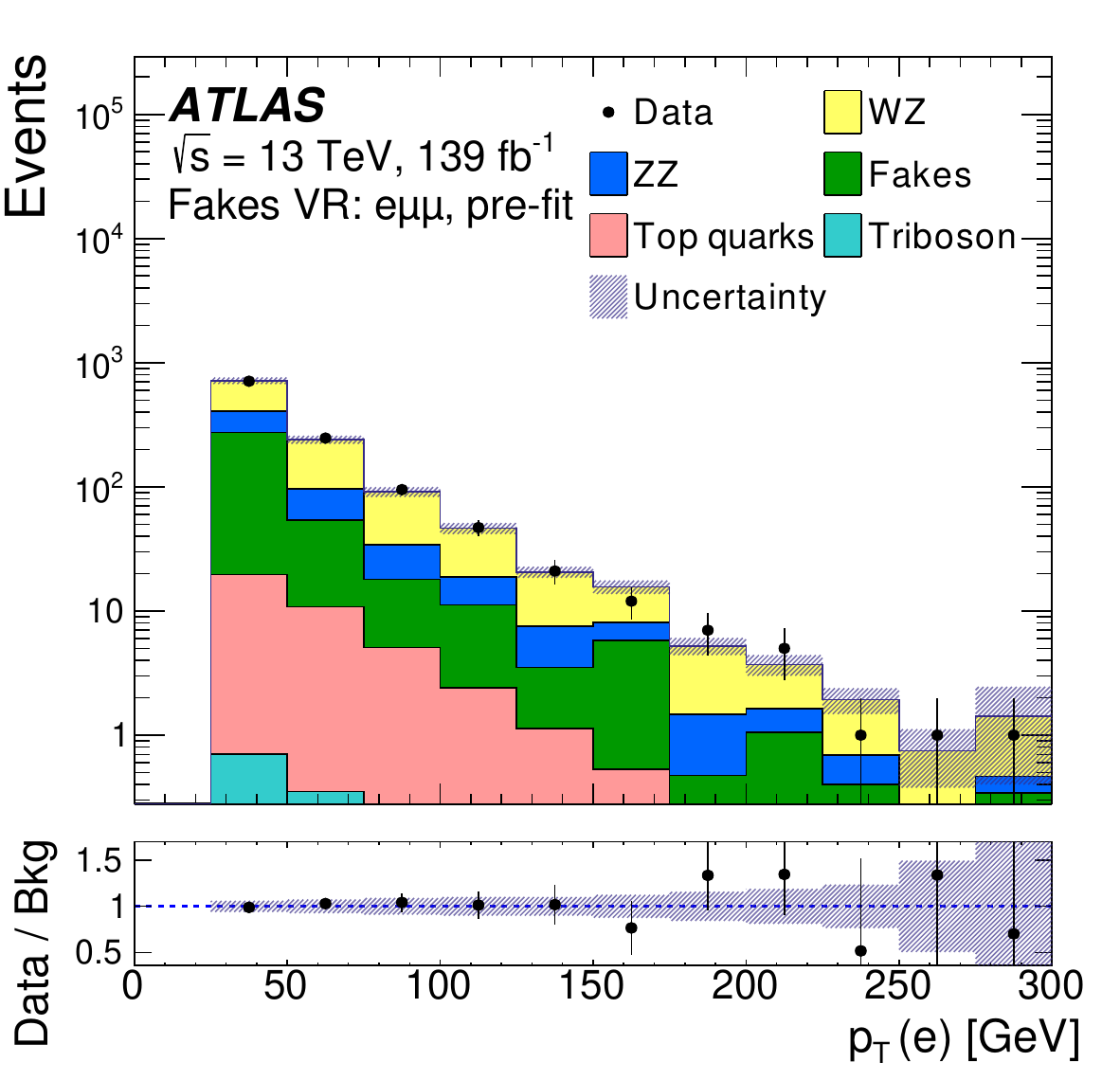}}
\subfloat[]{\includegraphics[width=0.48\textwidth,keepaspectratio]{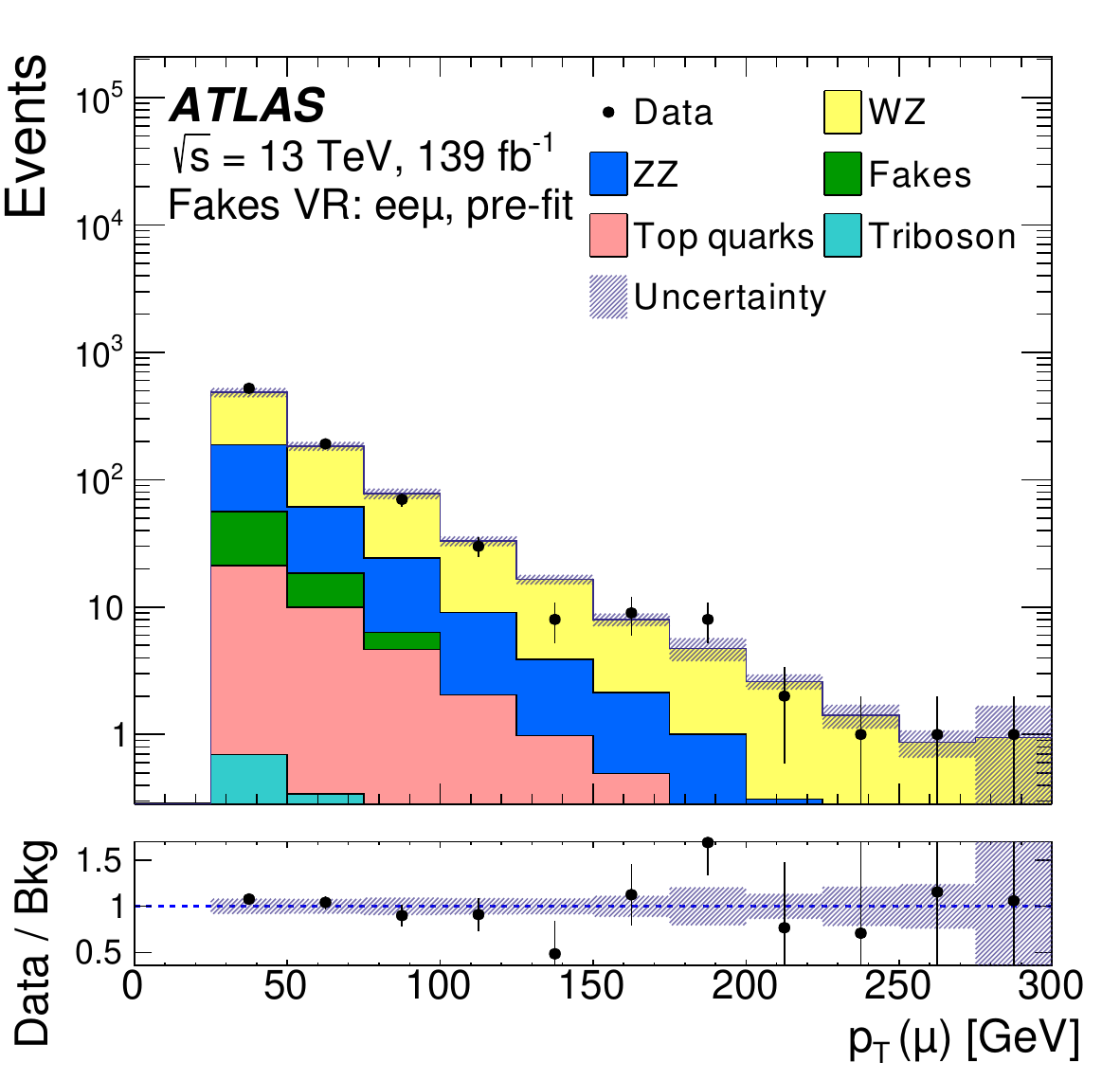}}
\caption{Comparison between data and prediction for the \pt distribution in the fakes validation region (which is the union of the on-\Zboson and off-\Zboson validation regions). Shown are~(a)~the electron
\pt in $e\mu\mu$ and~(b)~the muon \pt in $ee\mu$ events.  All
uncertainties, systematic and statistical, are included. The rightmost bin is inclusive and contains all events with $\pt > \SI{275}{\GeV}$. The hatched grey area in these figures shows the total uncertainty.  Further details are available in Ref.~\cite{EXOT-2019-36}.}
\label{fig:ml_fit_vr_pt}
\end{figure}

The variables of primary importance for this analysis are the invariant mass of all lepton candidates in the event~(\minv) and the \met. The signal regions are separated according to
the values of these quantities, as discussed in Ref.~\cite{EXOT-2019-36}. The \minv distributions in the two validation regions are shown in Figure~\ref{fig:ml_fit_vr}, while the \met
distributions are shown in Figure~\ref{fig:ml_fit_vr_met}. Lastly, the electron and muon candidate \pt distributions are shown in the fakes validation region in
Figure~\ref{fig:ml_fit_vr_pt}.
The comparisons in Figure~\ref{fig:ml_fit_vr} were also presented in Ref.~\cite{EXOT-2019-36} (albeit in logarithmic scale) and thus include fitted normalisation factors for the \WZjets and \ZZjets backgrounds from their respective control regions~(\enquote{post-fit}). On the other hand, the complementary distributions shown in Figures~\ref{fig:ml_fit_vr_met} and~\ref{fig:ml_fit_vr_pt} were  obtained independently of this statistical analysis, and thus employ the unconstrained SM background normalisations and uncertainties~(\enquote{pre-fit} distributions).  The background estimate is consistent with the data within the statistical and systematic uncertainties.

\FloatBarrier

\FloatBarrier

\section{Conclusions}

For physics analyses exploring signatures with one or more prompt leptons, background contributions due to fake/non-prompt leptons are often difficult to estimate in simulations.
Therefore, data-driven methods are commonly used. Three related methods have been adopted by the ATLAS Collaboration as recommended tools: the asymptotic matrix method, the Poisson likelihood
matrix method, and the fake-factor method. All three approaches depend on defining two categories of leptons, one of which~(\enquote{tight}) is subject to the same identification and
selection criteria as are used in the analysis. The other category~(\enquote{loose}) adds additional lepton candidates with less stringent selection requirements. The union of the two
sets is called the \enquote{baseline} sample. The criteria are typically defined such that the probability for a real baseline lepton to satisfy the tight criteria is substantially higher
than the corresponding probability for a fake/non-prompt lepton. Then, the relative numbers of loose and tight leptons in the analysis sample can be used to estimate the contribution
of fake/non-prompt leptons, either inclusively or differentially in any variables of interest.
 
Despite their similarities, the methods each have their own strengths and drawbacks. The asymptotic matrix method and fake-factor method provide a fake/non-prompt lepton weight for each event,
which is convenient for analyses. However, these methods are subject to large uncertainties if the efficiency for baseline fake/non-prompt leptons to satisfy the tight criteria is large in
parts of the analysis phase space. The Poisson likelihood matrix method returns a smaller uncertainty in such cases, and avoids any possibility of producing a negative estimate for the event
yield, but does not provide a per-event weight, introducing difficulties e.g.\ for differential estimations. The fake-factor method uses simulation rather than data to incorporate the
contribution from events where all leptons are tight.  This method can therefore be employed while the signal region for an analysis is fully blinded, although it may induce additional
simulation-related uncertainties in the background estimate.
 
The systematic uncertainties for all three methods arise from similar sources, with the largest contributions related to the extrapolation of the efficiencies measured in the control
samples to events in the analysis sample. Differences in the fake/non-prompt lepton composition in the samples must be accounted for and appropriate uncertainties must be assigned
to this extrapolation.
 
The performance of the Poisson likelihood matrix method and the fake-factor method has been demonstrated in a differential \ttZ cross section measurement and in a model-independent search
for BSM phenomena in multi-lepton final states, respectively. In both cases, the chosen method was shown to provide a reliable estimate of the fake/non-prompt lepton background, as
measured in validation regions.

 \FloatBarrier
 
\section*{Acknowledgements}
We thank CERN for the very successful operation of the LHC, as well as the support staﬀ from our institutions without whom ATLAS could not be operated effciently.
We acknowledge the support of ANPCyT, Argentina; YerPhI, Armenia; ARC, Australia; BMWFW and FWF, Austria; ANAS, Azerbaijan; SSTC, Belarus; CNPq and FAPESP, Brazil; NSERC, NRC and CFI, Canada; CERN; ANID, Chile; CAS, MOST and NSFC, China; Minciencias, Colombia; MEYS CR, Czech Republic; DNRF and DNSRC, Denmark; IN2P3-CNRS and CEA-DRF/IRFU, France; SRNSFG, Georgia; BMBF, HGF and MPG,Germany; GSRI, Greece; RGC and Hong Kong SAR, China; ISF and Benoziyo Center, Israel; INFN, Italy; MEXT and JSPS, Japan; CNRST, Morocco; NWO, Netherlands; RCN, Norway; MEiN,Poland; FCT, Portugal; MNE/IFA, Romania; JINR; MES of Russia and NRC KI, Russian Federation; MESTD, Serbia; MSSR, Slovakia; ARRS and MIZ\v{S}, Slovenia; DSI/NRF, South Africa; MICINN, Spain; SRC and Wallenberg Foundation, Sweden; SERI, SNSF and Cantons of Bern and Geneva, Switzerland; MOST, Taiwan; TENMAK, T\"urkiye; STFC, United Kingdom; DOE and NSF, United States of America.

In addition, individual groups and members have received support from BCKDF, CANARIE, Compute Canada and CRC, Canada; COST, ERC, ERDF, Horizon 2020 and Marie Sk{\l}odowska-Curie Actions, European Union; Investissements d’Avenir Labex, Investissements d’Avenir Idex and ANR, France; DFG and AvH Foundation, Germany; Herakleitos, Thales and Aristeia programmes co-financed by EU-ESF and the Greek NSRF, Greece; BSF-NSF and GIF, Israel; Norwegian Financial Mechanism 2014-2021,
Norway; NCN and NAWA, Poland; La Caixa Banking Foundation, CERCA Programme Generalitat de Catalunya and PROMETEO and GenT Programmes Generalitat Valenciana, Spain; G\"{o}ran Gustafssons Stiftelse, Sweden; The Royal Society and Leverhulme Trust, UnitedKingdom.

The crucial computing support from all WLCG partners is acknowledged gratefully, in particular from CERN, the ATLAS Tier-1 facilities at TRIUMF (Canada), NDGF (Denmark, Norway, Sweden), CC-IN2P3 (France), KIT/GridKA (Germany), INFN-CNAF (Italy), NL-T1 (Netherlands), PIC (Spain), ASGC (Taiwan), RAL (UK) and BNL (USA), the Tier-2 facilities worldwide and large non-WLCG resource providers. Major contributors of computing resources are listed in Ref.~\cite{ATL-SOFT-PUB-2023-001}.

\printbibliography
\clearpage
 
\begin{flushleft}
\hypersetup{urlcolor=black}
{\Large The ATLAS Collaboration}

\bigskip

\AtlasOrcid[0000-0002-6665-4934]{G.~Aad}$^\textrm{\scriptsize 102}$,
\AtlasOrcid[0000-0002-5888-2734]{B.~Abbott}$^\textrm{\scriptsize 120}$,
\AtlasOrcid[0000-0002-7248-3203]{D.C.~Abbott}$^\textrm{\scriptsize 103}$,
\AtlasOrcid[0000-0002-1002-1652]{K.~Abeling}$^\textrm{\scriptsize 55}$,
\AtlasOrcid[0000-0002-8496-9294]{S.H.~Abidi}$^\textrm{\scriptsize 29}$,
\AtlasOrcid[0000-0002-9987-2292]{A.~Aboulhorma}$^\textrm{\scriptsize 35e}$,
\AtlasOrcid[0000-0001-5329-6640]{H.~Abramowicz}$^\textrm{\scriptsize 151}$,
\AtlasOrcid[0000-0002-1599-2896]{H.~Abreu}$^\textrm{\scriptsize 150}$,
\AtlasOrcid[0000-0003-0403-3697]{Y.~Abulaiti}$^\textrm{\scriptsize 117}$,
\AtlasOrcid[0000-0003-0762-7204]{A.C.~Abusleme~Hoffman}$^\textrm{\scriptsize 137a}$
% ...
\bigskip
\\

% ...
$^{29}$Physics Department, Brookhaven National Laboratory, Upton NY; United States of America.\\
% ...
$^{35}$$^{(a)}$Facult\'e des Sciences Ain Chock, R\'eseau Universitaire de Physique des Hautes Energies - Universit\'e Hassan II, Casablanca;$^{(b)}$Facult\'{e} des Sciences, Universit\'{e} Ibn-Tofail, K\'{e}nitra;$^{(c)}$Facult\'e des Sciences Semlalia, Universit\'e Cadi Ayyad, LPHEA-Marrakech;$^{(d)}$LPMR, Facult\'e des Sciences, Universit\'e Mohamed Premier, Oujda;$^{(e)}$Facult\'e des sciences, Universit\'e Mohammed V, Rabat;$^{(f)}$Institute of Applied Physics, Mohammed VI Polytechnic University, Ben Guerir; Morocco.\\
% ...
$^{55}$II. Physikalisches Institut, Georg-August-Universit\"{a}t G\"ottingen, G\"ottingen; Germany.\\
% ...
$^{102}$CPPM, Aix-Marseille Universit\'e, CNRS/IN2P3, Marseille; France.\\
$^{103}$Department of Physics, University of Massachusetts, Amherst MA; United States of America.\\
% ...
$^{117}$Department of Physics, New York University, New York NY; United States of America.\\
% ...
$^{120}$Homer L. Dodge Department of Physics and Astronomy, University of Oklahoma, Norman OK; United States of America.\\
% ...
$^{137}$$^{(a)}$Departamento de F\'isica, Pontificia Universidad Cat\'olica de Chile, Santiago;$^{(b)}$Millennium Institute for Subatomic physics at high energy frontier (SAPHIR), Santiago;$^{(c)}$Instituto de Investigaci\'on Multidisciplinario en Ciencia y Tecnolog\'ia, y Departamento de F\'isica, Universidad de La Serena;$^{(d)}$Universidad Andres Bello, Department of Physics, Santiago;$^{(e)}$Instituto de Alta Investigaci\'on, Universidad de Tarapac\'a, Arica;$^{(f)}$Departamento de F\'isica, Universidad T\'ecnica Federico Santa Mar\'ia, Valpara\'iso; Chile.\\
% ...
$^{150}$Department of Physics, Technion, Israel Institute of Technology, Haifa; Israel.\\
$^{151}$Raymond and Beverly Sackler School of Physics and Astronomy, Tel Aviv University, Tel Aviv; Israel.\\
% ...
$^{a}$ Also Affiliated with an institute covered by a cooperation agreement with CERN.\\
$^{b}$ Also at An-Najah National University, Nablus; Palestine.\\
$^{c}$ Also at Borough of Manhattan Community College, City University of New York, New York NY; United States of America.\\
$^{d}$ Also at Bruno Kessler Foundation, Trento; Italy.\\
$^{e}$ Also at Center for High Energy Physics, Peking University; China.\\
$^{f}$ Also at Center for Interdisciplinary Research and Innovation (CIRI-AUTH), Thessaloniki ; Greece.\\
% ...

\end{flushleft}


\clearpage

\end{document}